\documentclass[preprint2]{aastex}

\begin{document}

\defcitealias{smith18}{Paper~I}



\title{The Hot Gas Exhaust of Starburst Engines in Mergers: \\ Testing
Models of Stellar Feedback and Star Formation Regulation}

\author{Beverly J. Smith}
\affil{Department of Physics and Astronomy, East Tennessee State University,
Johnson City TN 37614, USA}

\author{Peter Wagstaff}
\affil{Department of Physics and Astronomy, East Tennessee State University
Johnson City TN 37614, USA}

\author{Curtis Struck}
\affil{Department of Physics and Astronomy, Iowa State University,
Ames IA 50011, USA}

\author{Roberto Soria}
\affil{College of Astronomy and Space Sciences, University of the Chinese Academy of Sciences, 
Beijing 100049, China;
Sydney Institute for Astronomy, School of Physics A28, The University of Sydney, Sydney, NSW 2006, Australia}

\author{Brianne Dunn}
\affil{Department of Physics and Astronomy, East Tennessee State University,
Johnson City TN 37614, USA;
Now at Department of Physics and Astronomy, Clemson University, Clemson South Carolina 29634}

\author{Douglas Swartz}
\affil{Astrophysics Office, NASA Marshall Space Flight Center, ZP12, Huntsville AL 35812, USA}

\author{Mark L. Giroux}
\affil{Department of Physics and Astronomy, East Tennessee State University
Johnson City TN 37614, USA}


\begin{abstract}

Using archival data from the Chandra X-ray telescope, we have 
measured the spatial extent of the hot interstellar gas in a sample
of 
49
nearby
interacting galaxy pairs, mergers, and merger remnants.
For systems with SFR $>$ 1~M$_{\sun}$~yr$^{-1}$,
the volume and mass of hot gas are strongly and linearly correlated with the
star formation rate (SFR). 
This supports 
the idea that stellar/supernovae feedback dominates the production
of hot gas in these galaxies.
We compared 
the mass of X-ray-emitting hot gas M$_{\rm X}$(gas) 
with the molecular and
atomic hydrogen interstellar gas masses in these galaxies
(M$_{\rm H_2}$ and M$_{\rm HI}$, respectively),
using published carbon
monoxide 
and 21 cm HI measurements.
Systems with higher SFRs
have larger M$_{\rm X}$(gas)/(M$_{\rm H_2}$ + M$_{\rm HI}$) ratios on average,
in agreement with recent numerical simulations of star formation
and feedback in merging galaxies.
The M$_{\rm X}$(gas)/(M$_{\rm H_2}$ + M$_{\rm HI}$) ratio also increases
with dust temperature on average.
The ratio M$_{\rm X}$(gas)/SFR is anti-correlated 
with the IRAS 60 $\mu$m to 100 $\mu$m flux ratio 
and with the Spitzer 3.6 $\mu$m to 24 $\mu$m color. 
These trends may be due to
variations in 
the 
spatial density of young stars,
the 
stellar age, 
the ratio of young to old stars, the initial mass function, 
and/or 
the 
efficiency of stellar feedback.
Galaxies with low SFR ($<$1 M$_{\sun}$~yr$^{-1}$) 
and high K band luminosities
may have an excess of hot gas relative to the relation for higher SFR
galaxies, while galaxies with low K band luminosities (and therefore
low stellar masses) may have a deficiency in hot gas, but 
our sample is not large enough for strong
statistical significance.

\end{abstract}



\vfill
\eject

\section{Introduction }

Feedback from stellar winds, radiation pressure, 
and supernovae 
play a major role in regulating star formation, by heating, 
ionizing, and accelerating the
interstellar gas and adding turbulence. 
However, the details of these processes are not well-understood.
Computer simulations are frequently used
to study stellar feedback and star formation,
using various prescriptions
to model the feedback.  
These processes are complicated to model because 
different feedback mechanisms 
help regulate star formation in different ways, 
multiple mechanisms operate simultaneously, and the different
mechanisms affect
each other.  Radiation pressure
from young stars
disrupts molecular clouds, 
decreasing the amount of dense gas and 
preventing too-rapid
gravitational collapse of clouds \citep{hopkins11, hopkins13b, hopkins14},
while shock heating
by supernovae and stellar winds
is responsible for most of the hot X-ray-emitting
gas in galaxies \citep{hopkins12b}.
Before supernovae begin in a young star-forming region,
radiation pressure and stellar winds 
clear out dense gas in star forming regions, heating and stirring
the gas; later
supernovae thus occur in lower density gas, causing the hot gas 
to survive longer and inhibiting subsequent star formation
\citep{hopkins12a, agertz13}.
The more efficient the
feedback, the lower
the efficiency of subsequent star formation \citep{cox06c, hopkins13b}.
Supernovae provide both thermal energy, heating the gas,
as well as kinetic feedback, which increases turbulence and thus
affects
later star formation
\citep{springel00, hopkins14}.
Another way feedback regulates
star formation is by removing gas from the main disk
of the galaxy, either temporarily or permanently (e.g., \citealp{muratov15}).
Supernova-driven winds may drive gas out into the halo; this 
hot halo material may then cool and fall back in on the galaxy, triggering
delayed star formation \citep{hopkins13a}.
Winds due to supernovae may remove gas from the galaxy entirely;
in some simulations 
the mass loss rate from supernovae-driven winds 
is greater than the star formation rate (SFR) \citep{hopkins12a, hopkins13a}.

The latest generation of simulations include multi-phase interstellar
gas, to follow both the dense cores of molecular clouds where star
formation occurs, the warmer atomic gas, and the hot intracloud
medium \citep{hopkins13b, hopkins14, renaud13, renaud14, 
renaud15, renaud19, sparre16, fensch17, 
moreno19}.
The results of such simulations sometimes 
depend upon the resolution 
of the simulation and the details of the calculations,
with higher resolution models 
producing more efficient 
star formation \citep{teyssier10, hopkins13a, hayward14, sparre16},
and 
the duration and intensity of a starburst 
depending 
upon the 
prescription for 
feedback assumed in the model \citep{hopkins12a, fensch17}.
How stellar feedback is implemented in these codes has profound cosmological
consequences.   Stellar feedback is needed in cosmological
simulations of galaxy formation and evolution to explain the
observed 
galaxy mass function \citep{keres9}, the
galaxy stellar mass-halo mass relation 
\citep{hopkins14,
agertz15, agertz16,
trujillo15} and the galaxian mass-metallicity relation \citep{
finlator08, ma16, torrey19}.
For cosmological models to reproduce the so-called galaxy
main sequence 
(a correlation between stellar
mass and star formation; 
\citealp{brinchmann04, noeske2007, salim07}) 
or the Kennicutt-Schmidt Law (a relation
between SFR and molecular gas mass; \citealp{schmidt59, kennicutt98, kennicutt12}), stellar feedback is 
necessary
\citep{hopkins14, orr18}.

To test these feedback models, X-ray observations are required.
With high resolution X-ray imaging, the distribution, temperature, and mass
of the hot gas within galaxies can be studied, and compared to
other properties of the galaxies.
In star-forming galaxies, the bulk of the hot gas is attributed to feedback
from Type II supernova and young stars
\citep{strickland00, strickland04a, strickland04b, grimes05, owen09, li13, mineo12, smith18}.
\citet{hopkins12a} 
model the X-ray production 
due to stellar feedback in different types of galaxies,
and conclude that for normal spirals and dwarf galaxies, supernovae
and stellar winds dominate, but in intense starbursts radiation
pressure dominates.
The soft X-rays from galactic winds originate from a small fraction of the total
hot gas; the bulk of the hot gas is such low density it is difficult to observe
directly \citep{strickland_stevens00}.
Freely-flowing hot gas produces little X-ray emission, in contrast to 
hot gas confined by surrounding cooler gas \citep{hopkins12a}.
Observational studies show that for star-forming
galaxies, the X-ray luminosity from hot gas,
L$_{\rm X}$(gas), is proportional to the SFR \citep{strickland04b, grimes05, mineo12, 
smith18}.  This is in contrast to some theoretical estimates,
which predict that 
L$_{\rm X}$(gas) should be proportional to SFR$^2$ 
\citep{chevalier85, zhang14}.
More modern theoretical calculations including gravitational
forces and improved radiative cooling 
are able to reproduce the observed
L$_{\rm X}$(gas) $\propto$ SFR 
relation for star-forming galaxies if the mass-loading 
factor (mass outflow rate/SFR) decreases as SFR increases 
\citep{bustard16, meiksin16}.   The 
recent cosmological hydrodynamical simulations 
of \citet{vandevoort16}
including feedback
find a constant
L$_{\rm X}$(gas)/SFR 
ratio for galaxies with 
halo masses between 10$^{10.5}$ $-$ 10$^{12}$ M$_{\sun}$, 
where the Milky Way has a halo
mass of $\sim$ 10$^{12}$ M$_{\sun}$.

Over timescales of many gigayears,
virialization of gas provided by stellar mass loss from older stars 
can contribute 
to the X-ray-emitting hot gas in galaxies, particularly 
in massive galaxies with low SFRs
(e.g., \citealp{ciotti91, pellegrini98, mathews03}).
In quiescent early-type galaxies, this contribution dominates,
as L$_{\rm X}$(gas) increases with mass rather than with SFR
(e.g., \citealp{osullivan01, kim13, su15, goulding16}).
The possible existence of this additional source of hot gas 
may need to be taken into account in interpreting X-ray data in
terms of stellar feedback, particularly in galaxies 
with low SFRs and high masses.

In the current study, our goal is to track the evolution of the hot gas 
in galaxies compared
to the other components of the galaxies, particularly the molecular
and atomic 
gas, and compare with expectations from theoretical models.
This study is a follow-up to 
our earlier archival Chandra study of 49 nearby major mergers
in a range of merger stages 
(\citealp{smith18}, hereafter Paper~I).
In the earlier study,
we removed the resolved point sources and 
extracted the spectrum of the diffuse X-ray emission.
We then separated this spectrum into a power law
and a thermal component and corrected for internal absorption.
Assuming the thermal component was due to hot gas, we compared
the thermal luminosity  
L$_{\rm X}$(gas) with the global SFR as derived from UV/optical data.
Although there is considerable system-to-system variation in the
L$_{\rm X}$(gas)/SFR ratio,
we did not see any trends of L$_{\rm X}$(gas)/SFR with merger stage,
active galactic nuclei (AGN) activity, or
SFR for galaxies with SFR $>$ 1~M$_{\sun}$~yr$^{-1}$.
These results suggest that in star-forming galaxies, stellar feedback
reaches an approximately steady-stage condition. 
In Paper~I, we concluded that, for star forming galaxies,
about 2\% of the total energy output from supernovae and stellar winds is
converted into X-ray flux; this result is in agreement with earlier
results from smaller samples of galaxies \citep{grimes05, mineo12}.

In the current study, we revisit the same sample of mergers,
and use the Chandra data to derive the spatial extent of the hot
gas in these galaxies and therefore 
the mass of hot X-ray-emitting gas M$_{\rm X}$(gas).
We compare M$_{\rm X}$(gas) with the amount of cold molecular
and atomic hydrogen gas in these galaxies, as obtained from published
carbon monoxide and 21 cm HI observations.  
Our goal is to better understand how interstellar gas cycles
between hot and cold phases due to star formation and stellar feedback,
and how this cycle affects the efficiency of star formation (SFE).

In Section 2 of this paper, we review the selection of the
sample and the available ultraviolet, infrared, and optical data.
In Section 3, we explain the molecular and atomic hydrogen gas data.
In Section 4, we determine the spatial extent
of the hot gas in the galaxies. We obtain the volume and mass 
of hot X-ray-emitting gas and the electron density in Section 5.
These values are then compared with other parameters of the systems
in Section 6.   The results 
are discussed in Section 7, and conclusions are provided in Section 8.

\section{Sample Selection and UV/IR/Optical Data}

The sample selection is described in detail in Paper~I.
Briefly, the sample includes 49 pre-merger interacting pairs, post-merger
remnants, and mid-merger systems 
in the nearby Universe (distance $<$ 180 Mpc).  
Initially, galaxies were chosen based on their morphologies
from the \citet{arp66} Atlas of Peculiar
Galaxies, or from other published surveys of mergers and merger remnants,
selecting 
approximately equal-mass interacting pairs or the remnants of 
the merger of such pairs.
The final sample was then selected based on the availability of 
suitable 
Chandra data.
See Paper~I for details.

The sample of galaxies is given in Table 1.
Table 1 also provides 
basic data on these systems from Paper~I, including
distances assuming a Hubble constant of 73 km~s$^{-1}$~Mpc$^{-1}$,
correcting for peculiar velocities due to the Virgo Cluster,
the Great Attractor, and the Shapley Supercluster.  
The median distance for our sample galaxies is 51.5 Mpc.
Table 1 also provides 
the far-infrared luminosity L$_{\rm FIR}$
and the near-infrared K band luminosity L$_{\rm K}$,
obtained 
from Infrared Astronomical Satellite (IRAS) and 2-micron All-Sky 
Survey (2MASS) data, respectively,
as described in Paper~I.
In addition, Table 1 includes SFRs, derived from a combination of Spitzer infrared
and GALEX UV photometry as described in Paper~I.
When available, the far-UV (FUV) is used; otherwise, near-UV (NUV) photometry
is used.
These SFRs correspond to the SFR averaged over a time period
of $\sim$100 Myrs \citep{kennicutt12}.
Table 1 also identifies the 13 galaxies in the sample that are classified
in the NASA Extragalactic Database 
(NED\footnote{http://ned.ipac.caltech.edu})
as Seyfert 1, Seyfert 2, or Low Ionization Nuclear Emission Region (Liner)
galaxies.
Detailed descriptions of the individual galaxies in the sample
are provided in the Appendix of Paper~I.

Based on their morphologies, in Paper~I we classified the systems
into seven merger stages. 
These stages are: (1) separated but interacting pair with
small or no tails, (2) separated pair with moderate or long tails,
(3) pair with disks in contact, (4) common envelope, two nuclei,
and tails, (5) single nucleus and two strong tails, 
(6), single nucleus but weak tails, and (7) disturbed
elliptical with little or no tails.
The staging is approximate, with an uncertainty of $\pm$1 stage.

In Figure 1, various properties 
of the galaxies 
(distance,
L$_{\rm FIR}$, the L$_{\rm FIR}$/L$_{\rm K}$ ratio, and the IRAS
60 $\mu$m to 100 $\mu$m flux ratio F$_{60}$/F$_{100}$)
are plotted against the merger stage.
The black open squares in
Figure 1 are the data for the individual galaxies; the blue filled diamonds that are offset slightly
to the left of the stage show the median value for that stage.  The errorbars on the blue diamonds show
the semi-interquartile range, equal to half the difference between
the 75th percentile and the 25th percentile.
As discussed in Paper~I, this sample is inhomogeneous because it
was selected based on the availability of archival Chandra data.
As illustrated in Figure 1, the sample has some biases.
The galaxies in the middle of the merger sequence tend to
be more distant and so tend to have higher FIR luminosities.
This means they have 
higher SFRs, since 
L$_{\rm FIR}$ is an approximate
measure of the SFR
for galaxies with high SFRs (e.g., \citealp{kennicutt98, kennicutt12}).
The late-stage mergers tend to be closer and have lower L$_{\rm FIR}$.
Late-stage mergers are difficult to identify at large distances, thus
confirmed
examples tend to be nearby.

The late-stage mergers 
also
tend to have lower L$_{\rm FIR}$/L$_{\rm K}$ ratios.
This ratio is an approximate measure of 
the specific SFR (sSFR), defined as the SFR/stellar mass, since 
the K band luminosity L$_{\rm K}$ is an 
approximate measure of the stellar mass
\citep{maraston98, bell00, into2013, andreani2018}, 
although it is affected by age 
and possible AGN contributions.
The mid-merger systems also tend to have higher
dust temperatures, as traced by the IRAS F$_{60}$/F$_{100}$ ratio
(last panel Figure 1).   
The uncertainty in the staging,
the biases in the sample, and the 
small number
of systems in each stage means trends
with merger stage 
are uncertain. 
As seen in Figure 1, the AGN tend to be mid-merger systems with
high L$_{\rm FIR}$ and F$_{60}$/F$_{100}$. 
Although AGN can contribute to the heating of interstellar dust in galaxies, 
for most of our 
AGNs published studies
of the IR spectra of the galaxies conclude that dust heating is dominated
by star formation rather than the AGN (see the detailed discussions
on the individual galaxies in the Appendix of Paper~I).

Figure 2 displays some well-known correlations between these basic
parameters.
The observed correlation between 
SFR and L$_{\rm K}$ (top left panel) or its
equivalent has been seen many times before
for star forming galaxies (e.g., \citealp{smith98, andreani2018}).
This relation is a consequence of 
the correlation 
between SFR and stellar mass, which  
is known as the `galaxy main sequence' for star forming
galaxies 
(e.g., \citealp{brinchmann04, salim07}).
For our sample, this correlation is only a weak correlation, because
of the inclusion of some systems with low SFRs compared to L$_{\rm K}$.
Galaxies with low SFR compared to the best-fit `galaxy main sequence' relation
are considered quenched, 
quenching, or post-starburst galaxies. 
In our sample, our post-starburst galaxies are all late-stage mergers,
and have low 
L$_{\rm FIR}$/L$_{\rm K}$ ratios.

Figure 2 shows that the SFR is correlated with 
both
L$_{\rm FIR}$/L$_{\rm K}$ 
and the Spitzer [3.6 $\mu$m] $-$ [24 $\mu$m] color
for our sample galaxies\footnote{
[3.6] $-$ [24] is defined as the 
magnitude in the 3.6 $\mu$m filter minus that in the 24 $\mu$m filter,
using zero magnitude flux densities of 277.5 Jy and 7.3 Jy, respectively.}.
The majority of our galaxies fall in a narrow range of 
L$_{\rm FIR}$/L$_{\rm K}$,
$-$1 $\le$ log L$_{\rm FIR}$/L$_{\rm K}$ $\le$ 0, but a handful have
lower 
L$_{\rm FIR}$/L$_{\rm K}$ ratios (the post-starburst systems with low SFRs) 
and a few have
higher 
L$_{\rm FIR}$/L$_{\rm K}$
ratios.
The [3.6] $-$ [24] color 
is an approximate measure of the ratio of the number of
young-to-old stars (e.g., \citealp{smith07}), 
increasing with increasing proportions of young stars.
This means that [3.6] $-$ [24] is another approximate measure of the
sSFR.

Figure 2 also shows that F$_{60}$/F$_{100}$ is weakly correlated
with SFR, with considerable scatter.
This relation or its equivalent has been noted
before (e.g., \citealp{soifer1987, smith1987}).
Higher F$_{60}$/F$_{100}$ ratios imply hotter
dust on average and more intense UV interstellar radiation
fields (ISRF)
(e.g., \citealp{desert90}), 
which are correlated but
not perfectly with the overall SFR of the galaxy.

\section{Atomic and Molecular Interstellar Gas}

In the current study, we compare the hot X-ray-emitting gas mass in these galaxies with 
the interstellar molecular and atomic hydrogen gas masses. 
We obtained published measurements of the 2.6 mm CO (1 $-$ 0) fluxes
of the sample galaxies from the literature, and used these to derive molecular gas masses.
Since there is some uncertainty as to the relation
between the CO luminosity and the molecular gas mass, we converted
the CO fluxes into
molecular gas masses M$_{\rm H_2}$ by two methods. First, 
we calculated
M$_{\rm H_2}$ for all galaxies 
assuming a constant conversion equal to the 
Galactic conversion factor between H$_2$ column
density 
N(H$_2$)(cm$^{-2}$) 
and CO intensity I(CO)
of 
N(H$_2$)(cm$^{-2}$) 
= 2.0 $\times$ 10$^{20}$ I(CO)(K km~s$^{-1}$)
\citep{dame01, bolatto13}.
The Galactic conversion is thought to be appropriate for most
galaxies, however, low metallicity systems may be deficient in CO relative
to H$_2$, 
while extreme starburst galaxies may have enhanced CO/H$_2$ ratios
(e.g., \citealp{downes98, bolatto13}).
Thus for comparison we made a second estimate of M$_{\rm H_2}$ using a variable
CO/H$_2$ ratio.
For galaxies with L$_{\rm FIR}$ $>$ 10$^{11}$ L$_{\sun}$
(e.g., extreme starbursts),
we used a lower conversion factor of 
4 $\times$ 10$^{19}$ cm$^{-2}$/(K~km~s$^{-1}$) 
(e.g., \citealp{Ueda2014}). For galaxies with low K band luminosities
(e.g., possible low metallicity systems),
L$_{\rm K}$ $<$ 10$^{10}$ L$_{\sun}$, we used an enhanced
ratio of 5 $\times$ 10$^{20}$ cm$^{-2}$/(K~km~s$^{-1}$)
(e.g., \citealp{bolatto13}).
For all other galaxies we used the standard Galactic value given above.
Since accurate metallicities are not available for all of the galaxies
in our sample and because there is some uncertainty
as to how the CO/H$_2$ ratio varies with metallicity, 
we do not use a more complicated metallicity-dependent conversion
in this study.
In Section 6 of this paper, we compare various properties of the
galaxies.  We do the correlation analysis with both CO/H$_2$ ratios,
to test whether our conclusions are influenced by our choice
of CO/H$_2$ conversion factors.

Molecular gas masses calculated
with a constant CO/H$_2$ ratio 
equal to the Galactic value
are provided in column 2 of Table 2.
Molecular gas masses calculated
with the variable CO/H$_2$ ratio are given in column 3 of Table 2.
The reference for the original CO measurement is given in column 4
of Table 2.
Note that molecular masses are not available for all of the 
galaxies in the sample.  In some cases, no CO observations have
been published.  In other cases, only measurements of the central
region have been made,
where the beamsize
is significantly smaller than the optical extent of the galaxy.
In those cases, we are not able to get reliable upper limits to the global molecular
gas content so no molecular gas mass is listed in Table 2.
Followup CO observations would be useful to complete the molecular gas census of the
sample galaxies.

In the bottom row of Figure 3, 
the star formation
efficiency,
which we define as the global SFR/M$_{\rm H_2}$
ratio for the galaxy\footnote{With this definition, the SFE is equal to 
1/$\tau$$_{\rm dep}$, where $\tau$$_{\rm dep}$
is the global depletion timescale, the time to use up
the molecular gas.}, 
is plotted 
against the 
merger stage.
The left panel of Figure 3 
has SFE calculated
with a constant CO/H$_{2}$ ratio and the right 
with the variable
CO/H$_{2}$ ratio.
These two determinations of the SFE are included 
in Table 2
in columns 5 and 6, respectively.
As in Figure 1, the black open squares in
Figure 3 are the data for the individual galaxies; the blue filled diamonds that are offset slightly
to the left of the stage show the median value for that stage.  The errorbars on the blue diamonds show
the semi-interquartile range, equal to half the difference between
the 75th percentile and the 25th percentile.

Systems in the middle merger stages tend to have higher SFEs
than those in the early stages.  This is consistent with
earlier surveys
that found 
that 
L$_{\rm FIR}$/M$_{\rm H_2}$ is enhanced 
near nuclear coalescence
\citep{casoli91, georgakakis00}.
Given the small numbers of galaxies per merger stage in our sample and the spread in the data
per merger stage, however, this result is uncertain for our sample, especially if one also takes into
account 
the uncertainties
in the CO/H$_2$ ratio, and the selection effects.
Because of these factors any trends with merger stage are uncertain
for our sample. 

We also scoured the literature for measurements of the global HI masses of our
galaxy sample.  These values are tabulated in column 7 of Table 2,
and the reference for the HI data is given in column 8.
In Figure 3, we provide 
plots of merger stage vs.\ quantities derived from
the CO and HI data.
The top row of Figure 3 
compares 
M$_{\rm HI}$/M$_{\rm H_2}$ with
the merger stage. 
In the left panel, we assume
a constant CO/H$_{2}$ ratio in calculating M$_{\rm H_2}$,
while in the right panel we use
the variable
CO/H$_{2}$ ratio.
An apparent increase in the HI gas fraction in the late stages
of the merger sequence (left panel) is weakened
when a variable CO/H$_2$ ratio is used (right panel).

The SFE is plotted against 
dust temperature as measured
by the IRAS 60 $\mu$m to 100 $\mu$m flux ratio 
in the two top panels of Figure 4, for the two
CO/H$_2$ conversion factors.
A trend is clearly visible, in that hotter dust is correlated
with more efficient star formation.  
This relation is well-known (e.g., \citealp{young86, Sanders1991}).
Note that the scatter is larger with the variable CO/H$_2$ ratio
than for the constant conversion factor.
In the bottom two panels of Figure 4, we compare the SFE with the SFR for the two
conversion factors.
There is a trend, in that systems with the highest SFRs have
high SFEs, however, there is a lot of scatter, and there are some low SFR
systems with high SFE.  A spread in the SFE for a given SFR has been
observed before (e.g., \citealp{young86, Sanders1991, Young1996, sanders96, daddi10}).

The scatter in the SFE vs.\ SFR correlation may be due to 
variations in the fraction of the CO-emitting gas involved
in star formation.
This would lead to variations in the SFE according to our definition,
SFR/M$_{\rm H_2}$,
where H$_2$ is derived from CO observations. 
Larger SFE may mean that a larger fraction of the CO-emitting cold molecular gas is in a dense
state, an idea that is supported by both observations 
\citep{solomon92, gao04, juneau09, wu10}
and simulations \citep{teyssier10, renaud14, renaud19, sparre16}.  
These simulations show that an increase in turbulent compression during
an interaction can cause the gas probability density
function to shift to higher densities, producing an increase in the amount
of very high density gas.   Thus the variations in SFE from galaxy to galaxy
may be caused by differences in the dynamical state of the galaxies.

\section{X-Ray Spatial Extent }

All of the sample galaxies were observed with the Chandra ACIS-S array,
and all of the galaxies fit well within the 
8\farcm3 $\times$ 8\farcm3 field of view of the 
S3 chip of this array.
Details of the individual observations, including exposure times
and ObsID numbers, are provided in Paper~I.
In Paper~I,
we identified point sources in the field using 
the {\it ciao}
software tool {\it wavdetect}.
The point sources themselves and their statistics were the subject
of another 
paper \citep{smith12}.

After removing the point sources, in Paper~I we 
used the Chandra Interactive Analysis of Observations ({\it ciao})
software routine {\it specextract} to 
extract the diffuse X-ray spectrum within
the optical B band 25 mag~arcsec$^{-2}$ isophote.
This optical isophote was measured on Sloan Digitized Sky Survey (SDSS) g
images using standard g-to-B conversion factors, or, if SDSS images
weren't available, equivalent levels on GALEX near-UV images
were used
(see Smith et al.\ 2018 for details).
In Paper~I,
we used the {\it xspec}\footnote{https://heasarc.gsfc.nasa.gov/xanadu/xspec/}
software to
fit the 0.3 $-$ 8.0 keV background-subtracted 
point-source-removed
spectrum within
the 
$\mu$$_{\rm B}$ = 25 mag~arcsec$^{-2}$
isophote (e.g., D25)
to a combination power law plus thermal (MEKAL) spectrum,
assuming a power law photon index of 1.8 and correcting for
both Galactic and internal absorption. 
The power law component is assumed to be caused
by faint unresolved point sources.
The absorption-corrected 0.3 $-$ 8.0 keV luminosities for the MEKAL
component 
are provided in Table 1 of the current paper; we 
assume that the MEKAL component is from hot gas.
These X-ray luminosities have been corrected for absorption
within the galaxies as described in Paper~I.

In the current study, we measured the spatial extent
of the diffuse soft X-ray flux in these galaxies, 
and use these estimates
to calculate
the electron densities and masses of the hot X-ray-emitting gas. 
Our procedure is as follows.
After initial processing and 
deflaring of the data as described in Paper~I,
we constructed 0.3 $-$ 1.0 keV maps of each galaxy.
We then made an initial estimate
of the spatial extent of the low energy diffuse X-ray emission by eye from
the 0.3 $-$ 1.0 keV maps assuming an elliptical distribution, estimating
the centroid of the emission, the radial extent, the ellipticity, 
and the position angle
of the emission.   For some
of the pre-merger systems, two distinct regions of
diffuse light are seen, associated with the two galaxies
in the pair, so two elliptical regions were marked and the two regions 
were treated separately.

We then divided these ellipses into a set of concentric elliptical
annuli,
and determined background-subtracted 0.3 $-$ 1.0 keV counts and 
photon flux surface brightness
in each annuli 
using the {\it ciao} routine
{\it dmextract}, excluding 
the point sources detected by the {\it wavdetect} software.
For background subtraction, we used large areas outside of the 
optical extent of the galaxies excluding bright point sources.
All of our target galaxies have small enough angular size(s) 
such that we can obtain sufficient background regions on the same
chip.  
The flux calibration was done using
a 0.8 keV monoenergetic exposure map.
When multiple datasets were available, each set was calibrated
individually and the results combined.
We then 
produced radial profiles
for each galaxy.

The derivation of the radial profiles was done iteratively, 
modifying the initial region on the sky 
and the annuli widths until good radial profiles were produced.
We started by dividing the initial preliminary ellipse into 10 radial
annuli, adding three more annuli
outside of the initial
radius for a total of 13 annuli.  
If there were too few counts to get a good radial
profile with 13 annuli, we divided the initial ellipse into only five annuli, and added 2 outside the initial region
for a total of 7 annuli.  In some low S/N cases, to get sufficient counts
it was necessary
to 
divide the initial ellipse into 
only
three annuli,  plus two additional annuli outside, for a total of five
annuli.
In total, we were able to derive radial profiles for 28 systems by this
method, with 16 using 13 annuli, six using seven annuli, and and six using five annuli.

The final background-subtracted
radial profiles of the diffuse emission as obtained from 
{\it dmextract} are displayed in Figures 5 - 7,
after conversion into 0.3 $-$ 1.0 keV surface brightness in units
of photons~s$^{-1}$~cm$^{-2}$~arcsec$^{-2}$.
In most cases, these radial profiles are centrally-peaked, but there are
a few exceptions, most notably Arp 244 and Arp 299 (see Figure 6).

As a check on these results, we also obtained radial profiles using
a different method.  Instead of {\it dmextract}, we used
the {\it ciao} routine {\it specextract} to extract
the soft (0.3 $-$ 1.0 keV) X-ray spectra for each annulus.
When multiple datasets are
available, the ``combine = yes" option was used, which calibrates
each dataset individually, then the weighted spectra were coadded.
The {\it ISIS} (Interactive Spectral Interpretation System) 
software\footnote{https://space.mit.edu/ASC/ISIS/}
\citep{houck00}
was then used to derive background-subtracted 0.3 $-$ 1.0 keV counts
and photon fluxes in each annuli, taking into account
the calibrated response function of the detector.
These two procedures give reasonably consistent radial profiles, with the
{\it dmextract} method giving lower fluxes by about a factor of 1.2
and somewhat smaller uncertainties.  
In the subsequent determination of the radial extent of the
X-ray emission in the galaxies and the following analysis, 
we used the {\it dmextract}-determined radial profiles. 

Our goal in this paper is to obtain the physical size of the 
hot gas distribution within the galaxies, to derive electron densities
and hot gas masses.
An issue, however, is that how far out in the galaxy the X-ray emission 
can be detected
depends upon the 
exposure time for the observations and the width of the annulus that is used.
For the same annulus width, more sensitive observations
can detect gas further out in the galaxy.  This would give 
larger radii for the hot gas extent,
although the
gas in the outskirts may contribute little to the overall X-ray luminosity of the galaxy.
This could lead to a bias in the analysis, producing larger
volumes of hot gas for longer observations, which will affect the derivation of the electron
densities and therefore the masses of hot gas.   Because this is an archival sample,
there is a large galaxy-to-galaxy variation in the observing times used.

To get around this issue, it is desirable to use a consistent definition for the radius from galaxy
to galaxy.
In past studies of the hot gas distribution of galaxies, a number of different methods have been used to 
determine the volume of hot gas.
For example, 
\citet{boroson11} and \citet{goulding16} measured the extent out 
to where the diffuse
emission equals the background.
\citet{mcquinn18} used a similar method, measuring the
extent out to when the diffuse emission is detected at a 2$\sigma$ level.
Other groups measured the emission within the optical D25 isophote or the
optical effective radius, and used
this extent as the X-ray size in deriving electron densities
\citep{mineo12, su15, gaspari2019}.
A third method was used by 
\citet{strickland04a} and 
\citet{grimes05}, who used the radius which encloses a given fraction of the total 0.3 $-$ 1.0 keV flux.
They find that the 90\% enclosed-light fraction corresponds to an 
0.3 $-$ 1.0 keV surface brightness
between approximately
$\sim$10$^{-9}$ $-$ 10$^{-8}$ photons~s$^{-1}$~cm$^{-2}$~arcsec$^{-2}$ for their sample galaxies.

Because our dataset is so heterogeneous, after some experimentation
we chose to measure the radial extent out to a consistent 0.3 $-$ 1.0 keV
surface brightness level for all of the sample galaxies.  To decide on this
level, 
we explored how the enclosed-light fraction varies with 
different surface brightness cutoffs,
assuming that the 
counts within the optical B band 25 mag~arcsec$^{-2}$ isophote 
is the `total' flux (this issue is
discussed further below).
Upon experimentation, we found that for most galaxies a 0.3 $-$ 1.0 keV surface brightness cutoff of 
3 $\times$ 10$^{-9}$ photons~s$^{-1}$~cm$^{-2}$~arcsec$^{-2}$ produced counts that agreed
with the total counts within 10\%.  This is consistent with the \citet{grimes05} 
and \citet{strickland04a} results for their 90\% enclosed-light fractions.

There were 18 systems that were
detected in the MEKAL component in Paper~I but had too
few counts for us to derive an acceptable radial profile.
For these galaxies, we derived approximate sizes by starting
with the initial by-eye elliptical regions, then iteratively
increasing the size of the ellipse by 30\% until the galaxy
is detected at the $\ge$2$\sigma$ level and the 0.3 $-$ 1.0 keV
counts in the expanded ellipse 
equaled 
those
in the $\mu$$_{\rm B}$ = 25 mag~arcsec$^{-2}$ isophote
within the uncertainties. 
For the widely separated pre-merger pairs with two distinct regions of hot 
gas within the two optical galaxies, the two galaxies in the pair were treated separately in
this procedure.

Four of the galaxies for which we
could not find a radial profile (Arp 163, Arp 235, Arp 243, and Arp 263)
were undetected in the MEKAL component
in Paper~I.
These four galaxies are not included in any of the subsequent plots 
in this paper which involve quantities derived from the spatial size
of the X-ray emission.
Another galaxy, UGC 02238, was nominally detected in the MEKAL
component at the 2.6$\sigma$ level in the spectral decomposition in Paper~I,
however, in the 0.3 $-$ 1.0 keV map it was not detected within the optical
extent of the galaxy at the 2$\sigma$ level.
It is also omitted from the subsequent analysis in the current paper.
Another system, UGC 05189, was undetected in
Paper~I in the MEKAL component, however, 
we detected the inner disks of
both galaxies in the pair at the 5$\sigma$ level in the 0.3 $-$ 1.0 keV map.
The area covered by the diffuse gas is considerably smaller than the optical
extent, which might explain the non-detection in the spectral decomposition.

Except for the five systems for which we could not derive radial profiles,
the Chandra 0.3 $-$ 1.0 keV maps of the sample galaxies are displayed
in the Appendix of this paper (Figures 20 - 27).
In Table 3, we provide the central coordinates, 
major and minor axis radii, and position
angles of the final ellipses derived using the methods described above,
with the {\it dmextract}-derived sizes
at 3 $\times$
10$^{-9}$ photons~s$^{-1}$~cm$^{-2}$~arcsec$^{-2}$ level.
Table 3 also gives the number of annuli used in the radial
profile (13, 7, 5, or 1).
For systems with two distinct
regions of diffuse emission, two ellipses are given in Table 3.
In those cases, the name of the specific galaxy in the pair
associated with the particular region 
is identified in the second column of Table 3.
When the X-ray flux only comes from one galaxy in a pair,
the name of that individual galaxy is listed in Table 3.
If both galaxies in a pair are covered by a single region
of diffuse emission, both names are given in the second column
of Table 3.   If there is only one galaxy in the system, the second
column gives an alternative name for the galaxy.
Table 3 also provides the point-source subtracted, background-subtracted
0.3 $-$ 1.0 keV counts in the final ellipse.
Table 3 does not include UGC 02238 or the four systems
without radial profiles that are undetected in the thermal component 
in Paper~I.
The final ellipses are superimposed on images of the galaxies in the Appendix
of the paper.

In Figure 8, we compare the background-subtracted
0.3 $-$ 1.0 keV counts
obtained within the 
$\mu$$_{\rm B}$ = 25 mag~arcsec$^{-2}$ isophote with those 
extracted within the Table 3 radial extents. 
The solid line on this plot is the one-to-one relation, and the dashed lines
mark $\pm$ 10\% differences.
The systems marked by green hexagons in Figure 8 are those for which we were
not able to find a radius using a set of concentric annuli.
For most of the galaxies in the sample, the two 
measurements of the X-ray counts 
agree within the uncertainties with the range marked by the dotted lines.
For only one system,
IRAS 17208-0014,
does our total counts 
in the 3 $\times$ 10$^{-9}$ photons~s$^{-1}$~arcsec$^{-2}$ isophote
exceed that
in the optical isophotes by 10\% or more, 
taking into account the uncertainties
(i.e., only one system lie below
the bottom dotted line).  
For IRAS 17208-0014, the X-ray radial extent in Table 3 exceeds 
the optical D25 size by a factor of 1.5, and the 0.3 $-$ 1.0 keV
counts within the Table 3 ellipse are about 2.2 times those within the optical
isophotes.

Taking into account the uncertainties,
four systems in our sample
have counts within the `best' radii that are 60\% $-$ 80\% of
the counts within
the optical extent, and one (UGC 05189) has counts within the `best'
radii that are 50\% of the counts in the optical isophotes.
These systems lie above the top dotted line in Figure 8.
Most of these systems are galaxy pairs 
which have two distinct regions of X-ray emission
within the 
$\mu$$_{\rm B}$ = 25 mag~arcsec$^{-2}$ isophote.  Very faint diffuse
emission outside of these regions may contribute 
to the total counts in the optical
extent.  This faint emission
likely doesn't contribute much to the overall mass of hot
gas in the system.

For all but one of our systems, in our {\it dmextract} radial profiles
we can measure X-ray emission beyond
the 3 $\times$ 10$^{-9}$ photons~s$^{-1}$~cm$^{-2}$~arcsec$^{-2}$ isophote.
For completeness, for these systems
we provide the full (2$\sigma$) extent of the X-ray emission in 
another table, Table 4.
In all but 11 of these cases,
the
2$\sigma$ extent of the diffuse X-ray emission 
is 20\% or more larger than the optical D25 size.
The most extreme case is Markarian 231
for which 
the ratio of the 2$\sigma$ X-ray radius divided by the 
maximum
$\mu$$_{\rm B}$ = 25 mag~arcsec$^{-2}$ radius
is 2.6. 
For Markarian 231, the counts within the 2$\sigma$ extent are about 1.5 times those in the D25 radius.
Although the measured 2$\sigma$ X-ray sizes are often larger than the D25 extent,
the 0.3 $-$ 1.0 keV counts within the 2$\sigma$ radius are generally less than or consistent with the counts
within the D25 radius.  
This means that the emission outside of D25 does not contribute
significantly to the total flux.

\section{Volume and Mass of Hot Gas, Electron Densities, and Filling Factor}

We calculated the volume of hot gas for each system in the sample,
assuming that
the hot gas distribution has an
ellipsoidal structure with the third dimension equal
to the average of the other two.
For these calculations, we use the X-ray sizes at 
the 3 $\times$ 10$^{-9}$ photons~s$^{-1}$~arcsec$^{-2}$ isophote
as discussed above (Table 3).
For the pre-merger systems for which we could measure two
distinct regions of hot gas, we calculated the sum of the two
volumes.   

The uncertainty in the geometry of the hot gas likely contributes scatter
to the relationships shown below.
Although the true 3-dimensional distribution of the hot gas
in a particular
galaxy is unknown, assuming random orientations in space
we can use the statistics of the observed ellipticities 
of the diffuse X-ray 
emission on the sky 
(Table 3)
to make a rough estimate of the average uncertainty 
in the volume.
The average major/minor axial ratio of the diffuse X-ray emission
on the sky is 1.50, with an rms of 0.39. 
We therefore assume
that the line-of-sight dimension on
average will range from 1.5 times bigger than the average
of the other two dimensions,
to 1.5 times smaller.
Thus we 
assume that our estimates of the volume are uncertain by a factor
of 1.5.

Using the derived volumes of hot gas,
we estimated electron densities in the hot gas as a function of filling
factor. 
For this calculation, we used the 
relation L$_{\rm X}$(gas) = $\Lambda$n$_{\rm e}$$^2$fV, where
L$_{\rm X}$(gas) 
is the absorption-corrected 0.3 $-$ 8.0 keV X-ray luminosity
of the hot gas 
from Paper~I (the MEKAL component), $\Lambda$ is the  
cooling function \citep{mckee77, mccray87},
V is the volume of gas, 
n$_{\rm e}$ is the electron density,
and f is the filling factor.
In this calculation,
we 
assumed that the 
number of hydrogen atoms $\sim$
n$_{\rm e}$.
The derived gas masses depend upon the temperature of the X-ray-emitting
gas.  Unfortunately, for only 15 systems were we able 
to obtain a fit for the gas 
temperature in Paper~I (see Table 5 in that paper).
For the remaining systems, we assumed 
a temperature of 0.3 keV.  In Section 6.5 of this paper,
we investigate how this assumption
affects our results.
In calculating n$_{\rm e}$, we neglect
X-ray emission outside of the 0.3 $-$ 8.0 keV Chandra bandpass,
however, 
emission outside of this range may also contribute to cooling the gas.
In Section 6.5, we discuss this approximation and how it
depends upon temperature.

From the X-ray luminosity, the volume, and the temperature we 
derive n$_{\rm e}$$\sqrt{f}$ for our sample galaxies;
we are not able to independently determine 
n$_{\rm e}$ and f. 
We find that n$_{\rm e}$$\sqrt{f}$ 
ranges from 1.1 $\times$ 10$^{-3}$ $-$ 2.2 $\times$ 10$^{-2}$ cm$^{-3}$,
similar to the values found by \citet{mineo12} for their spirals.
Accounting for the uncertainty in volume
and conservatively
assuming a factor of two uncertainty
in L$_{\rm X}$(gas) 
(due in part to uncertainties in separating the thermal and non-thermal
emission; see Section 5.3 
in Paper I),
propagation of errors implies that our estimates
of n$_{\rm e}$$\sqrt{f}$ are uncertain by a factor of 
1.8 on average.

The radiative cooling times for the hot gas (i.e., total thermal energy
divided by L$_{\rm X}$(gas)) in these galaxies range from
16 to 700 Myrs, with a median time of about 60 Myrs.
These are similar to the \citet{mineo12} estimates for disk galaxies.
We then calculated the mass of the hot X-ray-emitting
gas M$_{\rm X}$(gas) 
= m$_{\rm p}$n$_{\rm e}$V, when m$_{\rm p}$ is the mass of a proton.
Accounting for the uncertainties in V and n$_{\rm e}$, we adopt
an uncertainty in our estimates of M$_{\rm X}$(gas) 
of a factor of two,
not including
the uncertainty in the filling factor.

\section{Correlation Analysis}

From the Chandra data we derived a set of parameters
for our sample galaxies, including 
the volume of hot gas, 
n$_{\rm e}$$\sqrt{f}$,
L$_{\rm X}$(gas), 
and
M$_{\rm X}$(gas).
From data at other wavelengths, we have another 
set of values for our galaxies, including
SFR, SFE, 
L$_{\rm FIR}$, L$_{\rm K}$, 
L$_{\rm FIR}$/L$_{\rm K}$, [3.6] $-$ [24], F$_{60}$/F$_{100}$,
M$_{\rm HI}$/M$_{\rm H_2}$, and the merger stage.
Combining these two sets, we derive additional parameters, including
L$_{\rm X}$(gas)/SFR,
M$_{\rm X}$(gas)/SFR, 
and 
M$_{\rm X}$(gas)/(M$_{\rm H_2}$+M$_{\rm HI}$).
In this section, we correlate these parameters against each other,
and calculate the best-fit linear 
log vs.\ log
relations for various
combinations of these parameters.   
In Paper~I, we found
that some trends change  at low SFRs, so 
we did these fits for two cases: the full range of SFRs
and the subset of systems with SFR $>$ 1 M$_{\sun}$~yr$^{-1}$.

For each relation, we calculated the root mean
square (rms) deviation from the best-fit line and the Spearman rank
order coefficient. 
These values are compiled in Table 5, along with the best-fit
parameters.
For comparison with the Spearman coefficients, Table 5 also provides
Pearson correlation coefficients, which assumes a linear relationship
between the two parameters.  The two types of correlation coefficients
agree fairly
well for our sample (see Table 5).
In Table 5, 
we classified the relations, into ``strong correlation",
``weak correlation", or ``no correlation".  We defined
a ``strong correlation" as one in which the 
Spearman coefficient is greater than 0.55 (i.e., $\le$0.1\%
likelihood of happening by chance), and a ``strong anti-correlation"
is one in which the Spearman coefficient is 
less than $-$0.55.
A ``weak correlation" is one in which the 
Spearman coefficient
is between 0.35 and 0.55, where 0.35 corresponds to 
a $\sim$5\% probability of happening by chance.
A ``weak anti-correlation" implies 
a Spearman coefficient
between $-$0.35 and $-$0.55,
and ``no correlation" means a Spearman coefficient between
$-$0.35 and 0.35.

The most important of the correlations
are plotted in Figures 9 $-$ 19.  For convenience,
when a plot is shown,
the number of the figure which displays each correlation
is provided in Table 5 along with the best-fit parameters and 
the Spearman/Pearson coefficients.
For clarity of presentation, we do not include errorbars
on the plots.  As discussed above, we estimate that our
values of 
M$_{\rm X}$(gas) 
are uncertain by about a factor of two.
This means that the uncertainty in log(M$_{\rm X}$(gas))
is about 0.3 dex.
The rms uncertainties on some of the fits 
involving log(M$_{\rm X}$(gas)) are close to or slightly
larger than this estimate (see Table 5), so the 
uncertainty in M$_{\rm X}$(gas) may be a
limiting factor in this analysis.
Because the uncertainty in the CO/H$_2$ ratio
is potentially an even larger factor, we
do the correlations for both CO/H$_2$ ratios.
This provides a test of whether
the results are biased by the choice of CO/H$_2$ ratios.

As another test, we also ran the correlation analysis 
using radii and volumes determined from the {\it specextract/ISIS} radial
profiles rather than {\it dmextract}. 
The best-fit relations and 
correlation coefficients changed slightly but the
basic conclusions of the paper were unchanged.
The relations given below were derived from the {\it dmextract} results.

To see
if our results are affected by the inclusion of Seyfert galaxies in the sample,
we also calculated the correlations excluding the AGNs.
The correlation coefficients tend to be somewhat smaller with the smaller sample,
however, the basic results do not change and the important correlations
discussed below still hold.

\subsection{Relations with Volume and with n$_{\rm e}$}

In the top row
of Figure 9, 
we plot the volume of hot gas
as a function of 
merger stage, L$_{\rm X}$(gas)/SFR,
and SFR.
The bottom row of panels in this figure
displays 
the ratio of 
the maximum radial extent of the X-ray emission to 
the maximum optical size as measured by the B band 25 mag~arcsec$^{-2}$ isophote
against 
merger stage, L$_{\rm X}$(gas)/SFR,
and SFR.
The first column of Figure 9 shows that 
stage 3 and stage 4 mergers tend to have large X-ray sizes
and large X-ray/optical size ratios.
This is a consequence of the bias towards higher SFRs
in the mid-merger stages.
The second column shows little correlation between the volume of hot gas 
and 
L$_{\rm X}$(gas)/SFR, 
or between the X-ray to optical
size and 
L$_{\rm X}$(gas)/SFR.
Systems in which the X-ray extent exceeds that in the
optical tend to have high SFRs (Figure 9; bottom right panel), but with a lot of scatter.

In Figure 9, the strongest correlation is seen 
in the top right panel: larger volumes of hot X-ray-emitting gas
are found in 
higher SFR systems.  
For the full set of galaxies,
the best-fit slope is less than one.
However, 
once low SFR
systems are excluded the slope 
is consistent with one.
Thus low SFR systems tend to have larger volumes than expected based
on their SFRs.

We have marked the location of the late-stage merger NGC 1700 on the two
upper right panels in Figure 9.   It stands out in the sample for having
a large volume of hot gas, compared to its SFR.  
This galaxy has a high L$_{\rm K}$ and a low SFR, with an 
elliptical-like
appearance and tidal debris.   It may be a system
for which virialized hot gas in the gravitational potential contributes
significantly to the observed diffuse X-ray-emitting gas.
This system is discussed further in Section 7.4.

In Figure 10, we plot 
the volume of hot gas (top row) and the X-ray/optical size ratio
(bottom row) against the SFE (first and second columns)
and the F$_{60}$/F$_{100}$ ratio (last column).
The first column utilizes a constant CO/H$_2$ ratio, while
the second uses the variable CO/H$_2$ ratio.
This Figure shows that the systems with large volumes 
tend to have large SFEs (when a variable
CO/H$_2$ ratio is used).
However, this is a very weak correlation
with a lot of scatter; some systems
with large SFEs have only moderate
volumes
and size ratios.   

No significant correlation is seen between
volume and F$_{60}$/F$_{100}$ (Figure 10, upper right),
in spite of the fact that volume is correlated
with SFR (Figure 9), and F$_{60}$/F$_{100}$ is 
correlated with SFR (Figure 2).  
However, the correlation
between F$_{60}$/F$_{100}$ and SFR is weak, with 
considerable scatter.  The sample galaxies only cover
a small range in log F$_{60}$/F$_{100}$ (0.3 dex)
while the SFR varies over two orders of magnitude.
This means that uncertainties in the IRAS fluxes
can make it difficult to detect a correlation.
Instead of being directly dependent on SFR itself,
theoretical models 
(e.g., \citealp{desert90})
suggest that
the F$_{60}$/F$_{100}$ 
ratio depends on the intensity
of the ISRF.  The average ISRF in a galaxy can vary greatly 
for a given global SFR of the galaxy,
depending upon the spatial density of young stars.
In contrast, the volume of hot gas in these
galaxies depends directly upon the total number of 
young stars rather than on the spatial density of those
stars.

The upper left panel of Figure 11 shows that  the volume of hot
gas is well-correlated with L$_{\rm X}$(gas).   
The slope in this log-log plot is close to one.
In the right panel of Figure 11, 
the volume is plotted against the derived n$_{\rm e}$$\sqrt{f}$.
There is a weak 
trend of decreasing n$_{\rm e}$$\sqrt{f}$ with increasing volume.

In the lower left panel of Figure 11, we show that the
volume is also correlated with L$_{\rm K}$.
However, the correlation is weaker than for 
volume vs.\ SFR, 
and the scatter is larger (Table 5).
This supports the idea that in this sample of galaxies the volume
of hot gas
is largely determined by the number of young stars, with the 
correlation of volume with L$_{\rm K}$ being a by-product of the 
SFR $-$ L$_{\rm K}$ correlation.

The volume of hot gas is weakly correlated with 
L$_{\rm FIR}$/L$_{\rm K}$
(bottom right panel of Figure 11)
and with [3.6] $-$ [24] (Table 5).
As noted earlier, both [3.6] $-$ [24] and 
L$_{\rm FIR}$/L$_{\rm K}$ are 
approximate measures of the 
sSFR.
This correlation may be a consequence of the 
correlation between volume and SFR, since sSFR tends to
increase with increasing SFR for this sample (see Figure 2
and Table 5).
Notice that NGC 1700 is particularly discrepant in these plots compared to 
the other galaxies, with a low
sSFR (i.e., low [3.6] $-$ [24] and low L$_{\rm FIR}$/L$_{\rm K}$) and a large volume of hot gas.

In 
Figure 12, 
SFR is plotted
against 
n$_{\rm e}$$\sqrt{f}$ (upper left panel),  
F$_{60}$/F$_{100}$ 
against 
n$_{\rm e}$$\sqrt{f}$ (upper right panel),
SFE with constant
CO/H$_2$ ratio 
vs.\ 
n$_{\rm e}$$\sqrt{f}$ 
(middle left), SFE with a variable
CO/H$_2$ ratio 
vs.\ 
n$_{\rm e}$$\sqrt{f}$ 
(middle right),
and 
[3.6] $-$ [24] 
vs.\ 
n$_{\rm e}$$\sqrt{f}$.  
No trends are seen in these five panels.

In the lower right panel of Figure 12, we plot the derived ratio
M$_{\rm X}$(gas)/L$_{\rm X}$(gas) for the 
15 galaxies with temperature measurements.
The blue solid line on this plot is the relation assuming a 
constant temperature of 0.3 keV.
From the equations given in Section 5,
M$_{\rm X}$(gas)/L$_{\rm X}$(gas) = m$_{\rm P}$/($\Lambda$n$_{\rm e}$f),
so the conversion
from L$_{\rm X}$(gas) to M$_{\rm X}$(gas) is a function of 
n$_{\rm e}$, with
M$_{\rm X}$(gas)/L$_{\rm X}$(gas) $\propto$ 
1/n$_{\rm e}$ 
if the temperature and filling factor are constant.  
The 15 data points lie above the blue line in this plot because they have 
temperatures higher than 0.3 keV (see Section 6.5), and $\Lambda$ decreases as temperature
increases.
The question of the assumed electron temperature is discussed further in
Section 6.5.

\subsection{M$_{\rm X}$(gas)/SFR vs.\ Other Properties}

The mass of hot gas is strongly correlated with SFR
(Figure 13, upper left).
For SFR $>$ 1~M$_{\sun}$~yr$^{-1}$, the slope of
log 
M$_{\rm X}$(gas) vs.\ log SFR is consistent with one.
However, the relationship flattens when lower SFR systems are included,
suggesting an excess of hot gas in low SFR systems.
Even when NGC 1700 is excluded, this flattening is seen.
We note that the other two galaxies with low SFR and high M$_{\rm X}$(gas)/SFR
in this figure, NGC 2865 and NGC 5018, both have moderately high K band luminosities
(Table 1).  As with NGC 1700, virialized gas in the gravitational potential may
be contributing to the observed hot gas in these galaxies
(see Paper I for detailed discussions of these galaxies).
Unfortunately, our sample only has a few low SFR, high L$_{\rm K}$ systems, 
so separating out this additional component to the hot gas is 
is uncertain.

M$_{\rm X}$(gas) is also correlated with 
L$_{\rm K}$ (Figure 13, upper right).
However, this relation has a lower correlation coefficient
than 
M$_{\rm X}$(gas) vs.\
SFR.
This suggests that the 
M$_{\rm X}$(gas)-L$_{\rm K}$ relation is a consequence of the
SFR-L$_{\rm K}$ correlation for our sample galaxies,
and the hot gas in most of our galaxies
is mainly due to SFR rather than older stars.

When both M$_{\rm X}$(gas) and SFR are normalized by a tracer
of stellar mass, L$_{\rm K}$, they still show a strong correlation
(Table 5).
This indicates that the relation between SFR and M$_{\rm X}$(gas)
is not simply a richness effect.
In contrast, when both M$_{\rm X}$(gas) and L$_{\rm K}$ are normalized
by SFR, the correlation is significantly weaker (Table 5).   
This again implies that 
M$_{\rm X}$(gas) is more closely tied to young stars than to old stars.

The correlation between M$_{\rm X}$(gas) and SFR is displayed
in another way in the bottom left panel of Figure 13, where we show
M$_{\rm X}$(gas)/SFR vs.\ SFR.
Although a weak anti-correlation is seen for 
the full sample, once systems with 
SFR $<$ 1 M$_{\sun}$~yr$^{-1}$ are removed no correlation is seen
and the rms scatter is relatively small (0.37 dex).
This is close to the expected scatter based on the uncertainty
in M$_{\rm X}$(gas) alone, which supports the contention that
processes associated with
a young stellar population
are the main factors responsible for the hot gas in these galaxies,
at least when low SFR systems are excluded.

M$_{\rm X}$(gas)/SFR is plotted against 
L$_{\rm K}$ in the lower right panel of Figure 13.
A very weak trend is seen when low SFR systems are
excluded. 
The post-merger NGC 1700 stands out as having a high 
M$_{\rm X}$(gas)/SFR.
After NGC 1700, the next two highest 
M$_{\rm X}$(gas)/SFR galaxies in this plot, NGC 2865 and NGC 5018, are
both stage 7 merger remnants with moderately high L$_{\rm K}$ and low
sSFR.
In contrast to these three galaxies, 
galaxies with
low K band luminosities ($\le$10$^{10}$ L$_{\sun}$)
have moderately low 
M$_{\rm X}$(gas)/SFR values, though not extreme.
In Paper~I, we found that low L$_{\rm K}$ systems have 
low L$_{\rm X}$(gas)/SFR.
Now, we are able to show that 
M$_{\rm X}$(gas)/SFR is also somewhat low for these systems.
This may indicate escape of hot gas from lower gravitational fields.
However, only a few galaxies fall in this range,
so the statistics are very uncertain.

In the upper left and upper middle panels of Figure 14, 
weak anti-correlations are seen
between M$_{\rm X}$(gas)/SFR and SFE, but these trends disappear
for the variable CO/H$_2$ ratio when low SFR systems are not included.
The fact that our dataset is incomplete in CO makes these conclusions
somewhat uncertain.

M$_{\rm X}$(gas)/SFR is anti-correlated with
the two tracers of sSFR, [3.6] $-$ [24] and 
L$_{\rm FIR}$/L$_{\rm K}$ (Figure 14, upper right and lower left panel,
respectively).
However,
when low SFR systems are excluded
the trend with [3.6] $-$ [24] weakens and the trend with
L$_{\rm FIR}$/L$_{\rm K}$
disappears.
This again suggests that low sSFR systems sometimes have excess hot gas.

M$_{\rm X}$(gas)/SFR is shown to be strongly anti-correlated with 
F$_{60}$/F$_{100}$ for the full sample
(Figure 14, lower middle panel).
This trend is weakened when only 
the high SFR sample is included, but is still detected.
The cause of this anti-correlation is uncertain; some possible
interpretations are discussed in Section 7.2.
In addition, a weak anti-correlation is visible between 
M$_{\rm X}$(gas)/SFR
and 
n$_{\rm e}$$\sqrt{f}$, 
particularly when low SFR systems are omitted
(lower right panel Figure 14).
In contrast, 
M$_{\rm X}$(gas)/SFR is not correlated with either
the volume or
the H~I-to-H$_2$ ratio 
(Table 5).

\subsection{Ratio of Mass Hot Gas to Mass Cold Gas vs.\ Other Properties}

In Figure 15, the ratio of the mass of hot X-ray-emitting gas
to the mass of cold gas (H~I + H$_2$) 
is plotted against SFR (top row) and SFE (bottom row).
The left panel in each
row was calculated using a constant CO/H$_2$ ratio, while
the right panel was calculated with a variable CO/H$_2$.
Figure 15 shows that 
M$_{\rm X}$(gas)/(M$_{\rm H_2}$ + M$_{\rm HI}$) 
increases with increasing
SFR, with a better correlation when a variable CO/H$_2$ ratio is used.
A higher Spearman coefficient and a steeper relation are found
when the low
SFR systems are omitted.
The slope is consistent with one when a variable CO/H$_2$ ratio
is used and low SFR systems are omitted.
The flatter relation when low SFR systems are included again
points to excess
M$_{\rm X}$(gas)
for low SFR systems.

A large amount of
scatter is evident in 
a plot of 
M$_{\rm X}$(gas)/(M$_{\rm H_2}$ + M$_{\rm HI}$) 
vs.\ SFE (Figure 15, bottom row),
but a reliable correlation
is present
when a variable CO/H$_2$ ratio is used.
The lack of a full set of CO data makes these results uncertain.

The scatter 
in M$_{\rm X}$(gas)/(M$_{\rm H_2}$ + M$_{\rm HI}$) 
may be due in part to
variations in the stellar mass.
In Figure 16, 
M$_{\rm X}$(gas)/(M$_{\rm H_2}$ + M$_{\rm HI}$) 
is plotted against 
L$_{\rm K}$ (top panel) and F$_{60}$/F$_{100}$ 
(bottom panel).  In the left column, the quantities 
were calculated using a constant CO/H$_2$ ratio, while
the right panel was calculated with a variable CO/H$_2$.
A weak correlation is visible in the upper right panel when low SFR 
galaxies are excluded and a variable CO/H$_2$ ratio is used.  The 
two lowest L$_{\rm K}$ systems have 
moderately
low 
M$_{\rm X}$(gas)/(M$_{\rm H_2}$ + M$_{\rm HI}$), 
and 
the highest
M$_{\rm X}$(gas)/(M$_{\rm H_2}$ + M$_{\rm HI}$) 
system, NGC 6240, has a very high
L$_{\rm K}$.
In the lower panels of Figure 16,
weak correlations between 
M$_{\rm X}$(gas)/(M$_{\rm H_2}$ + M$_{\rm HI}$), 
and
F$_{60}$/F$_{100}$ are seen, but only if low SFR systems are excluded.

Correlations are visible between 
M$_{\rm X}$(gas)/(M$_{\rm H_2}$ + M$_{\rm HI}$)
and our two tracers of sSFR (Figure 17),
especially when low SFR systems are excluded and a variable
CO/H$_2$ ratio is used.
The steepening of the slope when low SFR systems are excluded
again signals possible excess of hot gas in low SFR systems.

\subsection{Merger Stage vs.\ Gas Properties}

We plot the inferred mass of X-ray-emitting gas
M$_{\rm X}$(gas) 
against merger stage
in the top left panel
of Figure 18.
The mid-merger stages have higher
quantities of hot gas, on average, than the early or late stages.
However, this is largely due to the fact that the mid-merger
galaxies tend to have higher SFRs.
When the mass of hot gas is normalized by the SFR
(Figure 18, top right), no strong trend is seen.
The stage 7 galaxy NGC 1700 stands out as having a high
M$_{\rm X}$(gas)/SFR. 
The next two highest 
M$_{\rm X}$(gas)/SFR systems, the stage 7 galaxies
NGC 2865 and NGC 5018, also have 
low sSFR.

The bottom row of Figure 18 compares the merger stage with the ratio
of hot gas to cold gas M$_{\rm X}$(gas)/(M$_{\rm H_2}$ + M$_{\rm HI}$), 
using the standard
Galactic CO/H$_2$ ratio (left panel) or the 
variable
CO/H$_2$ ratio (right panel). Stages 3 and 4 tend to have proportionally more
hot gas.  
This is likely a consequence of the fact that galaxies in those stages
tend to have higher SFRs.
Because of the inhomogeneity of the sample, the small
number of systems in each merger stage, and the 
lack of a full set of CO data, trends
with merger stage in our sample are uncertain.
In these plots, the galaxy with the highest
M$_{\rm X}$(gas)/(M$_{\rm H_2}$ + M$_{\rm HI}$)
is NGC 6240.  NGC 1700 is not plotted in the bottom row of Figure 18
because it lacks a full set of CO data.

\subsection{Gas Temperature}

As mentioned earlier, our derivations of electron density
and 
M$_{\rm X}$(gas) depend upon the assumed temperature, and 
temperatures are available for only 15 of our sample galaxies
from the X-ray spectra (see Paper~I).  
For the remaining galaxies, we assumed a constant temperature of 0.3 keV.
For comparison, for the 15 systems for which temperatures
are available in Paper~I, kT ranges from
0.37 keV to 1.0 keV.  
In some cases,
we were able to use a two-temperature model for the hot gas; in those cases,
we used the luminosity-weighted temperature in the subsequent analysis.
For comparison, \citet{mineo12} found lower temperatures
on average for their sample galaxies (mean of 0.24 keV for single-temperature
models). 
The derived temperatures depend upon the assumptions
used in modeling the X-ray spectrum, 
including how the power law component is modeled, so they
are somewhat uncertain (see Paper~I).
Because our sample is an archive-selected sample,
there is a selection bias in the subset of galaxies with derived 
temperatures.
Compared to the galaxies in the sample without measured temperatures,
the galaxies with temperatures
tend to have longer, more sensitive exposures, and they
tend to be more extreme 
systems with higher luminosities.
In contrast, the 
\citet{mineo12} sample, with lower temperatures on average, 
contains more normal spiral galaxies and irregulars
as well as some mergers.   We therefore assume the more modest
temperature of 0.3 keV for our galaxies without temperature measurements,
assuming that they are less extreme than the other systems.
However, this is quite uncertain.

To test whether our conclusions are affected by our assumption of 0.3 keV
for the galaxies without derived temperatures,
we re-ran our correlation analysis with four alternative assumptions.
First, we re-ran the analysis 
using a constant kT = 0.3 keV for all the galaxies,
even those for which we have a direct measure of the temperature.
Second, we did the calculations assuming a constant kT = 0.6 keV
for all galaxies.

Third, we re-ran the analysis assuming that the temperature is
correlated with SFR.
In Paper
I we did not find any correlation of temperature with SFR.  
\citet{mineo12} also did not find a correlation between temperature
and SFR for their sample of star-forming galaxies.  
However, \citet{grimes05} noted that
the ULIRGs in their sample tend to have higher temperatures, up to
about 0.8 keV.   
Therefore, as a limiting case to investigate how temperature may
potentially affect our results, we assume that log T$_{\rm X}$ increases
linearly with log SFR,
and we set kT = 0.2 keV for the systems with the lowest 
SFRs (0.1 M$_{\sun}$~yr$^{-1}$), increasing to 1.0 keV for 
systems with SFR = 100 M$_{\sun}$~yr$^{-1}$.

As a fourth test, we investigated how our results changed if
we assumed that the temperature depends upon L$_{\rm X}$(gas) rather
than on SFR.   
In contrast to actively star-forming galaxies,
ellipticals show a steep relation between L$_{\rm X}$(gas) and temperature
of L$_{\rm X}$(gas) $\propto$ T$_{\rm X}$$^{4.5}$ \citep{goulding16}.
As a limiting case, we assumed that
L$_{\rm X}$(gas)
$\propto$ T$_{\rm X}$$^{4.5}$ as found for ellipticals \citep{goulding16}.
Assuming a temperature of 0.2 keV for the galaxies with the
lowest 
L$_{\rm X}$(gas) in our sample, this gives 1.0 keV for the highest
L$_{\rm X}$(gas) system.
This is a more extreme range than typically found for star-forming
galaxies, thus it is a limiting case.

For each of the above cases,
we also explored how our results change when we include
a correction from the observed 0.3 $-$ 8.0 keV
L$_{\rm X}$(gas) to the bolometric luminosity of the gas including light outside of the
0.3 $-$ 8.0 keV Chandra window.
This conversion is a function
of temperature.  Using the PIMMS\footnote{Portable Interactive Multi-Mission Simulator;
http://asc.harvard.edu/toolkit/pimms.jsp}
software, we find that 
L$_{\rm bol}$(gas)/L$_{\rm X}$(gas)(0.3 $-$ 8.0 keV) drops from 
2.39 at 0.3 keV to 1.39 at 
1.0 keV.   

In re-running the correlation analysis, we
find that the basic conclusions of this paper do not change dramatically with these different
assumptions about the temperature.   The Spearman coefficients
and the best-fit relations change slightly with different assumptions
about the temperature, but the basic conclusions remain the same.
For a few of the relations that have correlation coefficients near our
`weak'/`none' boundary or our `strong'/`weak' boundary, 
small changes in
the correlation coefficient 
may reclassify the relation.  The most notable case is the very weak
correlation between M$_{\rm X}$(gas)/SFR and L$_{\rm K}$, which drops below
the cutoff for a `weak' correlation for some of these alternative cases,
but increases slightly in significance for the linear log(T$_{\rm X}$)-log(SFR) case
including the correction for light outside of the Chandra bandpass.
This emphasizes that the
M$_{\rm X}$(gas)/SFR $-$ L$_{\rm K}$
correlation is very marginal, and more data is needed to
confirm or refute it.  

For most of the relations discussed above, however,
although the correlation coefficients change slightly with different
assumptions about the temperatures, the classification
of the relation does not change.  Thus the conclusions of this paper are not 
strongly influenced by our lack of temperature measurements.

\section{Discussion}

We calculated the volume, mass, and electron
density of the hot X-ray-emitting gas in our sample galaxies,
and compared with other properties of the galaxies, including the
SFR, L$_{\rm K}$, the mass of cold gas, and the SFE.
We have searched for correlations between a large number of
variables, and discovered 
several new 
correlations and anti-correlations in our data.  
These, and many apparent non-correlations, are listed in Table 5.

\subsection{Volume and M$_{\rm X}$(gas) vs.\
SFR and L$_{\rm K}$ }

Some of the most important correlations are:

\renewcommand{\labelitemi}{}
\begin{itemize} 
\item (1) 
The volume of hot gas increases as the SFR goes up,
with a high correlation
coefficient (Figure 9).
When galaxies with SFR $<$ 1~M$_{\sun}$~yr$^{-1}$
are excluded, the slope of the 
best fit
log volume $-$ log SFR line is 0.97 $\pm$ 0.15 (Figure 9).
Including low SFR systems flattens this relation.

\item (2) The volume of hot gas is also correlated with L$_{\rm K}$,
but with a smaller correlation coefficient (Figure 11).

\item (3) The volume of hot gas also correlates
with SFE, 
L$_{\rm FIR}$/L$_{\rm K}$, 
and 
[3.6] $-$ [24], 
but 
only weakly (Figures 10 and 11, and Table 5).

\item (4) There is a strong correlation between 
M$_{\rm X}$(gas) 
and SFR (Figure 13).  
The slope of the log-log plot is 0.88 $\pm$ 0.10
when low SFR galaxies are excluded,  
consistent with a simple M$_{\rm X}$(gas) $\propto$ SFR
relation.
This relation flattens when
low SFR systems are included.  

\item (5) 
M$_{\rm X}$(gas) is also correlated with L$_{\rm K}$ (Figure 13),
but with a lower correlation coefficient than 
M$_{\rm X}$(gas) 
and SFR.

\item (6) As the SFR increases, 
M$_{\rm X}$(gas)/(M$_{\rm H_2}$ + M$_{\rm HI}$) 
goes up (Figure 15),
especially when a variable CO/H$_2$ ratio is used and when low SFR systems are excluded.
For the latter case, the correlation
is strong and the slope of the log-log plot is consistent with one.

\item (7) M$_{\rm X}$(gas)/(M$_{H_2}$ + M$_{HI}$) is weakly
correlated with L$_{\rm K}$ when a variable CO/H$_2$ ratio
is used (Figure 16).

\item (8) There is a trend of increasing
M$_{\rm X}$(gas)/(M$_{\rm H_2}$ + M$_{\rm HI}$) 
ratio with increasing SFE, especially when
a variable CO/H$_2$ ratio is used (Figure 15).  
This trend is weaker
than the relation with SFR.

\item (9) 
M$_{\rm X}$(gas)/(M$_{\rm H_2}$ + M$_{\rm HI}$) 
is weakly correlated with F$_{60}$/F$_{100}$
(Figure 16).

\end{itemize}

For high SFR systems, 
the linear relations
between volume and SFR, 
and between
M$_{\rm X}$(gas) and SFR,
can be explained 
in a straight-forward manner: 
a larger SFR means more supernovae and
more stellar winds, which produce a larger volume of hot gas and a larger
M$_{\rm X}$(gas).
For galaxies with SFR $>$ 1~M$_{\sun}$~yr$^{-1}$, hot gas 
associated with star formation dominates M$_{\rm X}$(gas), 
and any contribution from processes associated with the older
stellar population is negligible. 
However, for galaxies with lower SFRs and high K band luminosities
(and therefore low sSFRs) we find evidence for excess hot
gas relative to the linear M$_{\rm X}$(gas)$-$SFR relation.  This may be due to
contributions to the X-ray-emitting hot gas from other sources, 
perhaps mass loss from older stars that has been virialized
in the gravitational potential. 

The weaker correlation between volume and SFE 
compared to volume vs.\ SFR
is accounted for by the
fact that some high SFE systems have only moderate SFRs, and it is the SFR
that controls the number of supernovae and the amount of stellar wind, not
the SFE.
The weakness of the correlation
between volume and the sSFR as measured by [3.6] $-$ [24] 
and L$_{\rm FIR}$/L$_{\rm K}$
may be explained in a similar manner.

The correlations between L$_{\rm K}$ and the hot gas mass, and between
L$_{\rm K}$ and 
the volume
of hot gas, may be indirect results of 
the correlation between SFR and 
L$_{\rm K}$.
The SFR-L$_{\rm K}$ correlation, in turn, is a consequence
of the fact that most of the galaxies in our
sample are star-forming galaxies on the galaxy main sequence. 
Because the volume$-$L$_{\rm K}$ and M$_{\rm X}$(gas)$-$L$_{\rm K}$ 
correlations are weaker than the volume$-$SFR and M$_{\rm X}$(gas)-SFR
correlations, we conclude that star formation
is more directly responsible for the
hot gas, not the older stellar population.

The strong correlation
between the hot-to-cold gas mass ratio and the SFR,
in contrast to the weak correlation between
the hot-to-cold gas mass ratio and
L$_{\rm K}$,
confirms that the younger stellar population 
is primarily responsible for the hot gas,
not older stars.
The amount of hot gas in our galaxies is small compared to the amount
of colder gas (see Table 2), so conversion of colder material into hot gas
affects the numerator in 
M$_{\rm X}$(gas)/(M$_{\rm H_2}$ + M$_{\rm HI}$) 
but not noticeably the denominator.
The higher the SFR, the more hot gas that is produced, 
thus the 
M$_{\rm X}$(gas)/(M$_{\rm H_2}$ + M$_{\rm HI}$) 
ratio
is directly correlated with the SFR.
The linear log M$_{\rm X}$(gas)/(M$_{\rm H_2}$ + M$_{\rm HI}$) vs.\ log SFR
relation for high SFR systems 
provides additional support for the
idea that the hot gas is mainly due to
young stars in these galaxies.  
The flattening of this relation at lower SFR again
indicates excess hot gas in low SFR, low sSFR systems.

The strong correlation
between M$_{\rm X}$(gas)/(M$_{\rm H_2}$ + M$_{\rm HI}$) and SFR
is consistent with the recent \citet{moreno19} simulations of 
star formation and
feedback in galaxy
mergers, in which they investigate the relative amounts of hot, warm,
cool, and cold-dense gas.  In their models, 
the interaction causes an increase in the amount of
cold ultra-dense interstellar gas by a factor of about three on average.
This
enhances the SFR.  
The amount of hot gas increases during the starburst (by about
400\%), 
while the total amount of cold and warm gas mass decreases only slightly or
remains constant. 
The net effect would be an increase in 
M$_{\rm X}$(gas)/(M$_{\rm H_2}$ + M$_{\rm HI}$) 
during the
burst, consistent with our correlation with SFR.
In the \citet{moreno19} models, 
the hot gas is produced solely by stellar/supernovae feedback;
they do not include AGN feedback or a pre-existing hot halo.

The larger scatter in the 
M$_{\rm X}$(gas)/(M$_{\rm H_2}$ + M$_{\rm HI}$) 
vs.\ SFE correlation  
and its weaker correlation compared to 
M$_{\rm X}$(gas)/(M$_{\rm H_2}$ + M$_{\rm HI}$) 
vs.\ SFR is likely due to some low SFR systems having
high SFEs; it is the SFR that directly controls the amount of hot gas
rather than the SFE.

The weak trend of increasing 
M$_{\rm X}$(gas)/(M$_{\rm H_2}$ + M$_{\rm HI}$) 
with increasing F$_{60}$/F$_{100}$ ratio 
may be another indirect consequence of the 
M$_{\rm X}$(gas)/(M$_{\rm H_2}$ + M$_{\rm HI}$) vs.\ SFR
correlation.  Since F$_{60}$/F$_{100}$ increases with 
SFR on average for our sample galaxies (Figure 2), 
galaxies with higher
M$_{\rm X}$(gas)/(M$_{\rm H_2}$ + M$_{\rm HI}$) and SFR
tend to have larger 
F$_{60}$/F$_{100}$. 

Another factor that may affect M$_{\rm X}$(gas) is
escape of hot gas from the gravitational field of the galaxy, particularly
in low mass systems.  Our data shows a hint of lower M$_{\rm X}$(gas)/SFR
for low L$_{\rm K}$ systems (Figure 13).  However, 
this is uncertain since our sample only includes a few low mass systems.
The majority of galaxies in our sample lie in only a small range of 
L$_{\rm K}$ (10$^{10}$ L$_{\sun}$ $-$ 
10$^{11.5}$ L$_{\sun}$), thus it is difficult
to find trends with L$_{\rm K}$ in our sample.
Low mass galaxies may have lower ratios of baryonic mass M$_{\rm baryon}$
to dynamical mass M$_{\rm dyn}$
compared to high mass systems
(e.g., \citealp{cote2000, torres2011}).
The lower M$_{\rm baryon}$/M$_{\rm dyn}$ in low mass systems has been attributed to
either mass loss from galactic winds (e.g., \citealp{vandenbosch2000, brook2012})
or less efficient infall into lower mass dark halos (e.g., \citealp{sales2017}).
A deficiency in hot gas in low mass systems, if confirmed, may point to increased
escape of baryons via winds.  A larger Chandra imaging survey 
including more low mass systems 
would be helpful to better characterize M$_{\rm X}$(gas)/SFR 
and the scatter in this ratio for low
mass galaxies.

To search for additional
evidence that the hot gas content in our sample galaxies is
affected by the mass and/or the older stellar population in addition to 
the SFR, we calculated the residuals from the best-fit linear relations for
log SFR vs. log L$_{\rm K}$, 
log M$_{\rm X}$(gas) vs. log SFR,
and 
log M$_{\rm X}$(gas) vs. log L$_{\rm K}$.  
We then searched for correlations
between these residuals (top three panels in Figure 19).  
Strong correlations between these residuals
might suggest the existence of a `fundamental plane' of
M$_{\rm X}$(gas)
vs.\ log SFR vs.\ log L$_{\rm K}$.  

A weak correlation 
is seen between the residuals of 
log M$_{\rm X}$(gas) vs. SFR
and those of 
M$_{\rm X}$(gas) vs. L$_{\rm K}$ (Figure 19, upper right).
A strong correlation 
is seen between the residuals of 
log M$_{\rm X}$(gas) vs. log L$_{\rm K}$
and 
those of 
log SFR vs. log L$_{\rm K}$ (Figure 19, left panel, middle row).
The most discrepant galaxies in this plot
are NGC 5018, NGC 2865, and Arp 222 (the three stage 7 mergers
in the lower left corner).  
All of these were identified
in Paper~I as post-starbursts, and all have low sSFR (i.e.,
have large negative residuals in 
the log SFR vs. log L$_{\rm K}$ relation).
They also have large negative residuals compared to the best-fit
log M$_{\rm X}$(gas) vs. log L$_{\rm K}$ relation.  
What is discrepant about these galaxies is their K band luminosities,
which are high relative to their SFRs.  
Figure 19 shows that 
NGC 1700 has a high relative M$_{\rm X}$(gas) compared to these other
low sSFR galaxies.

In the bottom row of Figure 19, we also compared these residuals with L$_{\rm K}$ and the SFR.
There is a positive correlation between the residuals
of the M$_{\rm X}$(gas) vs.\ L$_{\rm K}$ relation, and the SFR 
(Figure 19, bottom left).
Galaxies with low SFRs tend to be deficient in M$_{\rm X}$(gas) compared to
the M$_{\rm X}$(gas) vs. 
L$_{\rm K}$ relation.  That is because they also tend to have low sSFRs,
and it is the SFR that determines the mass of hot gas, not L$_{\rm K}$.
The galaxies in the lower corner of that plot have low SFRs compared to
their K band luminosities, and therefore they have 
low M$_{\rm X}$(gas) compared
to their L$_{\rm K}$.

We also see a positive correlation between the residuals
of the M$_{\rm X}$(gas) vs.\ SFR relation, and L$_{\rm K}$
(Figure 19, bottom right panel).
NGC 1700 stands out as having excess hot gas, while 
systems with low K band luminosities tend to have less hot gas
relative to the M$_{\rm X}$(gas) vs.\ SFR relation.
This suggests that some gas
may have been lost from these systems.
Unfortunately, our sample only contains a few galaxies with
low K band luminosities, and only a few galaxies with low sSFR,
so this result is uncertain.

We conclude that only a few galaxies in our sample deviate from 
a straight line in the log M$_{\rm X}$(gas) 
$-$ log SFR $-$ log L$_{\rm K}$ plane.
To better understand these deviations, it would be helpful to increase
the number of post-starburst galaxies in our sample, as well
as the number of low mass galaxies.

\subsection{Trends with M$_{\rm X}$(gas)/SFR and 
n$_{\rm e}$$\sqrt{f}$ }

In addition to the strong positive correlations discussed above,
some weak anti-correlations are also seen in the data:

\begin{itemize} 

\item (1) There are weak anti-correlations between
M$_{\rm X}$(gas)/SFR and F$_{60}$/F$_{100}$, and between
M$_{\rm X}$(gas)/SFR and [3.6] $-$ [24] (Figure 14).
These anti-correlations hold even when low SFR systems are excluded.

\item (2) As 
M$_{\rm X}$/SFR goes up,
n$_{\rm e}$$\sqrt{f}$ goes down, 
even when low SFR systems are excluded (Figure 14).
This is also a weak trend.

\item (3) There is a 
weak
trend of decreasing 
volume 
with increasing 
n$_{\rm e}$$\sqrt{f}$ 
(Figure 11).

Some parameters are neither correlated or anti-correlated:

\item (1) 
M$_{\rm X}$(gas)/SFR is not correlated with SFR, 
if low SFR systems are excluded
(Figure 13).
M$_{\rm X}$(gas)/SFR 
is not correlated with 
for a variable CO/H$_2$ ratio
(Figure 14).

\item (2) The SFR, [3.6] $-$ [24],
and F$_{60}$/F$_{100}$ do not correlate with 
n$_{\rm e}$$\sqrt{f}$ 
(Figure 12).

\item (3) No significant correlations are found
between M$_{\rm X}$(gas)/SFR and volume (Table 5), 
and no correlation between
M$_{\rm X}$(gas)/SFR and L$_{\rm FIR}$/L$_{\rm K}$
when low SFR systems are excluded
(Figure 14).

\end{itemize}

Although M$_{\rm X}$(gas)/SFR is anti-correlated with
F$_{60}$/F$_{100}$ and with [3.6] $-$ 24], 
M$_{\rm X}$(gas)/SFR 
is not 
correlated (either positively or negatively) with SFR,
in spite of the fact that F$_{60}$/F$_{100}$ and
[3.6] $-$ [24] are both (weakly) correlated with SFR.
Furthermore, although volume and 
n$_{\rm e}$$\sqrt{f}$ are weakly anti-correlated,
and 
SFR is correlated with volume, 
n$_{\rm e}$$\sqrt{f}$ is not correlated with SFR.

These results suggest that another factor contributes to the
observed variations in
M$_{\rm X}$(gas)/SFR 
and 
n$_{\rm e}$$\sqrt{f}$ 
besides SFR.
One possibility is differences in timescale; variations in the
age of an on-going starburst or the time since the end
of a starburst may affect 
n$_{\rm e}$$\sqrt{f}$ and  
M$_{\rm X}$(gas)/SFR, as well as other parameters of the system. 
Numerical simulations show that 
interaction-triggered 
starbursts can last for extended periods ($\ge$100 Myrs;
\citealp{lotz00, dimatteo08, bournaud11, fensch17}).
This timescale is similar to the radiative cooling times for the gas
(median of 60 Myrs, see Section 5); it is also similar
to the timescale over which the UV data is measuring the SFR ($\sim$100 Myrs;
\citealp{kennicutt12}).  If the cooling time is less than 
the timescale over which the SFR is measured, and if the cooling time is less than 
the age of the burst,
then late in a burst the M$_{\rm X}$(gas)/SFR
may decrease (i.e., some hot gas has cooled, but the UV-bright stars contributing
to our SFR estimate have not yet died). 
The sSFR as measured by [3.6] $-$ [24] and L$_{\rm FIR}$/L$_{\rm K}$ may also vary with time
during a burst.
Presumably 
the 
electron density and/or filling factor also evolve with time
during a burst, along
with F$_{60}$/F$_{100}$, the volume of hot gas, and 
M$_{\rm X}$(gas)/SFR.   Further theoretical
modeling is needed to better understand
the relationships between these parameters in evolving starbursts.

A second factor that may contribute to variations in
M$_{\rm X}$(gas)/SFR 
and 
n$_{\rm e}$$\sqrt{f}$ 
may be the
efficiency of early feedback.
According to numerical simulations,
stellar winds and radiation pressure early in a starburst
disrupt molecular
clouds, making it easier for subsequent supernovae to produce hot gas
\citep{hopkins12a, agertz13, hopkins13b}.
The efficiency of early feedback might be related to the
spatial density of star formation; more concentrated distributions
of young stars may have more early UV radiation per volume, allowing
quicker destruction of molecular gas.
This may lead to easier escape for hot gas from the region,
and thus less diffuse X-ray emission. 
More concentrated distributions of
young stars would presumably lead to more intense 
UV interstellar radiation fields and therefore
hotter dust and 
higher F$_{60}$/F$_{100}$ ratios (e.g., \citealp{desert90}).
The F$_{60}$/F$_{100}$ ratio
is weakly anti-correlated
with 
M$_{\rm X}$(gas)/SFR, consistent with this scenario.
The [3.6] $-$ [24] color may also increase with higher spatial concentrations
of young stars, and
[3.6] $-$ [24] 
is also weakly anti-correlated with 
M$_{\rm X}$(gas)/SFR.   
Further study is needed to investigate how all of these
parameters vary with the density of OB stars in a galaxy. 

A third factor that might 
affect
M$_{\rm X}$(gas)/SFR is the 
initial mass function (IMF). 
A top-heavy IMF may lead to an increase in
supernovae compared to lower mass stars, which might produce a larger
M$_{\rm X}$(gas)/SFR when the SFR is derived from the UV continuum.
It has been suggested that high SFR and/or high SFE galaxies 
may have IMFs skewed to high mass stars 
\citep{rieke80, elbaz95, brassington07, koppen07, weidner13, brown19}.  
Thus one might expect higher 
M$_{\rm X}$(gas)/SFR for higher SFR or higher SFE systems.
However, 
we do not see a correlation between
M$_{\rm X}$(gas)/SFR 
and SFR, or 
between M$_{\rm X}$(gas)/SFR 
and
SFE.
This means that either IMF variations are not responsible for the spread 
in 
M$_{\rm X}$(gas)/SFR, or the IMF is not correlated with SFR
or SFE.

Another factor that might 
affect
M$_{\rm X}$(gas)/SFR 
and 
n$_{\rm e}$$\sqrt{f}$ 
is metallicity.
A number of studies have concluded that
the SFR of star-forming galaxies depends upon metallicity in addition
to stellar mass;
for the same stellar mass, lower metallicity systems have higher SFRs
(\citealp{ellison08, mannucci10, lara10, hirschauer18}, but 
see \citealp{izotov14,izotov15}).
This result has been explained by infall of low metallicity gas,
fueling star formation.
Our 
M$_{\rm X}$(gas)/SFR 
values may be artificially skewed by metallicity, since
the value of L$_{\rm X}$(gas) that is derived from the Chandra spectra
is affected by metallicity (see Paper~I).  In addition, 
the fraction of the supernovae and stellar wind energy converted
into X-ray flux may be a function of metallicity.
A larger sample of galaxies including more low metallicity systems
would be helpful to investigate this issue further.

Unfortunately, we do not have a measure
of the volume filling factor of the hot gas, f, independently of
n$_{\rm e}$, 
to determine whether f varies
significantly from system to system.  Based on theoretical
arguments and/or hydrodynamical simulations, 
for a range of systems 
f has variously been estimated to be 
70$-$80\% \citep{mckee77},  20$-$40\% 
\citep{breitschwerdt12},
30$-$40\% \citep{kim17},
or anywhere between 10$-$90\% depending upon the supernovae rate
and the average gas density
\citep{li15}.   In general, according to simulations 
the higher the density of star formation,
the larger the expected hot gas filling factor \citep{breitschwerdt12, li15}.
One might expect higher SFRs
to produce faster winds, as has been found for the warm ionized
medium (e.g., \citealp{heckman15}).  A faster wind may lead to lower
n$_{\rm e}$ values.   If the filling factor increases with
SFR but 
n$_{\rm e}$ decreases, this might explain 
the lack of a trend 
between
n$_{\rm e}$$\sqrt{f}$ and SFR.
Independent determinations of 
n$_{\rm e}$ and f (e.g., \citealp{kregenow06, jo19}) 
are needed to test this possibility.

\subsection{The Scatter in M$_{\rm X}$(gas)}

One of the major conclusions of the current paper is that, 
excluding low SFR systems, M$_{\rm X}$(gas)/SFR is constant
with SFR 
with an rms spread of only 0.34 dex.  
A number of factors may contribute to this scatter,
in addition to age, metallicity, or IMF differences.
First, 
as discussed in Section 5, the decomposition of the X-ray spectrum into a thermal
and a non-thermal component introduces some uncertainty, adding uncertainty
to our determination of L$_{\rm X}$(gas) (see Paper~I).
Second, as also discussed in Section 5, the unknown extent of the hot gas along our line of sight leads to
uncertainties in the volume of the hot gas, which contributes to the 
scatter in the parameters derived from the volume.
Systematic variations in the geometry of the hot gas may
further affect the observed relations.  For example, systems with lower 
SFR may have disk-like distributions of cold gas, with coronal
gas extending 
out of the galactic plane, while higher SFR systems, which are more
likely to be in the midst of a merger, may have gas distributions which
are more spherical.
Another factor that may contribute to the scatter 
are system-to-system variations in the gravitational masses of the galaxies,
which likely affect outflow rates and potential loss of hot gas. 
We found that galaxies with low K band luminosities tend to have lower
M$_{\rm X}$(gas)/SFR ratios compared to other galaxies (Figure 16), suggesting
that low mass galaxies may lose some hot gas.   
The large-scale environment may also affect the 
M$_{\rm X}$(gas)/SFR ratio, however,  
L$_{\rm X}$(gas)/SFR is not correlated with local galaxy density (Paper~I);
M$_{\rm X}$(gas)/SFR and M$_{\rm X}$(gas)/(M$_{\rm H_2}$+M$_{HI}$)
are also not 
correlated with 
local galaxy density.
Another factor that may contribute to the observed scatter in these plots is 
our assumption of a temperature of kT = 0.3 keV for the hot gas
in the
systems without an X-ray determination of temperature.
Longer Chandra exposures would be useful
to spectroscopically determine the temperature of the gas in more of the galaxies.

\subsection{NGC 1700 }

As noted several times in this paper, the late-stage merger NGC 1700 does not fit
some of the strong relations seen in this study.  NGC 1700 has a large X-ray size 
relative to its SFR. It also has a high X-ray luminosity and a large mass of hot X-ray-emitting gas. 
This suggests that either NGC 1700 is in a special
evolutionary state compared to the other systems in our sample, or it acquired its hot gas via a different
process.  
Maybe NGC 1700 was a pre-existing elliptical that already had a large amount 
of hot gas, which then swallowed a gas-rich galaxy.  It is sometimes
difficult to distinguish between the remnant of a spiral-spiral merger and the remnant of an elliptical-spiral merger.
In appearance, NGC 1700 is an elliptical-like galaxy surrounded by tidal debris,
but its merger history is uncertain.
It was classified as the remnant
of a spiral-spiral major merger by \citet{schweizer92} and \citet{brown00}, however, \citet{statler96} and \citet{kleineberg11}
conclude that it is the result of the merger of at least three galaxies, two large spirals and a third smaller galaxy.
If 
NGC 1700 is the product of a single major merger, perhaps it is in a later stage in the 
conversion from a major merger to an elliptical than the other post-merger galaxies in our sample.
Theory suggests that ellipticals produced by major mergers can build a large quantity of hot gas by
the virialization 
of gas lost from red giants
in the gravitational potential well, 
with possible heating by Type Ia supernovae and/or 
AGN feedback \citep{ciotti91, ciotti17, pellegrini98, mathews03}.  This process is expected to be very slow, with timescales of many gigayears.
Expanding our sample to include more galaxies like NGC 1700 would be 
helpful to better understand how hot gas grows in such systems.
More generally, increasing the number of low sSFR galaxies in our
sample is needed to investigate how the hot gas in galaxies evolves
as star formation fades in a quenched or quenching galaxy.

\section{Summary}

We have 
measured the spatial extent of the hot interstellar gas in a sample
of 
49
interacting and merging galaxies in the nearby Universe.
For systems with SFR $>$ 1 M$_{\sun}$~yr,
we found strong near-linear correlations between 
the volume of hot gas and the SFR,
and between M$_{\rm X}$(gas) and SFR.
This supports
the idea that supernovae and stellar winds are responsible for
the hot gas.
As expected,
the M$_{\rm X}$(gas)/(M$_{\rm H_2}$ + M$_{\rm HI}$) 
ratio also increases linearly with increasing SFR for high SFR systems.
These results are consistent with
recent hydrodynamical
simulations of interactions including feedback.
The M$_{\rm X}$(gas)/(M$_{\rm H_2}$ + M$_{\rm HI}$) ratio also increases
with dust temperature on average, perhaps due to a larger proportion of
dust associated with the hot gas.

In low SFR, low sSFR systems,
we find evidence for an excess of hot gas relative to the relations
for higher SFR systems.
This excess may be associated with mass loss from older stars.
However, our sample only includes a few galaxies with low sSFR rates, 
so this result
is uncertain.   In addition,
we see a possible deficient of hot gas in low
mass systems, perhaps due to escape from the gravitational field
of the galaxy.  However, this result is also uncertain due to the
small number of low mass systems in our sample.

The M$_{\rm X}$(gas)/SFR
is weakly anti-correlated with F$_{60}$/F$_{100}$, [3.6] $-$ [24],
and n$_{\rm e}$$\sqrt{f}$.
The inferred electron density decreases
with increasing volume of hot gas assuming a constant filling factor.
These results may be 
a consequence of variations in the spatial density of young stars, age
of the stars, metallicity, IMF,
and/or efficiency of feedback in these galaxies.

\acknowledgments

This research was supported by NASA Chandra archive grant AR6-17009X, issued
by the Chandra X-ray Observatory Center, which is operated by the Smithsonian Astrophysical Observatory
for and on behalf of NASA under contract NAS8-03060.
Support was also provided by National Science Foundation Extragalactic Astronomy Grant 
ASTR-1714491.
The scientific results reported in this article are based
on data obtained from the Chandra Data Archive.
This research has also
made use of the NASA/IPAC Extragalatic Database (NED), 
which is operated by the Jet Propulsion Laboratory, 
California Institute of Technology,
under contract with NASA.
This work also utilizes archival data from
the Spitzer Space Telescope, which is operated by
the Jet Propulsion Laboratory (JPL), California Institute
of Technology under a contract with NASA.
This study also uses archival data from the NASA Galaxy
Evolution Explorer (GALEX), which was operated for NASA
by the California Institute of Technology under
NASA contract NAS5-98034.

\appendix{\bf APPENDIX: Chandra 0.3 $-$ 1.0 keV Maps}

For the 44 systems for which we can
measure X-ray radial profiles, 
the unsmoothed Chandra 0.3 $-$ 1.0 keV maps 
are displayed
in the right panels of Figures 20 $-$ 27.
When only one Chandra dataset is available for the galaxy,
the {\it ciao} command {\it fluximage} was used to 
convert into units of photons~s$^{-1}$~cm$^{-2}$~arcsec$^{-2}$, 
using an 
exposure-correction map
with a 0.8 keV effective energy.
When multiple Chandra datasets are available for one system, 
the datasets
have been merged together using the {\it ciao} command {\it merge$\_$obs},
which also does the exposure correction and flux calibration.
The left panels of Figures 20 $-$ 27 show either the SDSS g band image 
(when available) or the GALEX NUV image. 
Contours of the X-ray surface brightness are overlaid on the Chandra images.
These have been lightly smoothed using the ds9 
software\footnote{SAOImageDS9 development has been made possible by funding from the Chandra X-ray Science Center (CXC) and the High Energy Astrophysics Science Archive Center (HEASARC) with additional funding from the JWST Mission office at Space Telescope Science Institute.}, with the
smooth parameter set to 6.

\vfill
\eject

\begin{figure}
\plotone{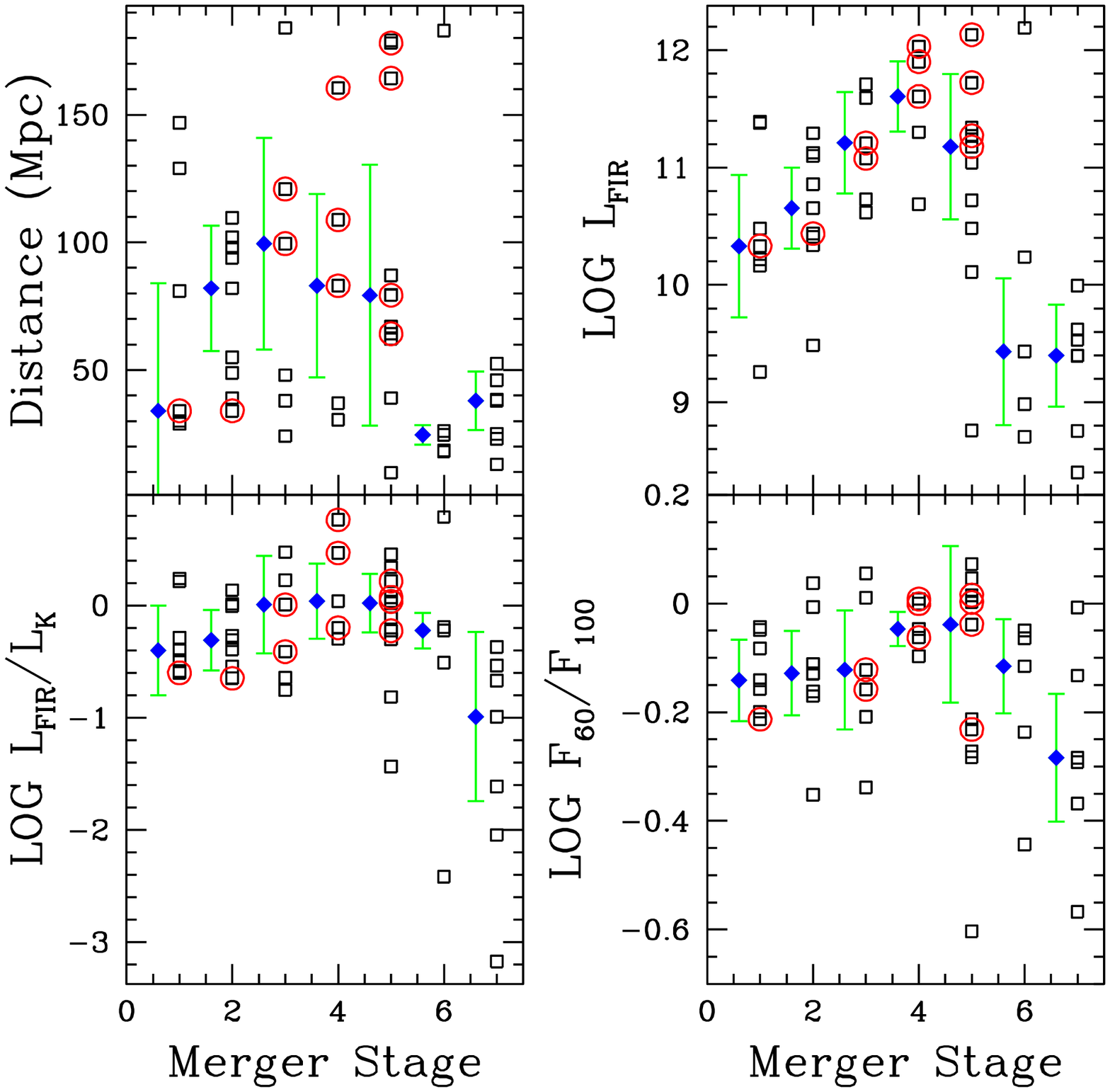}
\caption{
Plots of basic galaxy properties (distance,
L$_{\rm FIR}$, L$_{\rm FIR}$/L$_{\rm K}$, and F$_{60}$/F$_{100}$) 
vs.\ merger stage.
Black open squares mark the sample galaxies, with those circled
by red circles being AGN.
The filled blue diamonds are the median values for each stage, slightly
offset to the left.
The errorbars plotted on the median values are the
semi-interquartile range, equal to half the difference between
the 75th percentile and the 25th percentile.
}
\end{figure}

\begin{figure}
\plotone{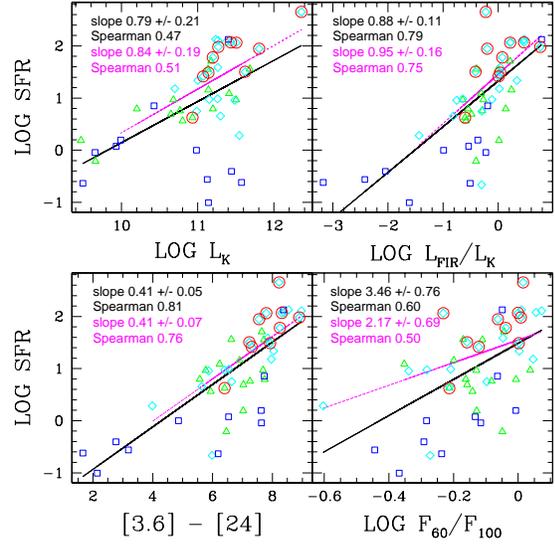}
\caption{
Correlations between basic galaxy properties (SFR vs.\ L$_{\rm K}$,
L$_{\rm FIR}$/L$_{\rm K}$, [3.6] $-$ [24], and F$_{60}$/F$_{100}$).
The best-fit line for the full sample is plotted as a solid black
line, while the best-fit line for systems with SFR $>$ 1.0 M$_{\sun}$~yr$^{-1}$
is given as a dotted line.
The best-fit slope and the Spearman rank correlation coefficient for the full
set is shown in black (on top), while the values for the high SFR
subset are shown in magenta (below).
Merger stages 1 and 2
systems are marked as open green triangles.  Merger stages 3, 4, and 5
are open cyan diamonds, and merger stages 6 and 7 are identified by
blue open squares.
AGN are identified by red circles.
}
\end{figure}

\begin{figure}
\plotone{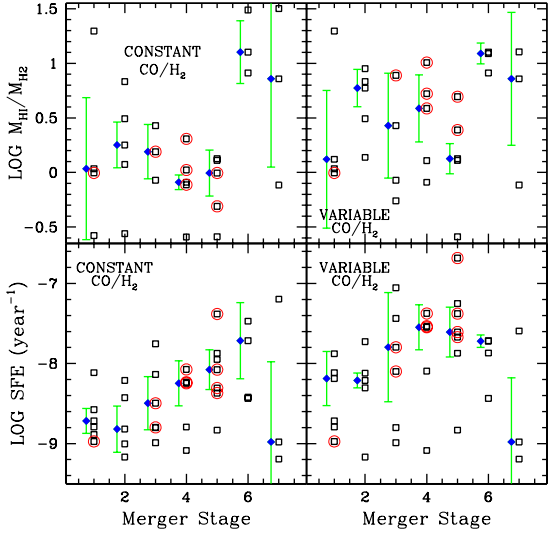}
\caption{
Top row:
plots of the HI mass/H$_2$ mass
against merger stage. 
Bottom row:
plots of the SFE 
against merger stage.  
In the left panels, the standard Galactic
CO/H$_2$ ratio is used for all galaxies.
In the right panels, a variable
CO/H$_2$ ratio is used.
Black open squares mark the sample galaxies, with those circled
by red circles being AGN.
The filled blue diamonds are the median values for each stage, slightly
offset to the left.
The errorbars plotted on the median values are the
semi-interquartile range, equal to half the difference between
the 75th percentile and the 25th percentile.
}
\end{figure}

\begin{figure}
\plotone{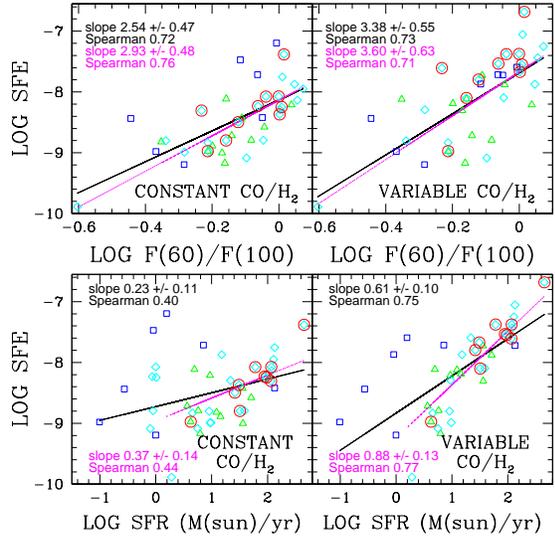}
\caption{
Comparisons between SFE and F$_{60}$/F$_{100}$ (top panels)
and SFE and SFR (bottom panels).  The left panels use
a constant CO/H$_2$ ratio, while the right use a variable CO/H$_2$ ratio.
The best-fit line for the full sample is plotted as a solid black
line, while the best-fit line for systems with SFR $>$ 1.0 M$_{\sun}$~yr$^{-1}$
is given as a dotted line.
The best-fit slope and the Spearman rank correlation coefficient for the full
set is shown in black (on top), while the values for the high SFR
subset are shown in magenta (below).
Merger stages 1 and 2
systems are marked as open green triangles.  Merger stages 3, 4, and 5
are open cyan diamonds, and merger stages 6 and 7 are identified by
blue open squares.
AGN are identified by red circles.
}
\end{figure}

\begin{figure}
\plotone{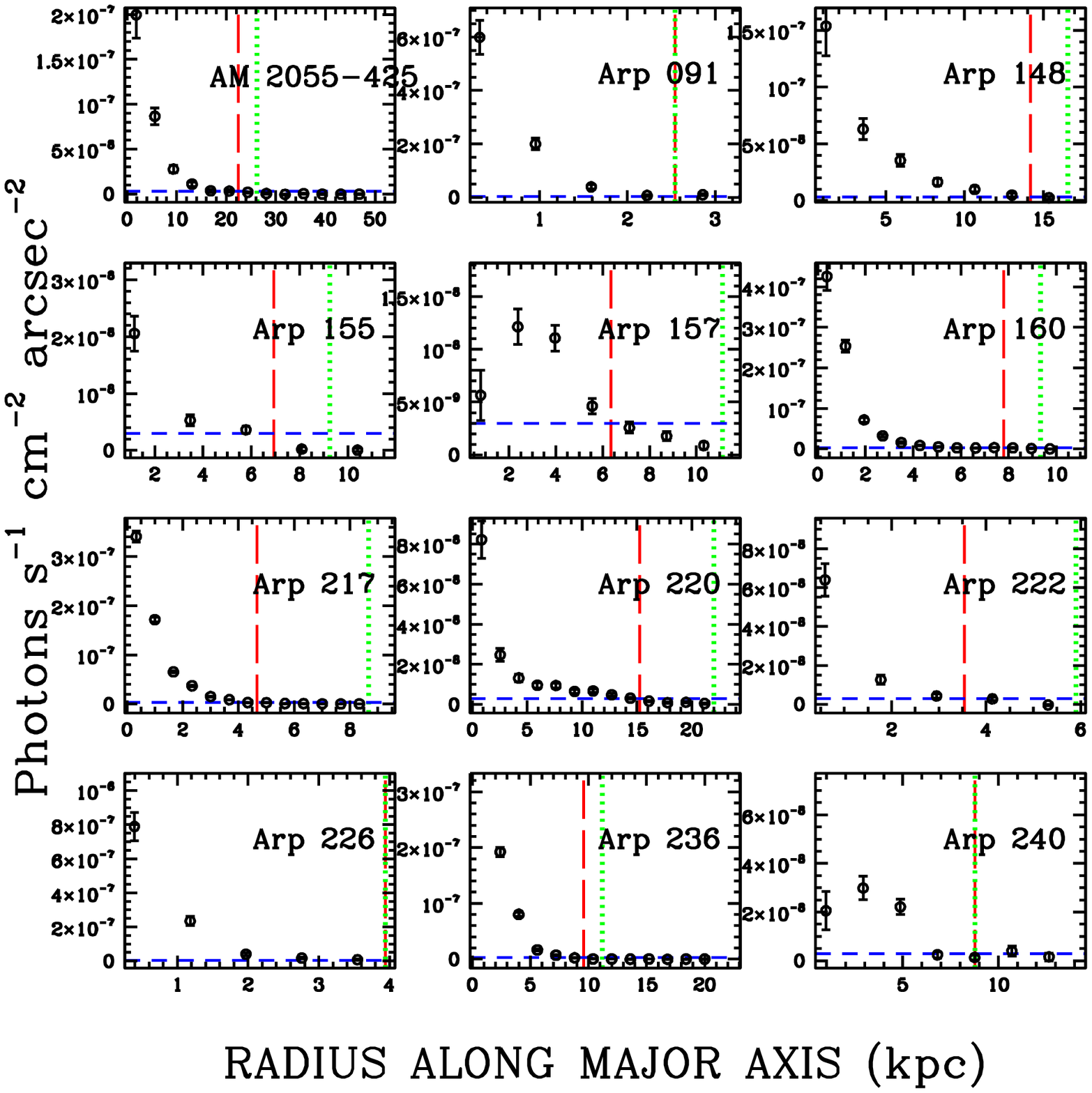}
\caption{
Montage of major axis radial profiles from elliptical annuli,
plotted against the distance along the major axis.
These were obtained using the {\it dmextract} software.
The blue horizontal line (short dashes) mark the nominal surface brightness cutoff
of 3 $\times$ 10$^{-9}$ photons~s$^{-1}$~cm$^{-2}$~arcsec$^{-2}$.
The red vertical line (long dashes) mark the `best' estimate of the radial extent of the
X-ray emission, used for the volume determination.
The green vertical line (dotted) mark the full 2$\sigma$ extent, for the galaxies
with high S/N observations.
See text for more details.
}
\end{figure}

\begin{figure}
\plotone{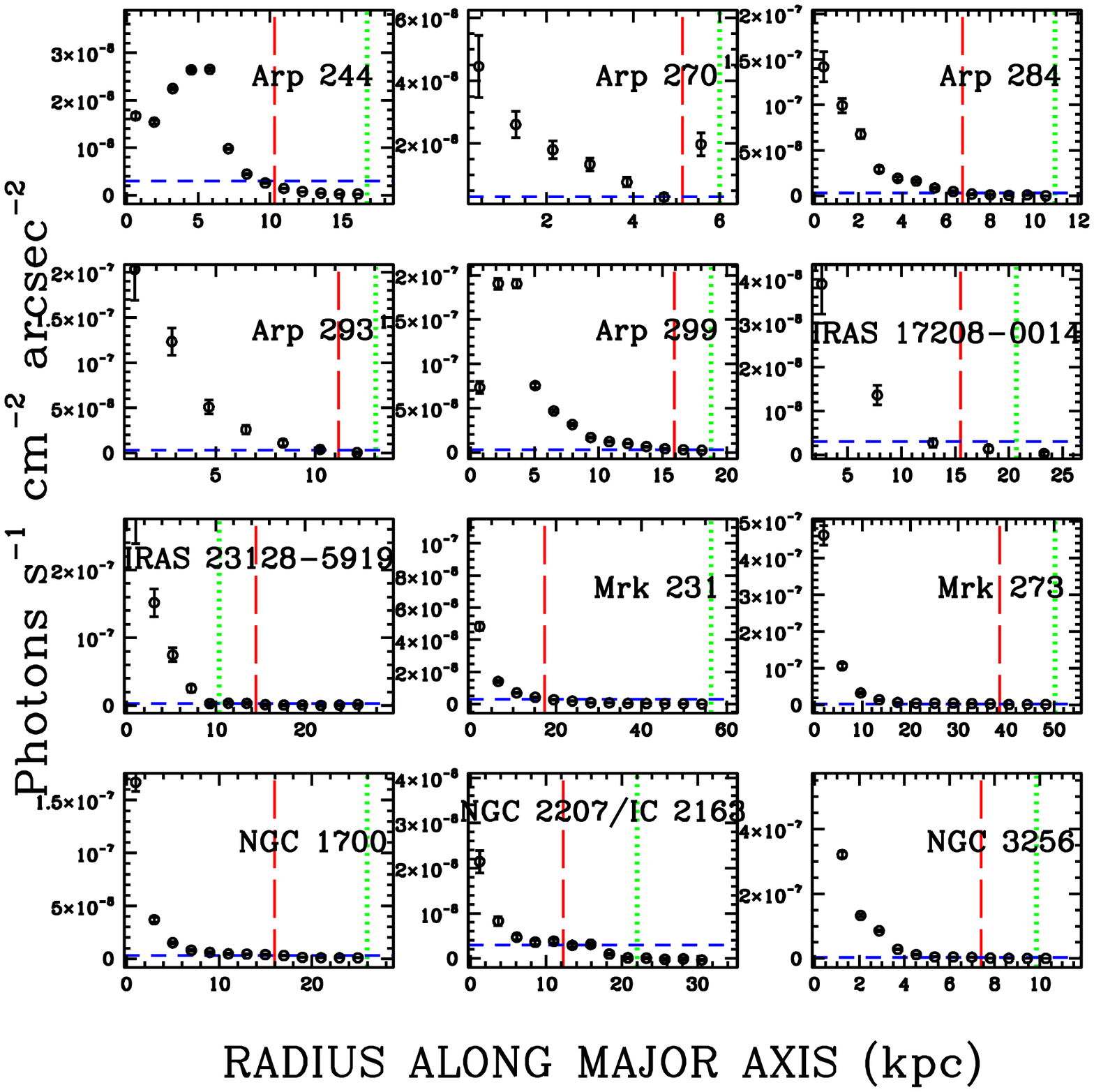}
\caption{
Montage of major axis radial profiles from elliptical annuli,
plotted against the distance along the major axis.
These were obtained using the {\it dmextract} software.
The blue horizontal line (short dashes) mark the nominal surface brightness cutoff
of 3 $\times$ 10$^{-9}$ photons~s$^{-1}$~cm$^{-2}$~arcsec$^{-2}$.
The red vertical line (long dashes) mark the `best' estimate of the radial extent of the
X-ray emission, used for the volume determination. 
The green vertical line (dotted) mark the full 2$\sigma$ extent, for the galaxies
with high S/N observations.
See text for more details.
}
\end{figure}

\begin{figure}
\plotone{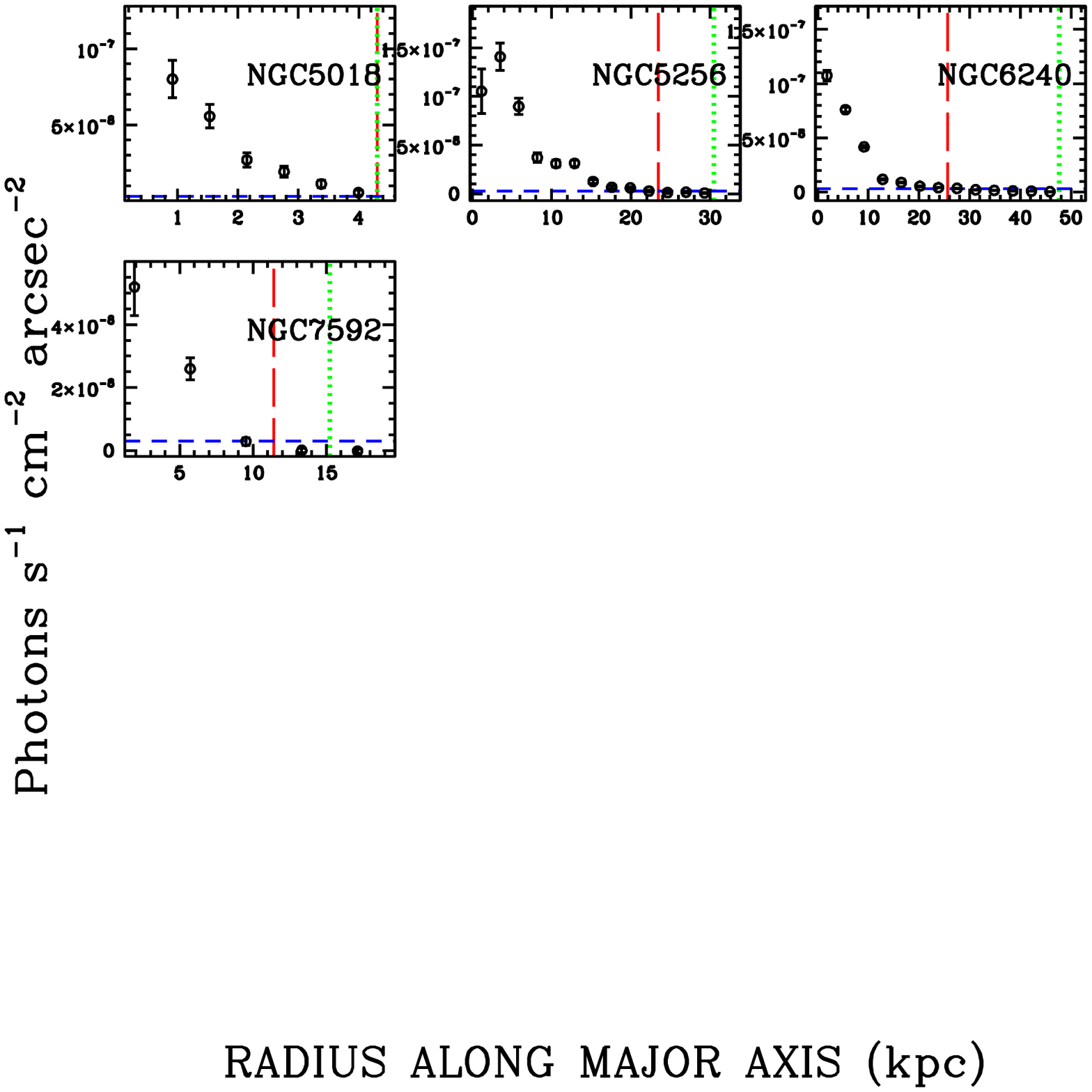}
\caption{
Montage of major axis radial profiles from elliptical annuli,
plotted against the distance along the major axis.
These were obtained using the {\it dmextract} software.
The blue horizontal line (short dashes) mark the nominal surface brightness cutoff
of 3 $\times$ 10$^{-9}$ photons~s$^{-1}$~cm$^{-2}$~arcsec$^{-2}$.
The red vertical line (long dashes) mark the `best' estimate of the radial extent of the
X-ray emission, used for the volume determination.
The green vertical line (dotted) mark the full 2$\sigma$ extent, for the galaxies
with high S/N observations.
See text for more details.
}
\end{figure}

\begin{figure}
\plotone{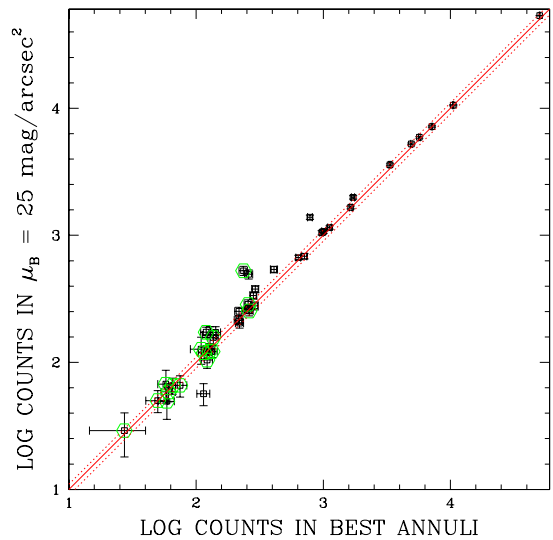}
\caption{
Comparison of soft (0.3 $-$ 1 keV) X-ray counts within
the $\mu$$_{\rm B}$ = 25 mag~arcsec$^{-2}$ isophote,
and within our best determination of the angular extent of the
X-ray emission.  The solid line is the one-to-one relation, while
the dotted lines represent $\pm$10\% variations.
Galaxies marked with open green hexagons are systems with low S/N for which
we used the single aperture method to get the sizes.
}
\end{figure}

\begin{figure}
\plotone{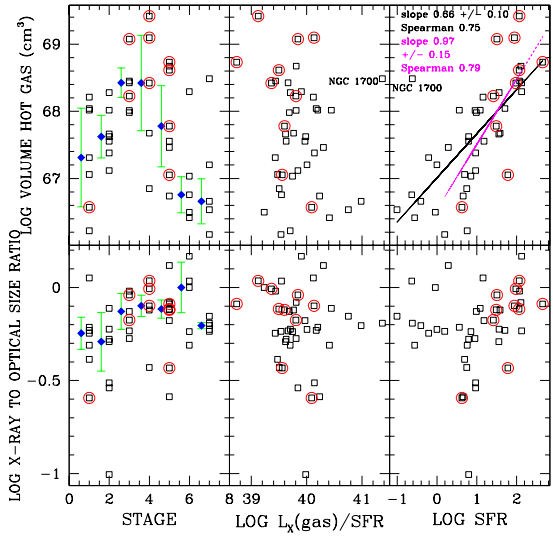}
\caption{
Plots of X-ray volume (top row) and X-ray/optical size ratio 
(bottom row)
vs.\ merger
stage (first column), L$_{\rm X}$/SFR (second column), and SFR
(third column).
The best-fit line is plotted in the top right panel.
Black open squares mark the sample galaxies, with those circled
by red circles being AGN.
In the left column of plots,
the filled blue diamonds are the median values for each stage, slightly
offset to the left.
The errorbars plotted on the median values are the
semi-interquartile range, equal to half the difference between
the 75th percentile and the 25th percentile.
}
\end{figure}

\begin{figure}
\plotone{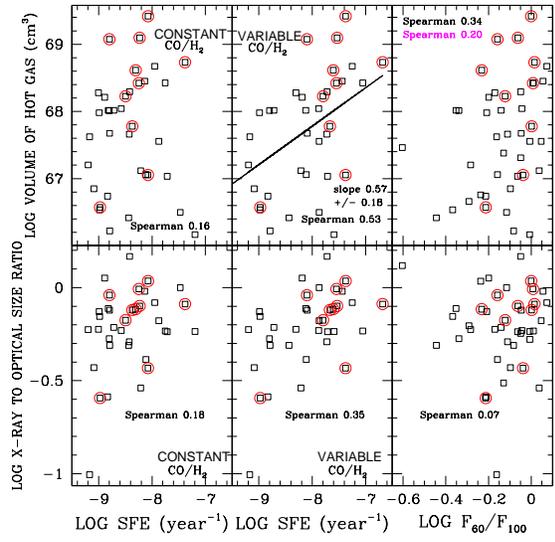}
\caption{
Plots of X-ray volume (top row) and X-ray/optical size ratio 
(bottom row)
vs.\ SFE
calculated with constant CO/H$_2$ ratio (first column), 
SFE calculated with variable CO/H$_2$ ratio (second column), and 
F$_{60}$/F$_{100}$ ratio
(third column).
AGN are identified by red circles.
}
\end{figure}

\begin{figure}
\plotone{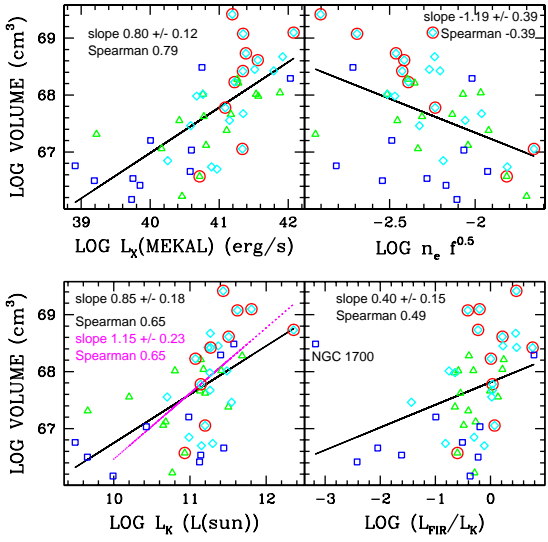}
\caption{
Top left:
the volume of the hot gas plotted against 
L$_{\rm X}$(gas).
The best-fit line is plotted.
Top right: the volume vs.\ 
n$_{\rm e}$$\sqrt{f}$,
where n$_{\rm e}$ is the 
electron density and f is the filling factor.
Bottom left: volume vs.\ L$_{\rm K}$.
Bottom right: volume vs.\ L$_{\rm FIR}$/L$_{\rm K}$.
Merger stages 1 and 2
systems are marked as open green triangles.  Merger stages 3, 4, and 5
are open cyan diamonds, and merger stages 6 and 7 are identified by
blue open squares.
AGN are identified by red circles.
}
\end{figure}

\begin{figure}
\plotone{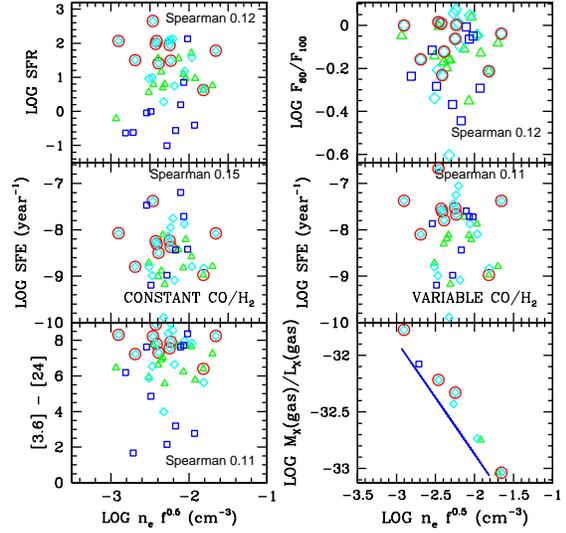}
\caption{
Upper left:
SFR plotted against 
n$_{\rm e}$$\sqrt{f}$.
Upper right:
F$_{60}$/F$_{100}$ vs.\ 
n$_{\rm e}$$\sqrt{f}$. 
The middle row compares 
SFE calculated with the two methods with 
n$_{\rm e}$$\sqrt{f}$. 
Bottom left: [3.6] $-$ [24] vs.\ 
n$_{\rm e}$$\sqrt{f}$. 
Bottom right:
M$_{\rm X}$(gas)/L$_{\rm X}$(gas) vs.\
n$_{\rm e}$$\sqrt{f}$ for the 15 systems with measured temperatures.
The solid blue line in the lower right plot is the assumed
relation assuming a constant temperature of 0.3 keV.
In these plots, n$_{\rm e}$ is the 
electron density and f is the filling factor.
Merger stages 1 and 2
systems are marked as open green triangles.  Merger stages 3, 4, and 5
are open cyan diamonds, and merger stages 6 and 7 are identified by
blue open squares.
AGN are identified by red circles.
}
\end{figure}

\clearpage

\begin{figure}
\plotone{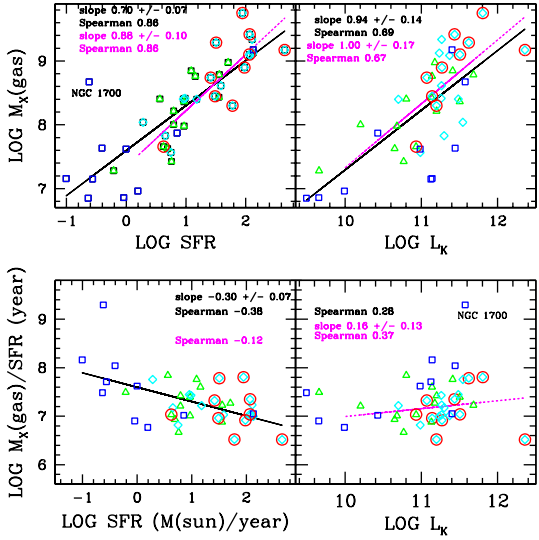}
\caption{
M$_{\rm X}$(gas) vs.\ SFR (top left),
M$_{\rm X}$(gas) vs.\ L$_{\rm K}$ (top right),
M$_{\rm X}$(gas)/SFR vs.\ SFR (bottom left),
and
M$_{\rm X}$(gas)/SFR vs.\ L$_{\rm K}$ (bottom right).
Merger stages 1 and 2
systems are marked as open green triangles.  Merger stages 3, 4, and 5
are open cyan diamonds, and merger stages 6 and 7 are identified by
blue open squares.
Spearman rank correlation coefficients for the full dataset are given in black,
while the Spearman coefficient for the subset of galaxies with
SFR $>$ 1~M$_{\sun}$~yr$^{-1}$ are given in magenta.
When reliable correlations are seen, the best-fit straight line
for the full dataset is shown as a solid black line,
and for the high SFR subset as a dashed magenta line.
AGN are identified by red circles.
}
\end{figure}

\begin{figure}
\plotone{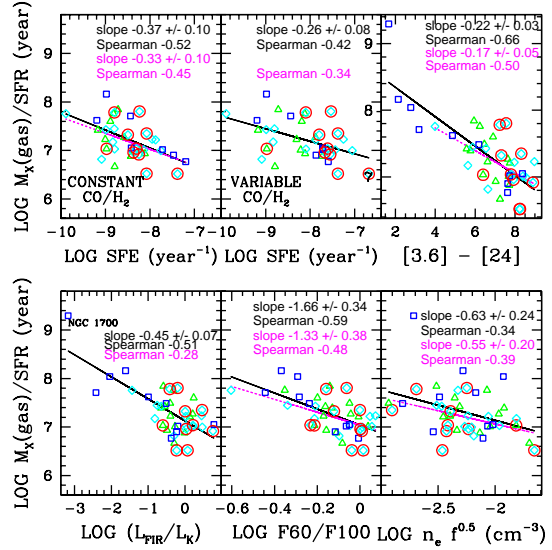}
\caption{
M$_{\rm X}$(gas)/SFR vs.\ the two estimates of SFE (first and second panels, 
top row),
M$_{\rm X}$(gas)/SFR vs.\ [3.6] $-$ [24] (top right),
M$_{\rm X}$(gas)/SFR vs.\ L$_{\rm FIR}$/F$_{\rm K}$ (bottom left),
M$_{\rm X}$(gas)/SFR vs.\ F$_{60}$/F$_{100}$ (bottom middle),
and
M$_{\rm X}$(gas)/SFR vs.\
n$_{\rm e}$$\sqrt{f}$ (bottom right).
Merger stages 1 and 2
systems are marked as open green triangles.  Merger stages 3, 4, and 5
are open cyan diamonds, and merger stages 6 and 7 are identified by
blue open squares.
Spearman rank correlation coefficients for the full dataset are given in black,
while the Spearman coefficient for the subset of galaxies with
SFR $>$ 1~M$_{\sun}$~yr$^{-1}$ are given in magenta.
When reliable correlations are seen, the best-fit straight line
for the full dataset is shown as a solid black line,
and for the high SFR subset as a dashed magenta line.
AGN are identified by red circles.
}
\end{figure}

\begin{figure}
\plotone{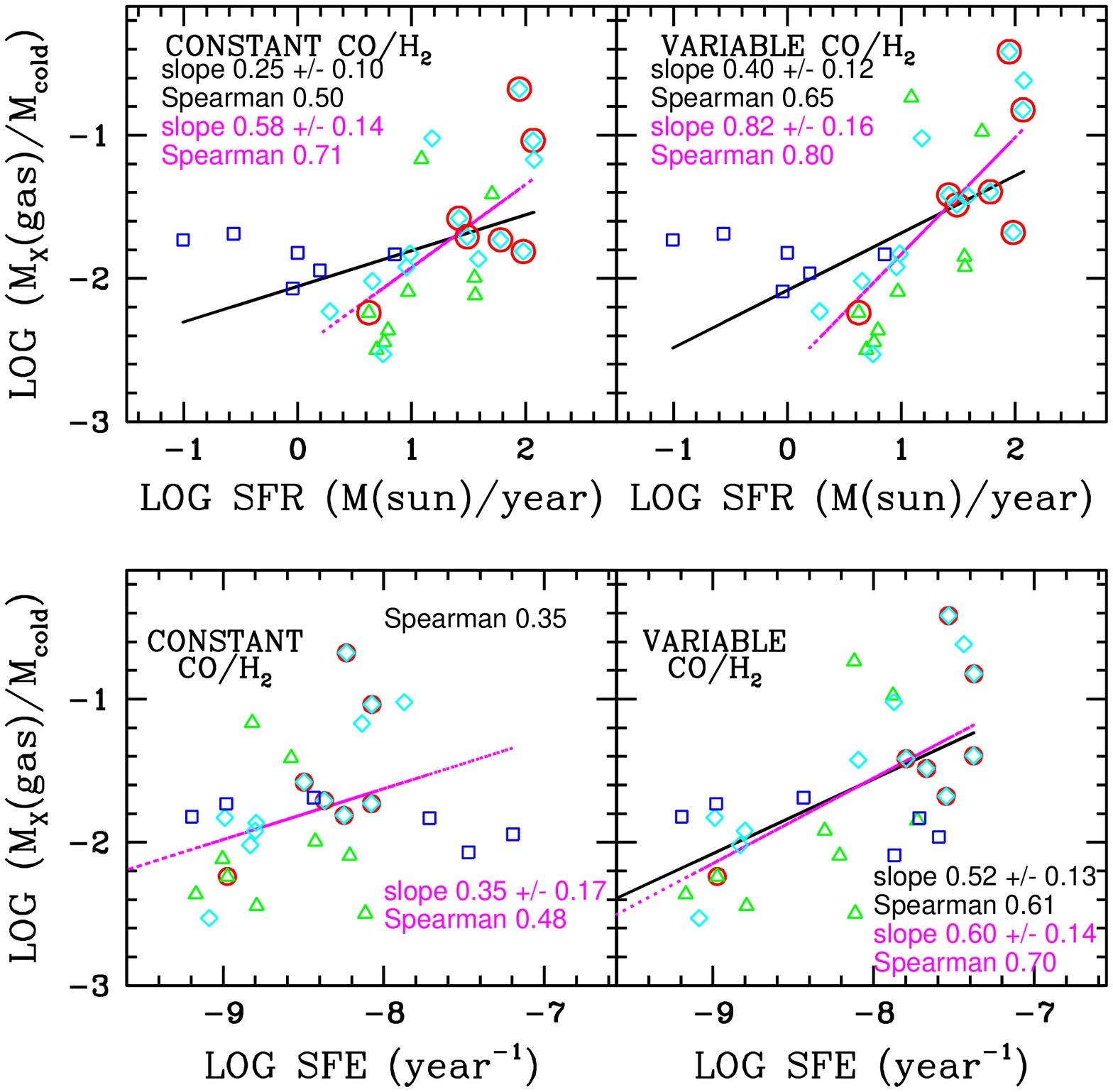}
\caption{
Plots of M$_{\rm X}$(gas)/(M$_{\rm HI}$+M$_{\rm H_2}$) vs.\
SFR (top panels) and SFE (bottom panels).
The left column uses a constant CO/H$_2$ ratio, while the right column
uses a variable ratio.
Merger stages 1 and 2
systems are marked as open green triangles.  Merger stages 3, 4, and 5
are open cyan diamonds, and merger stages 6 and 7 are identified by
blue open squares.
AGN are identified by red circles.
Spearman rank correlation coefficients for the full dataset are given in black,
while the Spearman coefficient for the subset of galaxies with
SFR $>$ 1~M$_{\sun}$~yr$^{-1}$ are given in magenta.
When reliable correlations are seen, the best-fit straight line
for the full dataset is shown as a solid black line,
and for the high SFR subset as a dashed magenta line.
}
\end{figure}

\begin{figure}
\plotone{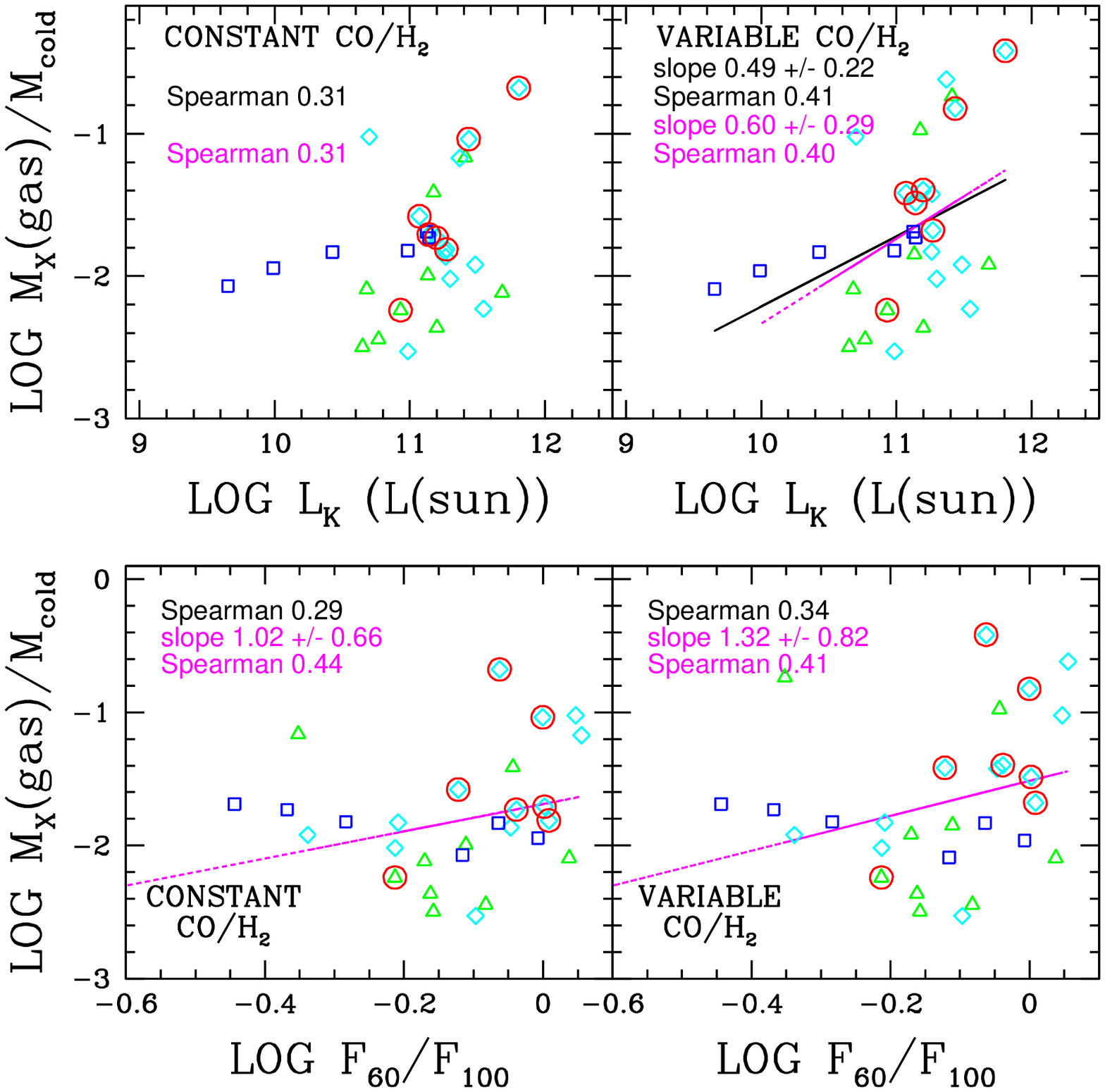}
\caption{
Plots of M$_{\rm X}$(gas)/(M$_{\rm HI}$+M$_{\rm H_2}$)
vs\
L$_{\rm K}$ (top row) and F$_{60}$/F$_{100}$ (bottom row).
The left column uses a constant CO/H$_2$ ratio, while the right column
uses a variable ratio.
Merger stages 1 and 2
systems are marked as open green triangles.  Merger stages 3, 4, and 5
are open cyan diamonds, and merger stages 6 and 7 are identified by
blue open squares.
AGN are identified by red circles.
Spearman rank correlation coefficients for the full dataset are given in black,
while the Spearman coefficient for the subset of galaxies with
SFR $>$ 1~M$_{\sun}$~yr$^{-1}$ are given in magenta.
When reliable correlations are seen, the best-fit straight line
for the full dataset is shown as a solid black line,
and for the high SFR subset as a dashed magenta line.
}
\end{figure}

\begin{figure}
\plotone{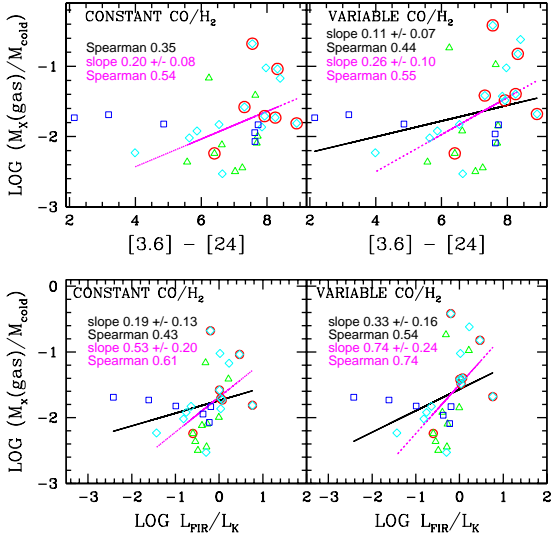}
\caption{
Comparisons between M$_{\rm X}$(gas)/(M$_{\rm HI}$+M$_{\rm H_2}$)
and
[3.6] $-$ [24] (top panels), and L$_{\rm FIR}$/L$_{\rm K}$ (bottom panels).
The best-fit line for the full sample is plotted as a solid black
line, while the best-fit line for systems with SFR $>$ 1.0 M$_{\sun}$~yr$^{-1}$
is given as a dotted line.
The best-fit slope and the Spearman rank correlation coefficient for the full
set is shown in black (on top), while the values for the high SFR
subset are shown in magenta (below).
Merger stages 1 and 2
systems are marked as open green triangles.  Merger stages 3, 4, and 5
are open cyan diamonds, and merger stages 6 and 7 are identified by
blue open squares.
AGN are identified by red circles.
NGC 1700 is not plotted.
When reliable correlations are seen, the best-fit straight line
for the full dataset is shown as a solid black line,
and for the high SFR subset as a dashed magenta line.
AGN are identified by red circles.
}
\end{figure}

\begin{figure}
\plotone{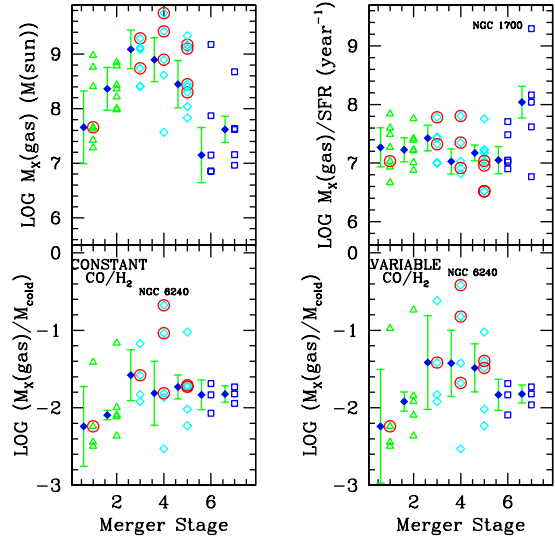}
\caption{Top left:
hot gas mass M$_{\rm X}$(gas) vs.\
merger stage.
Top right:
M$_{\rm X}$(gas)/SFR vs. merger stage.
Bottom row: 
M$_{\rm X}$(gas)/M$_{cold}$ = M$_{\rm X}$(gas)/(M$_{\rm H_2}$+M$_{\rm HI}$),
calculated with a constant
CO/H$_2$ ratio (left panel)
and the variable
CO/H$_2$ ratio (right panel).
Merger stages 1 and 2
systems are marked as open green triangles.  Merger stages 3, 4, and 5
are open cyan diamonds, and merger stages 6 and 7 are identified by
blue open squares.
AGN are identified by red circles.
NGC 1700 is not plotted in the bottom panels due to the lack of a complete
set of CO data.
The filled blue diamonds are the median values for each stage, slightly
offset to the left.
The errorbars plotted on the median values are the
semi-interquartile range, equal to half the difference between
the 75th percentile and the 25th percentile.
}
\end{figure}

\begin{figure}
\plotone{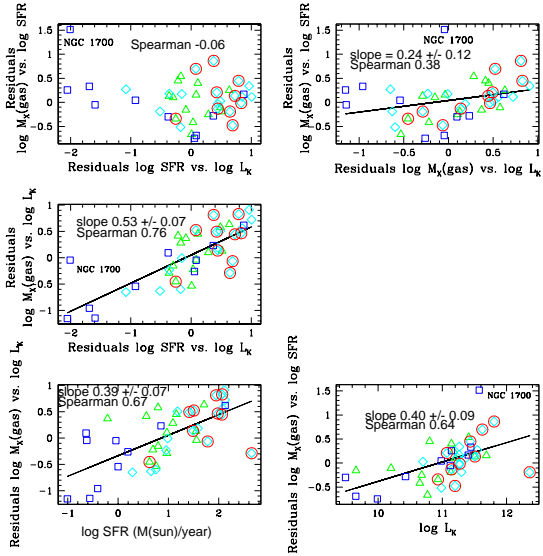}
\caption{A search for correlations between residuals of best-fit plots
for various quantities.
Residuals are defined as observed - (best-fit value),
i.e., a negative residual in the log SFR vs.\ log L$_{\rm K}$ plot
means that it has a low SFR relative to L$_{\rm K}$
(that is, a low sSFR or `post-starburst').
Merger stages 1 and 2
systems are marked as open green triangles.  Merger stages 3, 4, and 5
are open cyan diamonds, and merger stages 6 and 7 are identified by
blue open squares.
AGN are identified by red circles.
}
\end{figure}

\begin{figure}
\plotone{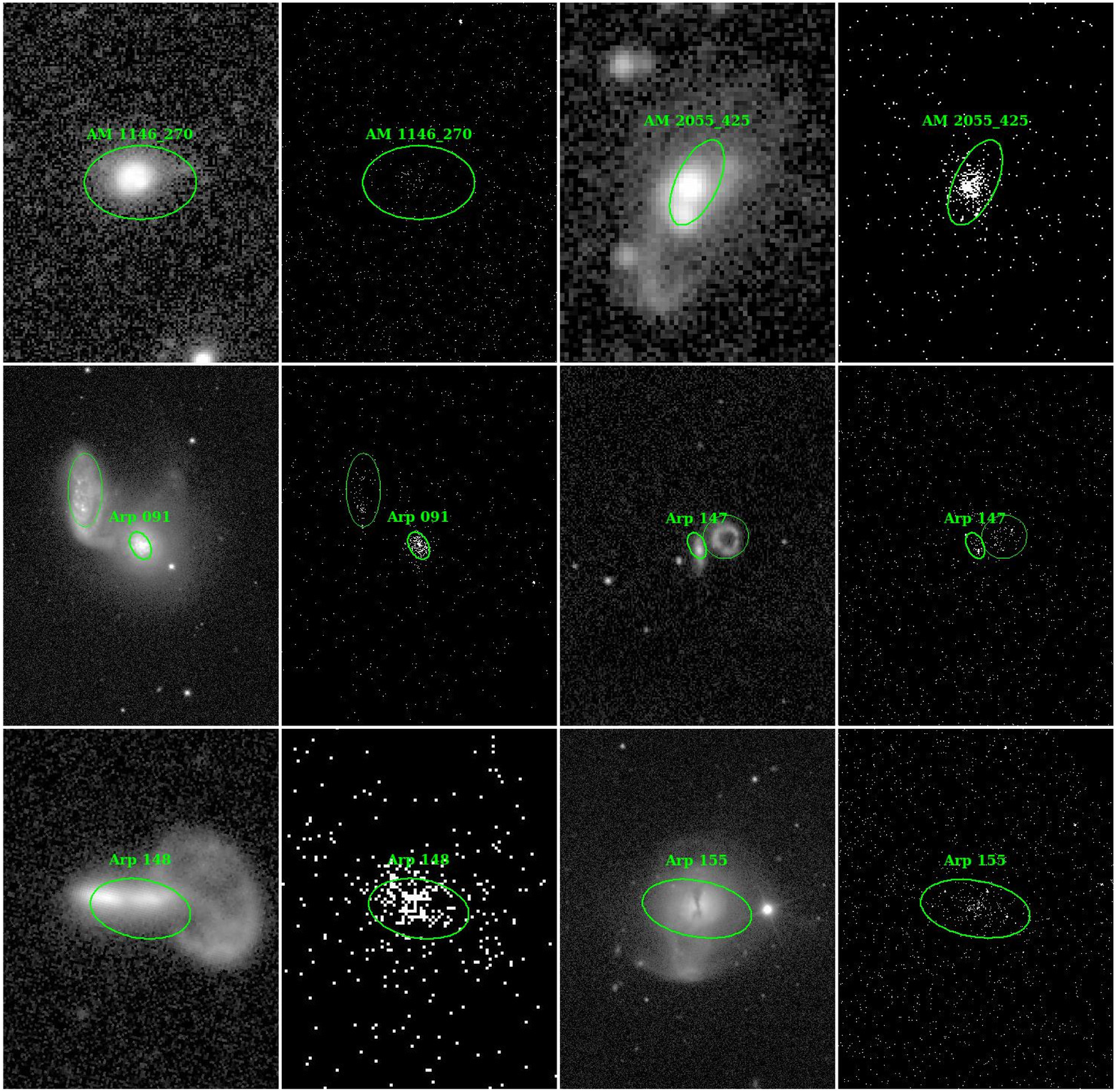}
\caption{
Montage of images of the galaxies.  The left panel is either the SDSS g image or the
GALEX NUV image (if no SDSS images exist).  
The right panel is the unsmoothed exposure-corrected Chandra 0.3 $-$ 1.0 keV
low energy map.  
Logarithmic contours are overlaid in white
on the Chandra image, with 
units of (3.0, 7.6, 33.5, 178, and 1000) $\times$ 10$^{-9}$ 
photons~s$^{-1}$~cm$^{-2}$~arcsec$^{-2}$. 
The final best-fit ellipse(s) are overlaid in green on both plots.
}
\end{figure}

\begin{figure}
\plotone{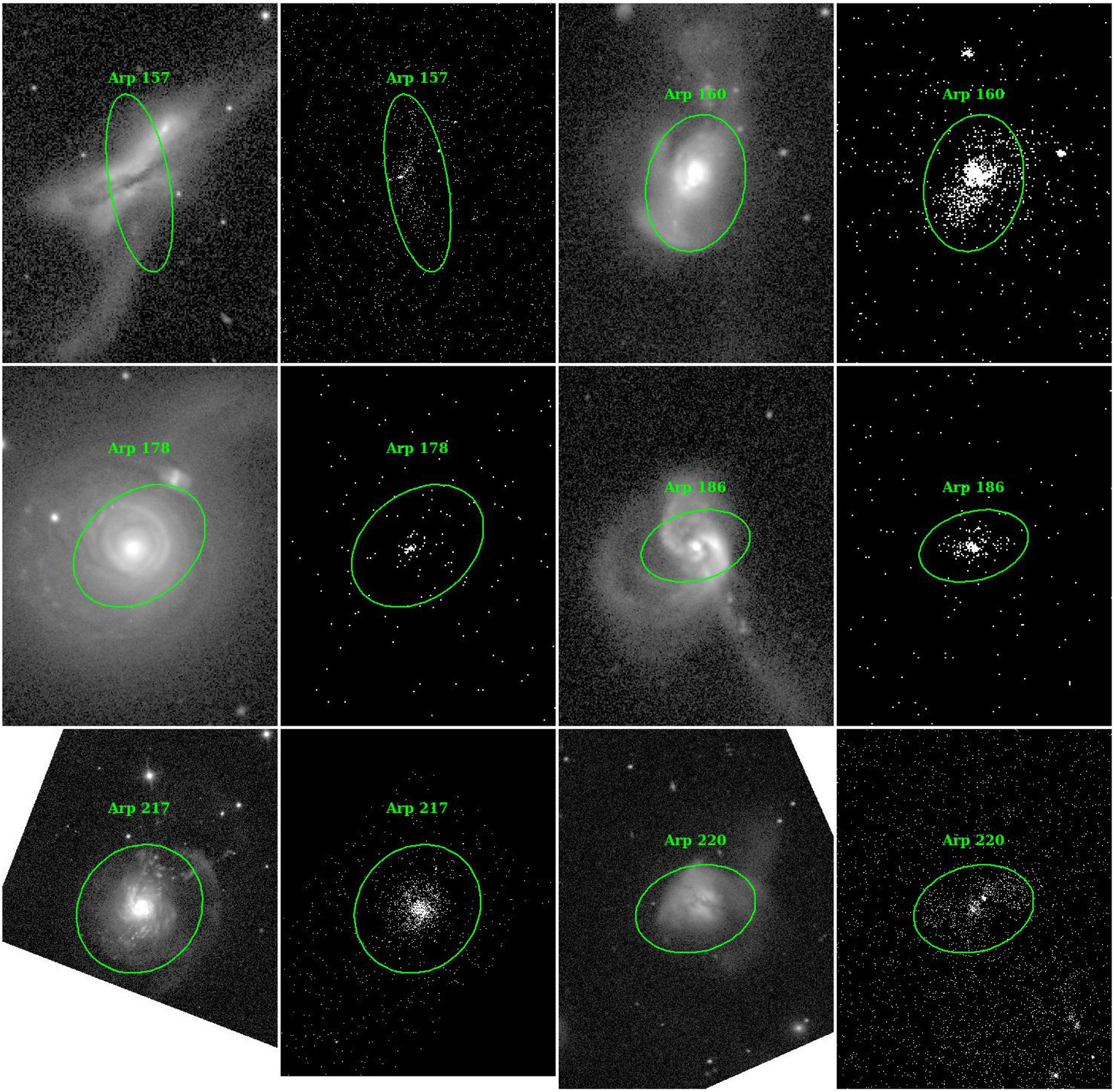}
\caption{
Montage of images of the galaxies.  
The left panel is either the SDSS g image or the
GALEX NUV image (if no SDSS images exist).  
The right panel is the unsmoothed exposure-corrected Chandra 0.3 $-$ 1.0 keV
low energy map.  
Logarithmic contours are overlaid in white
on the Chandra image, with 
units of (3.0, 7.6, 33.5, 178, and 1000) $\times$ 10$^{-9}$ 
photons~s$^{-1}$~cm$^{-2}$~arcsec$^{-2}$. 
The final best-fit ellipse(s) are overlaid in green on both plots.
}
\end{figure}

\begin{figure}
\plotone{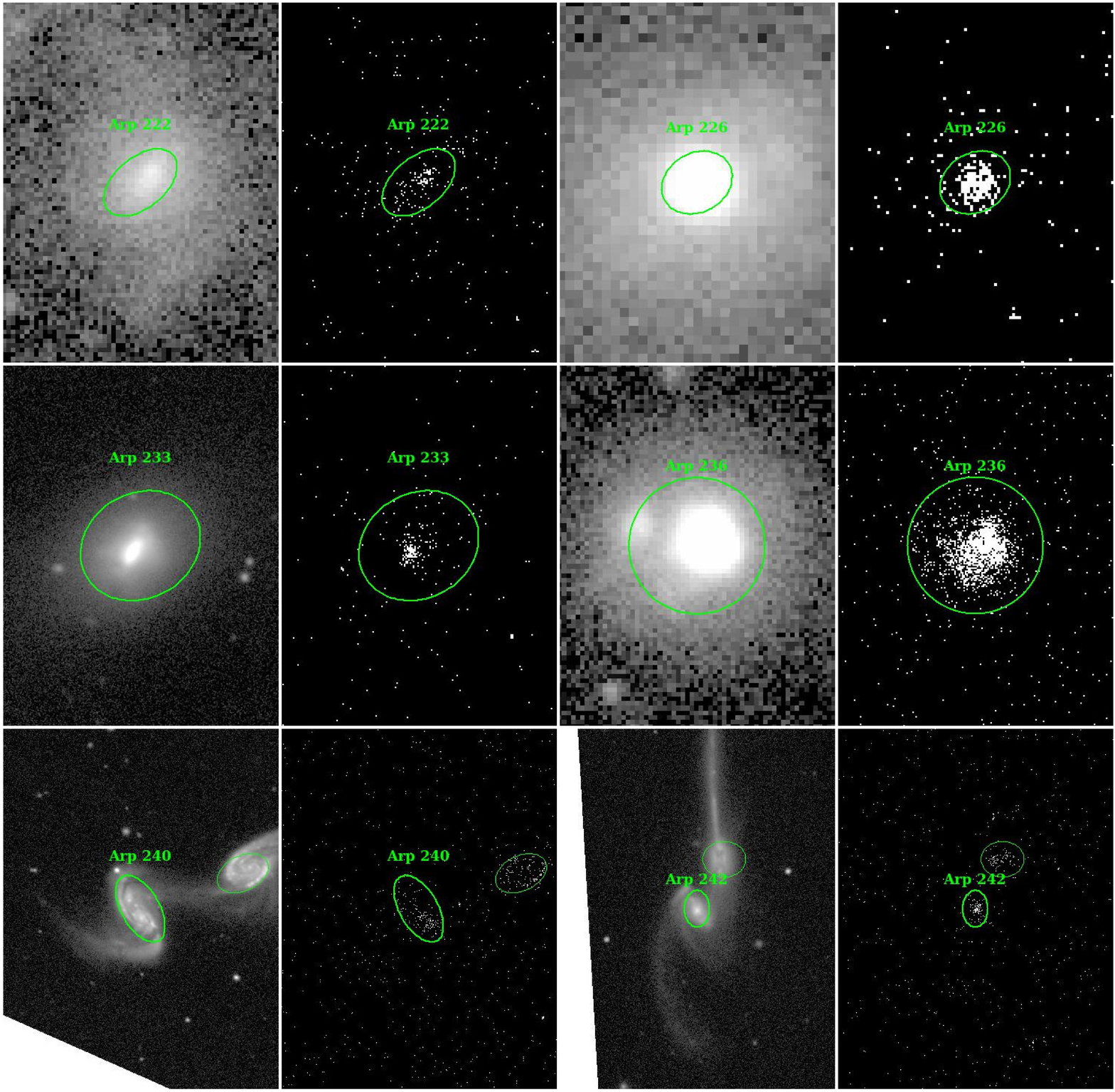}
\caption{
Montage of images of the galaxies.  
The left panel is either the SDSS g image or the
GALEX NUV image (if no SDSS images exist).  
The right panel is the unsmoothed exposure-corrected Chandra 0.3 $-$ 1.0 keV
low energy map.  
Logarithmic contours are overlaid in white
on the Chandra image, with 
units of (3.0, 7.6, 33.5, 178, and 1000) $\times$ 10$^{-9}$ 
photons~s$^{-1}$~cm$^{-2}$~arcsec$^{-2}$. 
The final best-fit ellipse(s) are overlaid in green on both plots.
}
\end{figure}

\begin{figure}
\plotone{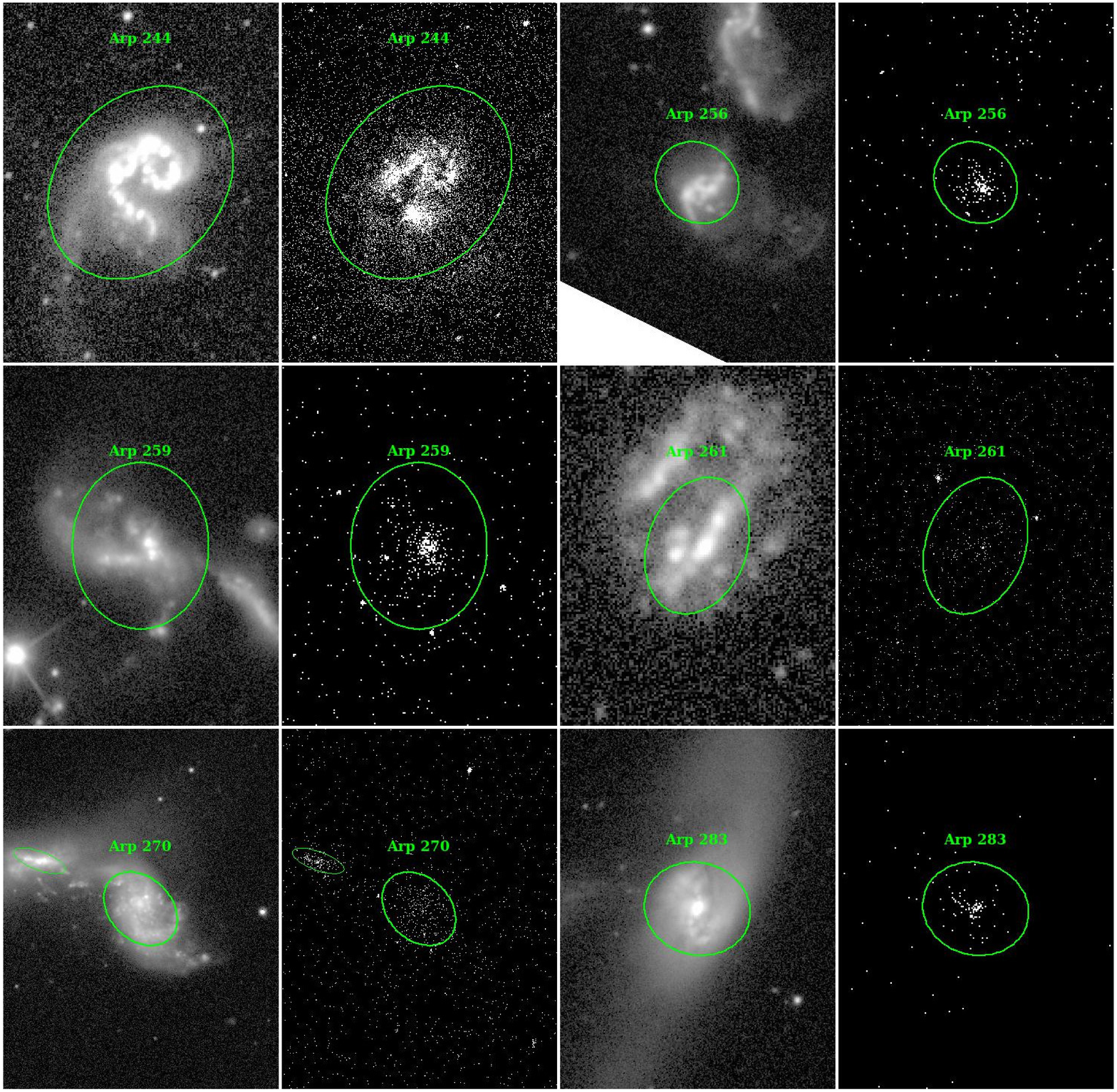}
\caption{
Montage of images of the galaxies.  
The left panel is either the SDSS g image or the
GALEX NUV image (if no SDSS images exist).  
The right panel is the unsmoothed exposure-corrected Chandra 0.3 $-$ 1.0 keV
low energy map.  
Logarithmic contours are overlaid in white
on the Chandra image, with 
units of (3.0, 7.6, 33.5, 178, and 1000) $\times$ 10$^{-9}$ 
photons~s$^{-1}$~cm$^{-2}$~arcsec$^{-2}$. 
The final best-fit ellipse(s) are overlaid in green on both plots.
}
\end{figure}

\begin{figure}
\plotone{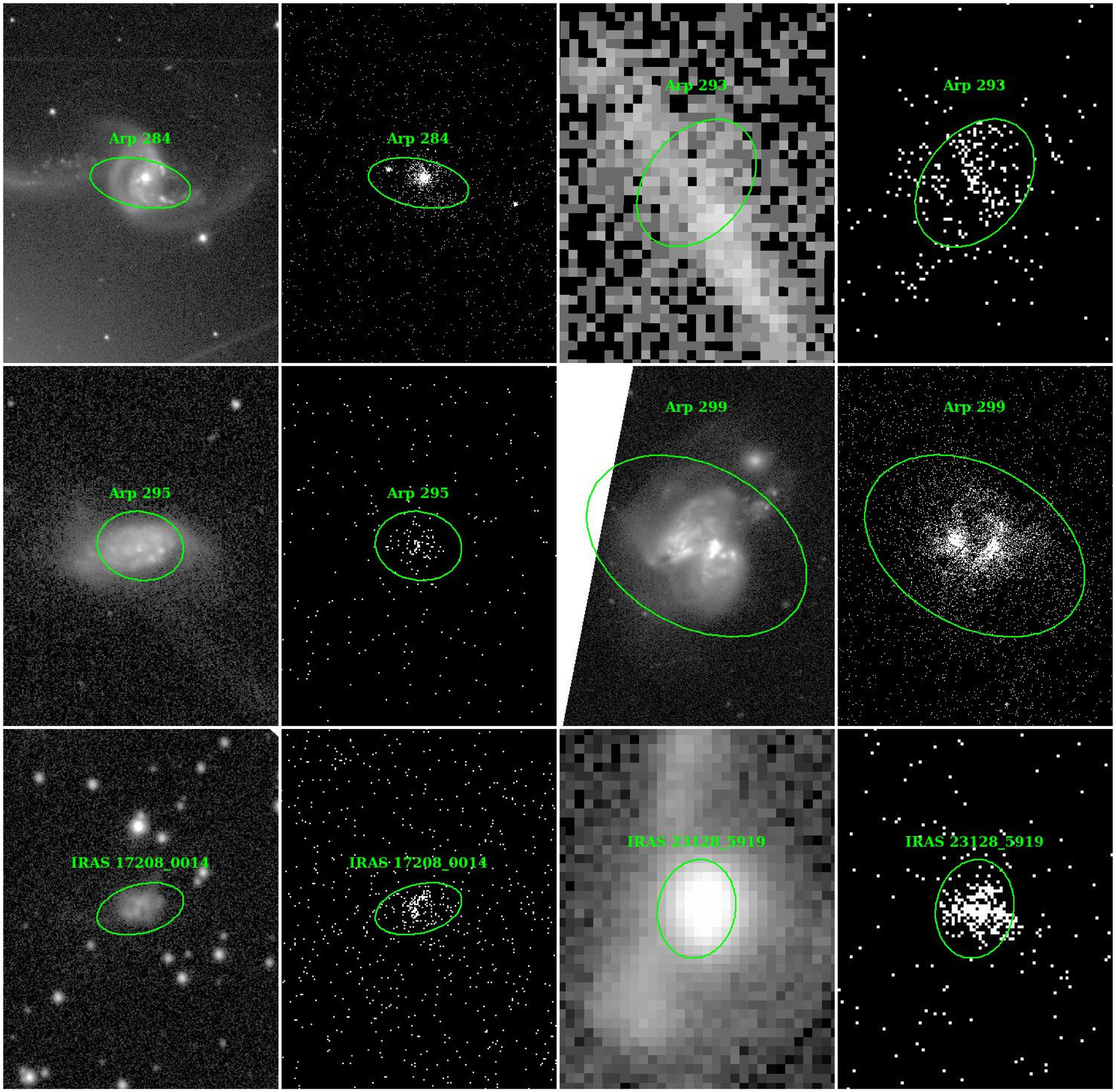}
\caption{
Montage of images of the galaxies.  
The left panel is either the SDSS g image or the
GALEX NUV image (if no SDSS images exist).  
The right panel is the unsmoothed exposure-corrected Chandra 0.3 $-$ 1.0 keV
low energy map.  
Logarithmic contours are overlaid in white
on the Chandra image, with 
units of (3.0, 7.6, 33.5, 178, and 1000) $\times$ 10$^{-9}$ 
photons~s$^{-1}$~cm$^{-2}$~arcsec$^{-2}$. 
The final best-fit ellipse(s) are overlaid in green on both plots.
}
\end{figure}

\begin{figure}
\plotone{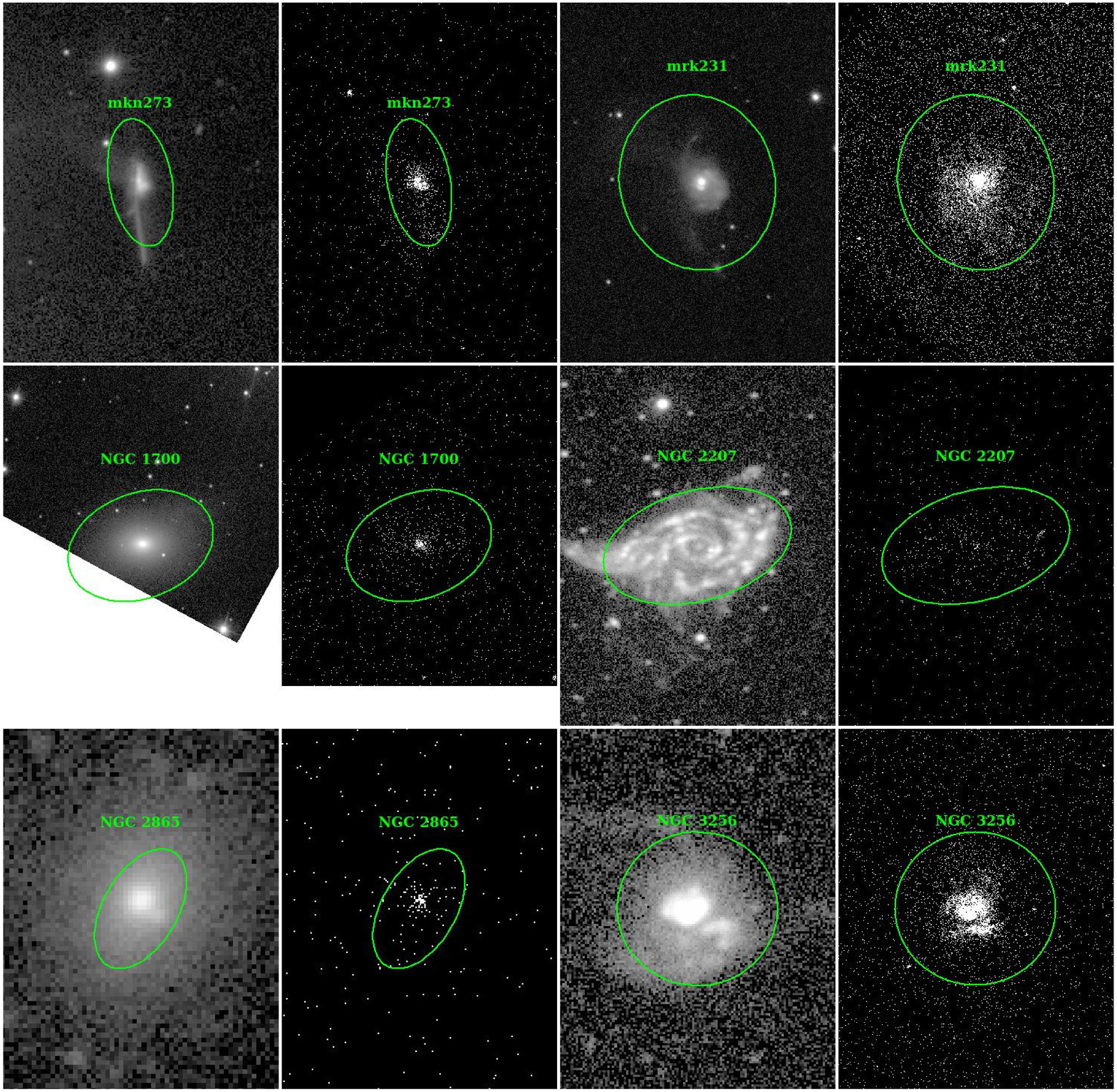}
\caption{
Montage of images of the galaxies.  
The left panel is either the SDSS g image or the
GALEX NUV image (if no SDSS images exist).  
The right panel is the unsmoothed exposure-corrected Chandra 0.3 $-$ 1.0 keV
low energy map.  
Logarithmic contours are overlaid in white
on the Chandra image, with 
units of (3.0, 7.6, 33.5, 178, and 1000) $\times$ 10$^{-9}$ 
photons~s$^{-1}$~cm$^{-2}$~arcsec$^{-2}$. 
The final best-fit ellipse(s) are overlaid in green on both plots.
}
\end{figure}

\clearpage

\begin{figure}
\plotone{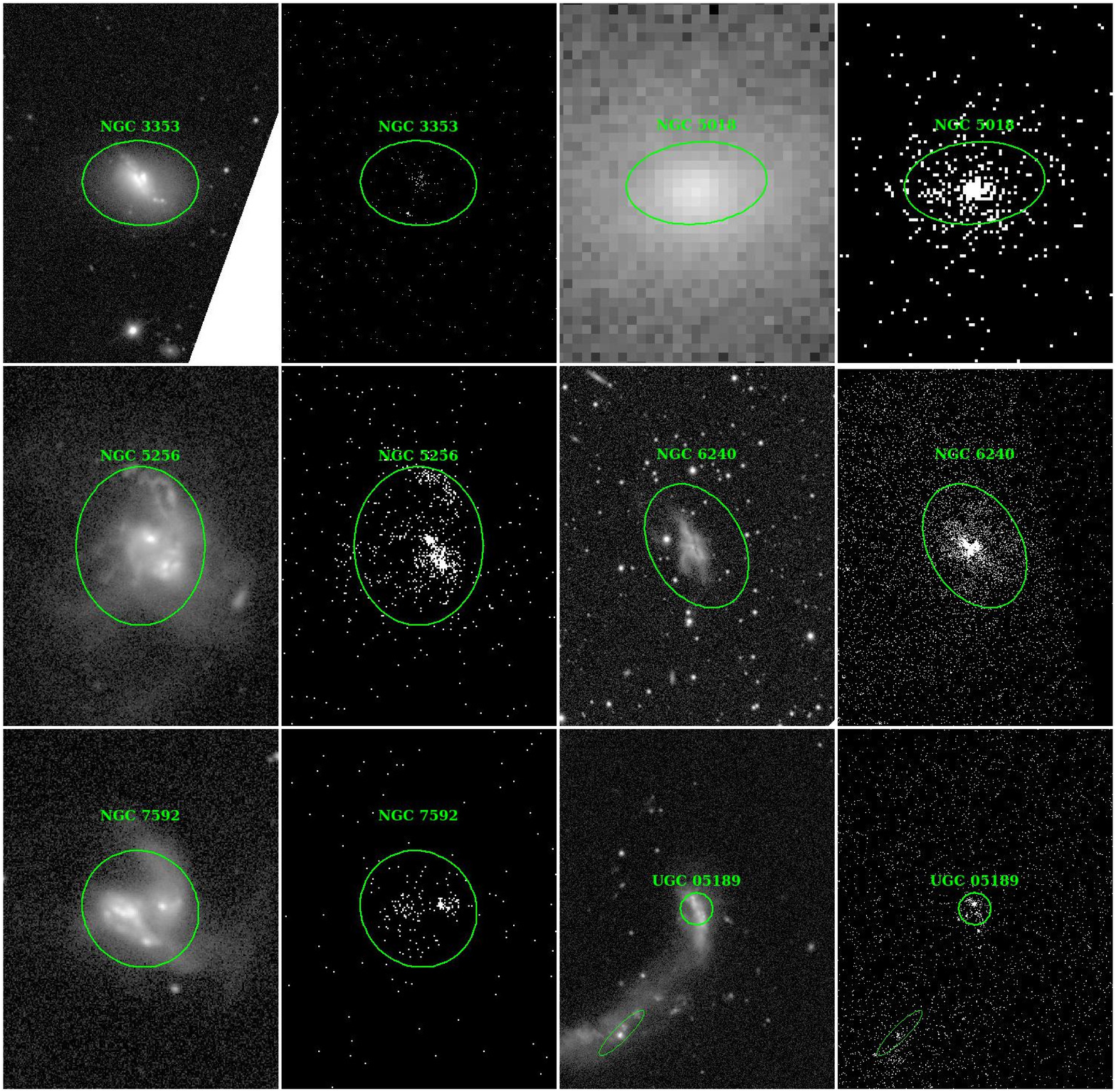}
\caption{
Montage of images of the galaxies.  
The left panel is either the SDSS g image or the
GALEX NUV image (if no SDSS images exist).  
The right panel is the unsmoothed exposure-corrected Chandra 0.3 $-$ 1.0 keV
low energy map.  
Logarithmic contours are overlaid in white
on the Chandra image, with 
units of (3.0, 7.6, 33.5, 178, and 1000) $\times$ 10$^{-9}$ 
photons~s$^{-1}$~cm$^{-2}$~arcsec$^{-2}$. 
The final best-fit ellipse(s) are overlaid in green on both plots.
}
\end{figure}

\clearpage

\begin{figure}
\plotone{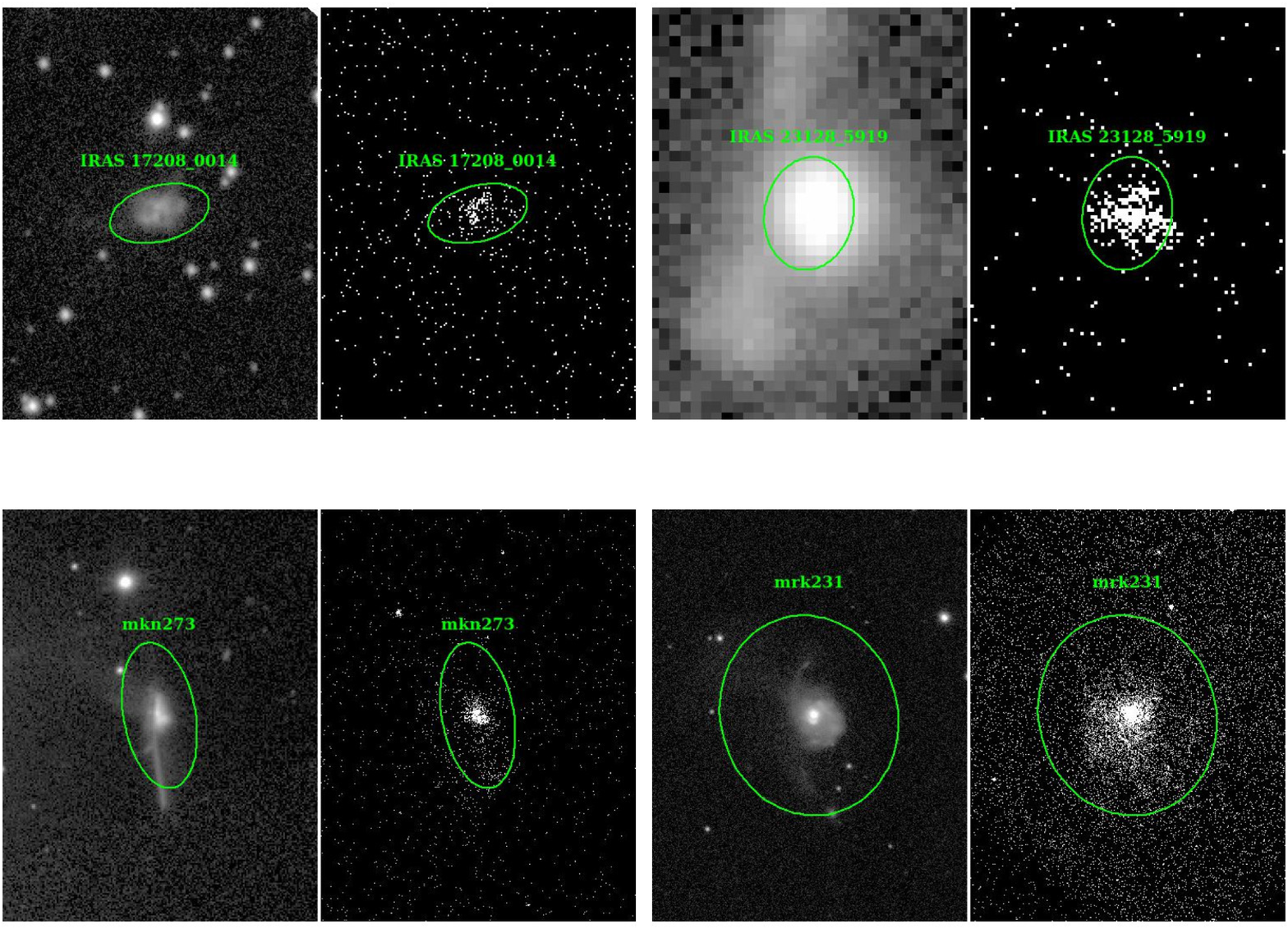}
\caption{
Montage of images of the galaxies.  
The left panel is either the SDSS g image or the
GALEX NUV image (if no SDSS images exist).  
The right panel is the unsmoothed exposure-corrected Chandra 0.3 $-$ 1.0 keV
low energy map.  
Logarithmic contours are overlaid in white
on the Chandra image, with 
units of (3.0, 7.6, 33.5, 178, and 1000) $\times$ 10$^{-9}$ 
photons~s$^{-1}$~cm$^{-2}$~arcsec$^{-2}$. 
The final best-fit ellipse(s) are overlaid in green on both plots.
}
\end{figure}

\begin{deluxetable}{ccrrrrrrrrc}
\tablecolumns{7}
\tablewidth{0pc}
\tablecaption{Basic Data on Sample Galaxies}
\tablehead{   
\colhead{Name} 
& \colhead{Stage}   
& \colhead{Distance}    
& \colhead{log L$_{\rm FIR}$} 
& \colhead{log L$_{\rm K}$} 
& \colhead{SFR} 
& \colhead{log L$_{\rm X}$(gas)}
& \colhead{AGN?}\\ 
\colhead{} 
& \colhead{}   
& \colhead{(Mpc)}    
& \colhead{(L$_{\sun}$)}  
&\colhead{(L$_{\sun}$)}
&\colhead{(M$_{\sun}$ yr$^{-1}$)}
& \colhead{(erg~s$^{-1}$)}
&
\\
}
\startdata
AM 1146-270 & 6 & 24.6  &  8.98 & 9.49 & 0.23 & 38.91  &  \\
AM 2055-425 & 5 & 179.1  &  11.72 & 11.27 & 128.23 & 41.92  &  \\
Arp 091 & 1 & 34  &  10.33 & 10.93 & 4.22 & 40.72  &  Sy2\\
Arp 147 & 1 & 129  &  10.22 & 10.80 & 3.65 & 40.76  &  \\
Arp 148 & 1 & 146.9  &  11.38 & 11.14 & 14.40 & 41.28  &  \\
Arp 155 & 7 & 46  &  9.99 & 10.98 & 1.00 & 40.01  &  \\
Arp 157 & 4 & 30.5  &  10.69 & 10.99 & 5.61 & 40.25  &  \\
Arp 160 & 5 & 39  &  10.72 & 10.70 & 15.14 & 41.15  &  \\
Arp 163 & 7 & 23.1  &  9.39 & 9.93 & 1.19 & $<$39.57  &  \\
Arp 178 & 5 & 62.1  &  10.11 & 11.54 & 1.92 & 40.59  &  \\
Arp 186 & 5 & 64.2  &  11.27 & 11.20 & 60.33 & 41.34  &  Sy2:HII\\
Arp 217 & 6 & 18.0  &  10.24 & 10.43 & 7.13 & 40.6  &  \\
Arp 220 & 4 & 83  &  12.03 & 11.27 & 95.65 & 41.34  &  Sy\\
Arp 222 & 6 & 26.1  &  8.71 & 11.12 & 0.27 & 39.85  &  \\
Arp 226 & 5 & 67  &  10.48 & 11.30 & 4.52 & 40.89  &  \\
Arp 233 & 7 & 25  &  9.62 & 9.99 & 1.57 & 39.73  &  \\
Arp 235 & 7 & 13  &  8.75 & 9.42 & 0.03 & $<$39.3  &  \\
Arp 236 & 1 & 81  &  11.39 & 11.18 & 51.02 & 41.88  &  \\
Arp 240 & 2 & 102  &  11.29 & 11.68 & 36.18 & 41.26  &  \\
Arp 242 & 2 & 98  &  10.65 & 11.20 & 6.23 & 40.78  &  \\
Arp 243 & 5 & 79.4  &  11.34 & 11.00 & 21.68 & $<$40.41  &  \\
Arp 244 & 3 & 24.1  &  10.62 & 11.26 & 9.70 & 40.67  &  \\
Arp 256 & 2 & 109.6  &  11.13 & 11.13 & 35.75 & 41.17  &  \\
Arp 259 & 2 & 55  &  10.34 & 10.20 & 6.20 & 40.42  &  \\
Arp 261 & 1 & 29  &  9.26 & 9.66 & 0.62 & 39.22  &  \\
Arp 263 & 5 & 9.8  &  8.76 & 9.06 & 0.22 & $<$38.76  &  \\
Arp 270 & 1 & 29  &  10.16 & 10.65 & 4.93 & 40.17  &  \\
Arp 283 & 1 & 30  &  10.48 & 10.77 & 5.74 & 40.46  &  \\
Arp 284 & 2 & 39  &  10.41 & 10.68 & 9.31 & 40.81  &  \\
Arp 293 & 2 & 82  &  11.10 & 11.41 & 12.24 & 41.54  &  Liner/HII\\
Arp 295 & 2 & 94  &  10.86 & 11.50 & 9.29 & 41.11  &  \\
Arp 299 & 3 & 48  &  11.60 & 11.37 & 119.01 & 41.51  &  \\
IRAS 17208-0014 & 6 & 183  &  12.19 & 11.40 & 133.97 & 42.03  &  Liner\\
IRAS 23128-5919 & 3 & 184  &  11.71 & 11.24 & 133.83 & 41.79  &  \\
Mrk 231 & 5 & 178.1  &  12.13 & 12.36 & 450.05 & 41.39  &  Sy1\\
Mrk 273 & 4 & 160.5  &  11.90 & 11.44 & 116.83 & 41.19  &  Sy2\\
NGC 034 & 5 & 79.3  &  11.18 & 11.14 & 30.78 & 41.09  &  Sy2\\
NGC 1700 & 7 & 52.5  &  8.40 & 11.57 & 0.24 & 40.75  &  \\
NGC 2207/IC 2163 & 3 & 38.0  &  10.73 & 11.48 & 9.03 & 40.76  &  \\
NGC 2865 & 7 & 37.9  &  9.53 & 11.14 & 0.10 & 39.75  &  \\
NGC 3256 & 4 & 37.0  &  11.30 & 11.26 & 38.50 & 41.35  &  \\
NGC 3353 & 6 & 18.5  &  9.43 & 9.66 & 0.91 & 39.19  &  \\
NGC 5018 & 7 & 38.4  &  9.40 & 11.44 & 0.39 & 40.58  &  \\
NGC 5256 & 3 & 120.9  &  11.21 & 11.62 & 32.02 & 41.35  &  Sy2\\
NGC 6240 & 4 & 108.8  &  11.61 & 11.81 & 88.46 & 42.08  &  Sy2/Liner\\
NGC 7592 & 3 & 99.5  &  11.08 & 11.07 & 26.08 & 41.22  &  Sy2\\
UGC 2238 & 5 & 87.1  &  11.04 & 11.15 & 9.41 & 40.97  &  Liner\\
UGC 5101 & 5 & 164.3  &  11.72 & 11.51 & 115.63 & 41.56  &  Sy1\\
UGC 5189 & 2 & 48.9  &  9.48 & 9.47 & 1.54 & $<$39.26  &  \\
\enddata
\end{deluxetable}

\begin{deluxetable}{crrcccrcccc}
\rotate
\tablecolumns{11}
\tablewidth{0pc}
\tablecaption{Global Molecular and Atomic Gas Mass in the Sample Galaxies}
\tablehead{   
\colhead{Name} 
& \colhead{LOG} 
& \colhead{LOG} 
& \colhead{CO} 
& \colhead{LOG} 
& \colhead{LOG} 
& \colhead{LOG} 
& \colhead{HI}
& \colhead{LOG} 
& \colhead{LOG} 
& \colhead{LOG} 
\\ 
& \colhead{M$_{\rm H_2}$$^\dagger$} 
& \colhead{M$_{\rm H_2}$$\ddagger$} 
& \colhead{REF$^{*}$} 
& \colhead{SFE$^\dagger$}
& \colhead{SFE$^\ddagger$}
& \colhead{M$_{\rm HI}$} 
& \colhead{REF$^{**}$}
& \colhead{M$_{\rm hot}$} 
& \colhead{M$_{\rm hot}$/M$_{\rm cold}$$^\dagger$} 
& \colhead{log M$_{\rm hot}$/M$_{\rm cold}$$^\ddagger$} 
\\ 
\colhead{} 
& \colhead{(M$_{\sun}$)} 
& \colhead{(M$_{\sun}$)} 
& \colhead{} 
& \colhead{(yr$^{-1}$)} 
& \colhead{(yr$^{-1}$)} 
& \colhead{(M$_{\sun}$)} 
& \colhead{} 
& \colhead{(M$_{\sun}$)} 
& \colhead{} 
& \colhead{}\\ 
}
\startdata
AM 1146-270 &  &   &   &  &  & 9.15 & 16 & 6.85 &  &  \\
AM 2055-425 & 10.06 & 9.36  &  1 & -7.94 & -7.24 &  &  & 9.34 &  &  \\
Arp 091 & 9.60 & 9.60  &  2 & -8.96 & -8.96 & 9.60 & 17 & 7.66 & -2.23 & -2.23 \\
Arp 147 & 9.28 & 9.28  &  3 & -8.71 & -8.71 &  &  & 8.40 &  &  \\
Arp 148 & 10.05 & 9.35  &  4 & -8.88 & -8.18 &  &  & 8.76 &  &  \\
Arp 155 & 9.19 & 9.19  &  5 & -9.18 & -9.18 & 9.08 & 17 & 7.62 & -1.81 & -1.81 \\
Arp 157 & 9.84 & 9.84  &  6 & -9.08 & -9.08 & 9.75 & 17 & 7.57 & -2.52 & -2.52 \\
Arp 160 & 9.05 & 9.05  &  6 & -7.86 & -7.86 & 9.18 & 17 & 8.40 & -1.01 & -1.01 \\
Arp 163 &  &   &   &  &  & 9.08 & 17 &  &  &  \\
Arp 178 & 10.17 & 10.17  &  2 & -9.88 & -9.88 & 9.58 & 17 & 8.04 & -2.22 & -2.22 \\
Arp 186 & 9.86 & 9.16  &  4 & -8.07 & -7.37 & 9.55 & 18 & 8.30 & -1.72 & -1.39 \\
Arp 217 & 8.57 & 8.57  &  7 & -7.7 & -7.7 & 9.67 & 17 & 7.87 & -1.82 & -1.82 \\
Arp 220 & 10.23 & 9.53  &  4 & -8.24 & -7.54 & 10.54 & 17 & 8.90 & -1.80 & -1.67 \\
Arp 222 & 7.87 & 7.87  &  8 & -8.43 & -8.43 & 8.79 & 17 & 7.15 & -1.68 & -1.68 \\
Arp 226 & 9.49 & 9.49  &  8 & -8.82 & -8.82 & 9.60 & 18 & 7.83 & -2.01 & -2.01 \\
Arp 233 & 7.39 & 7.79  &  9 & -7.19 & -7.58 & 8.89 & 17 & 6.96 & -1.93 & -1.95 \\
Arp 235 &  &   &   &  &  & 8.85 & 17 &  &  &  \\
Arp 236 & 10.28 & 9.59  &  4 & -8.57 & -7.87 & 9.71 & 19 & 8.98 & -1.40 & -0.97 \\
Arp 240 & 10.56 & 9.86  &  10 & -8.99 & -8.29 & 10.64 & 17 & 8.79 & -2.11 & -1.91 \\
Arp 242 & 9.96 & 9.96  &  10 & -9.16 & -9.16 & 10.46 & 17 & 8.22 & -2.35 & -2.35 \\
Arp 243 & 9.63 & 8.93  &  6 & -8.28 & -7.59 & 9.50 & 20 &  &  &  \\
Arp 244 & 9.98 & 9.98  &  10 & -8.98 & -8.98 & 9.90 & 21 & 8.41 & -1.82 & -1.82 \\
Arp 256 & 9.98 & 9.28  &  10 & -8.42 & -7.72 & 10.23 & 17 & 8.43 & -1.98 & -1.84 \\
Arp 259 &  &   &   &  &  & 10.21 & 17 & 8.00 &  &  \\
Arp 261 &  &   &   &  &  & 9.75 & 17 & 7.28 &  &  \\
Arp 263 &  &   &   &  &  & 9.25 & 17 &  &  &  \\
Arp 270 & 8.81 & 8.81  &  2 & -8.1 & -8.1 & 10.10 & 17 & 7.63 & -2.49 & -2.49 \\
Arp 283 & 9.55 & 9.55  &  7 & -8.78 & -8.78 & 9.58 & 17 & 7.43 & -2.43 & -2.43 \\
Arp 284 & 9.18 & 9.18  &  6 & -8.2 & -8.2 & 10.01 & 17 & 7.98 & -2.08 & -2.08 \\
Arp 293 & 9.91 & 9.21  &  4 & -8.81 & -8.11 & 9.35 & 22 & 8.85 & -1.15 & -0.73 \\
Arp 295 &  &   &   &  &  & 10.40 & 23 & 8.36 &  &  \\
Arp 299 & 10.21 & 9.51  &  10 & -8.13 & -7.43 & 9.25 & 17 & 9.09 & -1.16 & -0.61 \\
IRAS 17208-0014 & 10.55 & 9.85  &  11 & -8.41 & -7.71 &  &  & 9.18 &  &  \\
IRAS 23128-5919 & 9.88 & 9.18  &  1 & -7.74 & -7.04 &  &  & 9.12 &  &  \\
Mrk 231 & 10.03 & 9.34  &  6 & -7.37 & -6.67 &  &  & 9.17 &  &  \\
Mrk 273 & 10.14 & 9.44  &  4 & -8.06 & -7.36 & 10.16 & 17 & 9.42 & -1.03 & -0.81 \\
NGC 034 & 9.86 & 9.16  &  8 & -8.36 & -7.66 & 9.85 & 26 & 8.45 & -1.70 & -1.47 \\
NGC 1700 &  &   &   &  &  &  &  & 8.67 &  &  \\
NGC 2207/IC 2163 & 9.76 & 9.76  &  15 & -8.79 & -8.79 & 10.19 & 17 & 8.40 & -1.91 & -1.91 \\
NGC 2865 & 7.97 & 7.97  &  12 & -8.97 & -8.97 & 8.83 & 24 & 7.16 & -1.72 & -1.72 \\
NGC 3256 & 10.38 & 9.68  &  8 & -8.78 & -8.08 & 9.79 & 25 & 8.61 & -1.85 & -1.41 \\
NGC 3353 & 7.43 & 7.83  &  14 & -7.46 & -7.86 & 8.92 & 17 & 6.86 & -2.06 & -2.08 \\
NGC 5018 &  &   &   &  &  & 8.82 & 17 & 7.64 &  &  \\
NGC 5256 & 10.30 & 9.60  &  13 & -8.79 & -8.09 &  &  & 9.29 &  &  \\
NGC 6240 & 10.18 & 9.48  &  6 & -8.22 & -7.52 & 10.07 & 17 & 9.75 & -0.67 & -0.41 \\
NGC 7592 & 9.91 & 9.21  &  10 & -8.49 & -7.79 & 10.10 & 17 & 8.74 & -1.57 & -1.41 \\
UGC 2238 & 9.97 & 9.27  &  4 & -8.98 & -8.28 & 9.98 & 17 &  &  &  \\
UGC 5101 & 10.37 & 9.67  &  4 & -8.3 & -7.6 &  &  & 9.10 &  &  \\
UGC 5189 &  &   &   &  &  & 10.14 & 17 &  &  &  \\
\enddata
\tablenotetext{\dagger}{Assuming the
standard Galactic CO/H$_2$ ratio. 
SFE in this
paper is defined as SFR/M$_{\rm H_2}$.}
\tablenotetext{\ddagger}{Using the variable
CO/H$_2$ ratio (see text for details).
SFE 
is defined as SFR/M$_{\rm H_2}$.}
\tablenotetext{*}{CO references:
1: \citealp{Mirabel1990};
2: \citealp{Zhu1999};
3: \citealp{Horellou1995};
4: \citealp{Larson2016};
5: \citealp{Wiklind1995};
6: \citealp{Sanders1991};
7: \citealp{Young1996};
8: \citealp{Ueda2014};
9: \citealp{Israel2005};
10: \citealp{Bushouse1999};
11: \citealp{Solomon1997};
12: \citealp{Georgakakis2001}; 13: \citealp{Papadopoulos2012};
14: \citealp{Sage1992};
15: \citealp{Elmegreen2016}.}
\tablenotetext{**}{HI references:
16: \citealp{Doyle2005}
17: \citealp{1989HRHI};
18: \citealp{Obreschkow2009};
19: \citealp{martin1991};
20: \citealp{bushouse87};
21: \citealp{Gordon2001}
22: \citealp{vanDriel2000};
23: \citealp{hibbard96};
24: \citealp{Cox2004};
25: \citealp{English2003};
26: \citealp{Fernandez2014};
}
\end{deluxetable}

\begin{deluxetable}{ccrrrrrrrcr}
\rotate
\tablecolumns{11}
\tablewidth{0pc}
\tablecaption{Final Ellipses Used for Volume Calculations:
At 0.3 $-$ 1.0 keV Surface Brightness of 3 $\times$ 10$^{-9}$ photons~s$^{-1}$~cm$^{-1}$~arcsec$^{-2}$ }
\tablehead{   
\colhead{Name} 
&\colhead{Individual}
& \colhead{RA}   
& \colhead{DEC}    
& \colhead{Major} 
& \colhead{Minor} 
& \colhead{Major} 
& \colhead{Minor} 
& \colhead{P.A.$^1$}
& \colhead{Number} 
& \colhead{Diffuse} 
\\
\colhead{} 
&\colhead{Galaxy} 
& \colhead{(J2000)}   
& \colhead{(J2000)}    
& \colhead{Axis} 
& \colhead{Axis} 
& \colhead{Axis} 
& \colhead{Axis} 
& \colhead{(deg)}
& \colhead{Annuli}
& \colhead{0.3 $-$ } 
\\
\colhead{} 
& \colhead{}   
& \colhead{}    
& \colhead{}    
& \colhead{Radius} 
& \colhead{Radius} 
& \colhead{Radius} 
& \colhead{Radius} 
& \colhead{}
& \colhead{}
& \colhead{1.0 keV} 
\\
\colhead{} 
& \colhead{}   
& \colhead{}    
& \colhead{}    
& \colhead{($''$)} 
& \colhead{($''$)} 
& \colhead{(kpc)} 
& \colhead{(kpc)} 
& \colhead{}
& \colhead{}
& \colhead{Counts} 
\\
}
\startdata
AM 1146-270 & ESO 504-G017 & 11 48 46.046 & -27 22 49.11 & 36\farcs74 & 24\farcs15 & 4.38 & 2.88 & 0 & 1 & 27 $\pm$ 13  \\
AM 2055-425 & ESO 286-IG019 & 20 58 26.554 & -42 38 58.11 & 25\farcs88 & 11\farcs80 & 22.48 & 10.25 & 65 & 13 & 289 $\pm$ 19  \\
Arp 091 & NGC 5953 & 15 34 32.396 & +15 11 36.55 & 15\farcs42 & 9\farcs90 & 2.54 & 1.63 & 300 & 5 & 233 $\pm$ 17  \\
 & NGC 5954 & 15 34 34.911 & +15 12 12.96 & 24\farcs17 & 11\farcs90 & 3.99 & 1.83 & 90 & 1 & 59 $\pm$ 8  \\
Arp 147 & IC 298 NED02 & 3 11 19.621 & +1 18 51.35 & 8\farcs95 & 5\farcs40 & 5.60 & 3.38 & 115 & 1 & 60 $\pm$ 7  \\
 & IC 298 & 3 11 18.367 & +1 18 57.12 & 15\farcs00 & 14\farcs40 & 9.38 & 8.77 & 20 & 1 & 56 $\pm$ 9  \\
Arp 148 &  NED01 & 11 03 53.839 & +40 50 58.26 & 19\farcs92 & 11\farcs51 & 14.19 & 8.20 & 350 & 7 & 216 $\pm$ 17  \\
Arp 155 & NGC 3656 & 11 23 38.541 & +53 50 30.51 & 31\farcs09 & 15\farcs81 & 6.94 & 3.53 & 350 & 5 & 129 $\pm$ 17  \\
Arp 157 & NGC 520 & 1 24 34.398 & +3 47 27.11 & 42\farcs92 & 14\farcs26 & 6.35 & 2.11 & 100 & 7 & 259 $\pm$ 19  \\
Arp 160 & NGC 4194 & 12 14 09.490 & +54 31 32.96 & 41\farcs17 & 29\farcs27 & 7.79 & 5.54 & 258 & 13 & 1007 $\pm$ 35  \\
Arp 178 & NGC 5614 & 14 24 07.420 & +34 51 32.48 & 23\farcs87 & 17\farcs33 & 7.19 & 5.22 & 38 & 1 & 58 $\pm$ 8  \\
Arp 186 & NGC 1614 & 4 34 00.068 & -8 34 44.83 & 18\farcs17 & 11\farcs25 & 5.66 & 3.50 & 195 & 1 & 122 $\pm$ 12  \\
Arp 217 & NGC 3310 & 10 38 45.964 & +53 30 08.73 & 53\farcs48 & 48\farcs04 & 4.67 & 4.19 & 230 & 13 & 4924 $\pm$ 73  \\
Arp 220 & IC 4553 & 15 34 57.578 & +23 30 06.07 & 37\farcs87 & 26\farcs55 & 15.25 & 10.69 & 15 & 13 & 711 $\pm$ 32  \\
Arp 222 & NGC 7727 & 23 39 53.876 & -12 17 34.61 & 27\farcs98 & 16\farcs53 & 3.54 & 2.09 & 40 & 5 & 143 $\pm$ 13  \\
Arp 226 & NGC 7252 & 22 20 44.774 & -24 40 41.77 & 12\farcs13 & 9\farcs67 & 3.94 & 3.14 & 30 & 5 & 217 $\pm$ 16  \\
Arp 233 & UGC 05720 & 10 32 31.589 & +54 24 05.25 & 20\farcs41 & 17\farcs20 & 2.48 & 2.09 & 30 & 1 & 127 $\pm$ 12  \\
Arp 236 & IC 1623 & 01 07 46.897 & -17 30 26.91 & 24\farcs45 & 24\farcs37 & 9.60 & 9.57 & 90 & 13 & 1645 $\pm$ 42  \\
Arp 240 & NGC 5258 & 13 39 57.690 & +0 49 56.55 & 24\farcs14 & 12\farcs50 & 11.94 & 6.18 & 300 & 1 & 130 $\pm$ 13  \\
 & NGC 5257 & 13 39 53.221 & +0 50 19.82 & 23\farcs63 & 15\farcs50 & 11.69 & 7.55 & 25 & 7 & 127 $\pm$ 14  \\
 & \\
Arp 242 & NGC 4676B & 12 46 11.236 & +30 43 22.77 & 11\farcs95 & 8\farcs28 & 5.68 & 3.93 & 90 & 1 & 69 $\pm$ 9  \\
 & NGC 4676A & 12 46 09.850 & +30 43 55.24 & 14\farcs04 & 11\farcs28 & 6.67 & 5.59 & 0 & 1 & 68 $\pm$ 9  \\
Arp 244 & NGC 4038/9 & 12 01 53.451 & -18 52 26.50 & 88\farcs32 & 69\farcs12 & 10.32 & 8.08 & 50 & 13 & 50307 $\pm$ 248  \\
Arp 256 &  NED01 & 0 18 51.025 & -10 22 34.62 & 14\farcs33 & 12\farcs67 & 7.62 & 6.73 & 140 & 1 & 120 $\pm$ 11  \\
Arp 259 &  NED03/04 & 5 01 38.006 & -4 15 29.55 & 27\farcs32 & 22\farcs31 & 7.29 & 5.95 & 90 & 1 & 262 $\pm$ 19  \\
Arp 261 &  NED01 & 14 49 31.021 & -10 10 32.23 & 46\farcs16 & 32\farcs49 & 6.49 & 4.57 & 250 & 1 & 110 $\pm$ 19  \\
Arp 270 & NGC 3395 & 10 49 50.028 & +32 58 55.07 & 36\farcs54 & 27\farcs19 & 5.14 & 3.82 & 315 & 7 & 274 $\pm$ 22  \\
 & NGC 3396 & 10 49 55.286 & +32 59 26.47 & 17\farcs91 & 6\farcs19 & 2.52 & 0.85 & 160 & 1 & 135 $\pm$ 12  \\
Arp 283 & NGC 2798 & 9 17 22.830 & +41 59 59.55 & 17\farcs55 & 15\farcs13 & 2.55 & 2.20 & 345 & 1 & 65 $\pm$ 9  \\
Arp 284 & NGC 7714 & 23 36 14.310 & +2 09 14.05 & 35\farcs65 & 16\farcs66 & 6.74 & 3.15 & 349 & 13 & 971 $\pm$ 33  \\
Arp 293 & NGC 6286 & 16 58 31.581 & +58 56 15.77 & 28\farcs09 & 19\farcs71 & 11.17 & 7.84 & 231 & 7 & 260 $\pm$ 17  \\
Arp 295 & ARP295B & 23 42 00.805 & -3 36 53.46 & 14\farcs22 & 11\farcs23 & 6.48 & 5.12 & 350 & 1 & 59 $\pm$ 8  \\
Arp 299 & NGC 3690 & 11 28 32.174 & +58 33 46.26 & 68\farcs30 & 45\farcs99 & 15.90 & 10.71 & 330 & 13 & 7161 $\pm$ 90  \\
IRAS 17208-0014 & IRAS F17207-0014 & 17 23 21.953 & -0 17 01.90 & 17\farcs49 & 9\farcs52 & 15.52 & 8.45 & 15 & 5 & 115 $\pm$ 13  \\
IRAS 23128-5919 & ESO 148-IG002 & 23 15 46.922 & -59 03 14.16 & 16\farcs25 & 12\farcs74 & 14.50 & 11.37 & 260 & 13 & 220 $\pm$ 16  \\
Mrk 231 & UGC 08058 & 12 56 14.547 & +56 52 24.67 & 20\farcs09 & 17\farcs68 & 17.35 & 15.27 & 105 & 13 & 3349 $\pm$ 63  \\
Mrk 273 & UGC 08696 & 13 44 42.002 & +55 53 12.22 & 49\farcs60 & 24\farcs25 & 38.61 & 18.88 & 100 & 13 & 1121 $\pm$ 36  \\
NGC 034 & NGC 17 & 0 11 06.714 & -12 06 27.79 & 20\farcs74 & 20\farcs02 & 7.98 & 7.70 & 120 & 1 & 61 $\pm$ 8  \\
NGC 1700 & MGC-01-13-038 & 04 56 56.452 & -04 51 59.57 & 62\farcs77 & 44\farcs22 & 15.98 & 11.26 & 20 & 13 & 1714 $\pm$ 49  \\
NGC 2207/IC 2163 & NGC 2207/IC 2163 & 06 16 22.134 & -21 22 21.18 & 66\farcs36 & 38\farcs06 & 12.23 & 7.01 & 15 & 13 & 786 $\pm$ 34  \\
NGC 2865 & ESO 498-G001 & 9 23 30.226 & -23 09 44.13 & 21\farcs50 & 12\farcs20 & 3.95 & 2.24 & 240 & 1 & 75 $\pm$ 10  \\
NGC 3256 & NGC 3256N/S & 10 27 51.096 & -43 54 09.37 & 41\farcs22 & 39\farcs40 & 7.40 & 7.07 & 0 & 13 & 5702 $\pm$ 80  \\
NGC 3353 & UGC 05860 & 10 45 22.210 & +55 57 35.61 & 38\farcs14 & 27\farcs79 & 3.42 & 2.49 & 357 & 1 & 50 $\pm$ 10  \\
NGC 5018 & UGCA 335 & 13 13 01.040 & -19 31 04.39 & 23\farcs09 & 13\farcs44 & 4.30 & 2.50 & 5 & 7 & 283 $\pm$ 18  \\
NGC 5256 &  NED01/02 & 13 38 18.085 & +48 16 38.52 & 40\farcs00 & 32\farcs57 & 23.45 & 19.10 & 90 & 13 & 637 $\pm$ 27  \\
NGC 6240 & UGC 10592 & 16 52 58.527 & +2 24 04.36 & 48\farcs65 & 33\farcs88 & 25.67 & 17.88 & 120 & 13 & 10518 $\pm$ 112  \\
NGC 7592 & NGC 7592A/B & 23 18 22.276 & -4 24 57.89 & 23\farcs67 & 22\farcs16 & 11.42 & 10.70 & 315 & 5 & 126 $\pm$ 12  \\
UGC 5101 & CGCG 289-001 & 9 35 51.539 & +61 21 11.26 & 24\farcs01 & 14\farcs18 & 19.13 & 11.30 & 90 & 1 & 134 $\pm$ 13  \\
UGC 5189 & UGC 5189 NED01 & 9 42 53.329 & +9 29 38.77 & 10\farcs48 & 10\farcs40 & 2.49 & 2.47 & 5 & 1 & 179 $\pm$ 15  \\
 & UGC 5189 NED02 & 9 42 56.660 & +9 28 17.30 & 20\farcs39 & 4\farcs40 & 4.84 & 1.11 & 45 & 1 & 57 $\pm$ 10  \\
\enddata
\tablenotetext{1}{The position angle of the major axis defined south of east as in the {\it ds9} software.}
\end{deluxetable}

\begin{deluxetable}{ccrrrrrrrr}
\rotate
\tablecolumns{10}
\tablewidth{0pc}
\tablecaption{Outer (2$\sigma$) Ellipses for Galaxies with high S/N Observations }
\tablehead{   
\colhead{Name} 
& \colhead{RA}   
& \colhead{DEC}    
& \colhead{Major} 
& \colhead{Minor} 
& \colhead{Major} 
& \colhead{Minor} 
& \colhead{P.A.$^1$}
& \colhead{Diffuse} 
& \colhead{0.3 $-$ 1.0 keV} 
\\
\colhead{} 
& \colhead{(J2000)}   
& \colhead{(J2000)}    
& \colhead{Axis} 
& \colhead{Axis} 
& \colhead{Axis} 
& \colhead{Axis} 
& \colhead{(deg)}
& \colhead{0.3 $-$ } 
& \colhead{Surface} 
\\
\colhead{} 
& \colhead{}    
& \colhead{}    
& \colhead{Radius} 
& \colhead{Radius} 
& \colhead{Radius} 
& \colhead{Radius} 
& \colhead{}    
& \colhead{1.0 keV}
& \colhead{Brightness} 
\\
\colhead{} 
& \colhead{}    
& \colhead{}    
& \colhead{($''$)} 
& \colhead{($''$)} 
& \colhead{(kpc)} 
& \colhead{(kpc)} 
& \colhead{}
& \colhead{Counts} 
& \colhead{($\times$ 10$^{-10}$} 
\\
\colhead{} 
& \colhead{}    
& \colhead{}    
& \colhead{} 
& \colhead{} 
& \colhead{} 
& \colhead{} 
& \colhead{}
& \colhead{} 
& \colhead{Photons s$^{-1}$} 
\\
\colhead{} 
& \colhead{}    
& \colhead{}    
& \colhead{} 
& \colhead{} 
& \colhead{} 
& \colhead{} 
& \colhead{}
& \colhead{} 
& \colhead{cm$^{-2}$ arcsec$^{-2}$)} 
\\
}
\startdata
AM 2055-425 & 20 58 26.554 & -42 38 58.11 & 30\farcs20 & 13\farcs76 & 26.23 & 11.96 & 65 & 277 $\pm$ 19 & $\le$36.0  \\
Arp 091 & 15 34 32.396 & +15 11 36.55 & 15\farcs42 & 9\farcs90 & 2.54 & 1.63 & 300 & 233 $\pm$ 17 & $\le$116.0  \\
Arp 148 & 11 03 53.839 & +40 50 58.26 & 23\farcs24 & 13\farcs43 & 16.56 & 9.57 & 350 & 267 $\pm$ 18 & $\le$49.8  \\
Arp 155 & 11 23 38.541 & +53 50 30.51 & 41\farcs45 & 21\farcs08 & 9.25 & 4.70 & 350 & 167 $\pm$ 19 & $\le$19.9  \\
Arp 157 & 1 24 34.398 & +3 47 27.11 & 75\farcs10 & 24\farcs96 & 11.11 & 3.69 & 100 & 349 $\pm$ 27 & $\le$22.0  \\
Arp 160 & 12 14 09.490 & +54 31 32.96 & 49\farcs41 & 35\farcs13 & 9.35 & 6.64 & 258 & 1086 $\pm$ 36 & $\le$24.7  \\
Arp 217 & 10 38 45.964 & +53 30 08.73 & 99\farcs33 & 89\farcs22 & 8.67 & 7.79 & 230 & 5270 $\pm$ 80 & $\le$10.9  \\
Arp 220 & 15 34 57.578 & +23 30 06.07 & 54\farcs70 & 38\farcs34 & 22.02 & 15.44 & 15 & 851 $\pm$ 38 & $\le$16.6  \\
Arp 222 & 23 39 53.876 & -12 17 34.61 & 46\farcs63 & 27\farcs55 & 5.90 & 3.49 & 40 & 128 $\pm$ 16 & $\le$33.5  \\
Arp 226 & 22 20 44.774 & -24 40 41.77 & 12\farcs13 & 9\farcs67 & 3.94 & 3.14 & 30 & 217 $\pm$ 16 & $\le$106.6  \\
Arp 236 & 01 07 46.897 & -17 30 26.91 & 28\farcs52 & 28\farcs43 & 11.20 & 11.17 & 90 & 1652 $\pm$ 42 & $\le$18.1  \\
Arp 244 & 12 01 53.451 & -18 52 26.50 & 143\farcs52 & 112\farcs32 & 16.78 & 13.13 & 50 & 55121 $\pm$ 292 & $\le$2.4  \\
Arp 270 & 10 49 50.028 & +32 58 55.07 & 42\farcs63 & 31\farcs73 & 60 & 4.46 & 315 & 319 $\pm$ 24 & $\le$32.4  \\
Arp 284 & 23 36 14.310 & +2 09 14.05 & 57\farcs93 & 27\farcs07 & 10.96 & 5.12 & 349 & 1024 $\pm$ 36 & $\le$25.8  \\
Arp 293 & 16 58 31.581 & +58 56 15.77 & 32\farcs78 & 22\farcs99 & 13.03 & 9.14 & 231 & 262 $\pm$ 18 & $\le$59.6  \\
Arp 299 & 11 28 32.174 & +58 33 46.26 & 80\farcs72 & 54\farcs35 & 18.79 & 12.65 & 330 & 7434 $\pm$ 94 & $\le$13.0  \\
IRAS 17208-0014 & 17 23 21.953 & -0 17 01.90 & 23\farcs32 & 12\farcs69 & 20.69 & 11.27 & 15 & 101 $\pm$ 15 & $\le$29.4  \\
IRAS 23128-5919 & 23 15 46.922 & -59 03 14.16 & 11\farcs61 & 9\farcs10 & 10.36 & 8.12 & 260 & 214 $\pm$ 16 & $\le$55.7  \\
Mrk 231 & 12 56 14.547 & +56 52 24.67 & 65\farcs29 & 57\farcs46 & 56.40 & 49.64 & 105 & 5375 $\pm$ 109 & $\le$8.6  \\
Mrk 273 & 13 44 42.002 & +55 53 12.22 & 64\farcs48 & 31\farcs53 & 50.19 & 24.54 & 100 & 1212 $\pm$ 38 & $\le$18.8  \\
NGC 1700 & 04 56 56.452 & -04 51 59.57 & 102\farcs00 & 71\farcs86 & 25.97 & 18.30 & 20 & 2090 $\pm$ 62 & $\le$11.9  \\
NGC 2207/IC 2163 & 06 16 22.134 & -21 22 21.18 & 119\farcs45 & 68\farcs50 & 22.01 & 12.63 & 15 & 1401 $\pm$ 52 & $\le$12.0  \\
NGC 3256 & 10 27 51.096 & -43 54 09.37 & 54\farcs95 & 52\farcs53 & 9.86 & 9.43 & 0 & 5918 $\pm$ 84 & $\le$16.6  \\
NGC 5018 & 13 13 01.040 & -19 31 04.39 & 23\farcs09 & 13\farcs44 & 4.30 & 2.50 & 5 & 283 $\pm$ 18 & $\le$53.9  \\
NGC 5256 & 13 38 18.085 & +48 16 38.52 & 51\farcs99 & 42\farcs34 & 30.49 & 24.82 & 90 & 672 $\pm$ 29 & $\le$29.9  \\
NGC 6240 & 16 52 58.527 & +2 24 04.36 & 90\farcs35 & 62\farcs91 & 47.67 & 33.20 & 120 & 12936 $\pm$ 141 & $\le$8.3  \\
NGC 7592 & 23 18 22.276 & -4 24 57.89 & 31\farcs56 & 29\farcs55 & 15.23 & 14.26 & 315 & 124 $\pm$ 13 & $\le$37.1  \\
\enddata
\tablenotetext{1}{The position angle of the major axis defined south of east as in the {\it ds9} software.}
\end{deluxetable}

\begin{deluxetable}{rcccccc}
\tabletypesize \scriptsize
\tablecolumns{7}
\tablewidth{0pc}
\tablecaption{Correlations and Anti-Correlations}
\tablehead{   
\colhead{Fig}
&\colhead{SFR} 
&\colhead{CO/H$_2$} 
&\colhead{Relation} 
& \colhead{rms} 
& \colhead{Spear/} 
& \colhead{Correl.} 
\\ 
\colhead{Num}
&\colhead{Range}
&\colhead{Ratio}
&\colhead{}
& \colhead{}
& \colhead{Pearson}
& \colhead{} 
\\ 
\colhead{}
&\colhead{}
&\colhead{}
&\colhead{}
& \colhead{}
& \colhead{Coeff.}
& \colhead{} 
}
\startdata
\multicolumn{6}{c}{Basic Relations} \\
\hline \\
2  &  all  &   & LOG SFR = (0.79 $\pm$ 0.21) LOG L$_{\rm K}$  $-$  (7.75 $\pm$ 2.29) & 0.76 & 0.47/0.51 & weak \\
2  &  $>$1 $\frac{M_{\sun}}{yr}$  &   & LOG SFR = (0.84 $\pm$ 0.19) LOG L$_{\rm K}$  $-$  (8.05 $\pm$ 2.09) & 0.48 & 0.51/0.62 & weak \\
2  &  all  &   & LOG SFR = (0.88 $\pm$ 0.11) LOG (L$_{\rm FIR}$/L$_{\rm K}$)  +  (1.33 $\pm$ 0.1) & 0.56 & 0.79/0.77 & strong  \\
2  &  $>$1 $\frac{M_{\sun}}{yr}$  &   & LOG SFR = (0.95 $\pm$ 0.16) LOG (L$_{\rm FIR}$/L$_{\rm K}$)  +  (1.46 $\pm$ 0.08) & 0.42 & 0.75/0.72 & strong  \\
2  &  all  &   & LOG SFR = (0.41 $\pm$ 0.05) ([3.6] $-$ [24])  $-$  (1.74 $\pm$ 0.33) & 0.53 & 0.81/0.80 & strong  \\
2  &  $>$1 $\frac{M_{\sun}}{yr}$  &   & LOG SFR = (0.41 $\pm$ 0.07) ([3.6] $-$ [24])  $-$  (1.62 $\pm$ 0.5) & 0.43 & 0.76/0.72 & strong  \\
2  &  all  &   & LOG SFR = (3.46 $\pm$ 0.76) LOG (F$_{60}$/F$_{100}$)  +  (1.49 $\pm$ 0.15) & 0.70 & 0.60/0.58 & strong  \\
2  &  $>$1 $\frac{M_{\sun}}{yr}$  &   & LOG SFR = (2.17 $\pm$ 0.69) LOG (F$_{60}$/F$_{100}$)  +  (1.55 $\pm$ 0.12) & 0.54 & 0.50/0.48 & weak \\
4  &  all  &  Const & LOG SFE = (2.54 $\pm$ 0.47) LOG (F$_{60}$/F$_{100}$)  $-$  (8.12 $\pm$ 0.09) & 0.42 & 0.72/0.67 & strong  \\
4  &  $>$1 $\frac{M_{\sun}}{yr}$  &  Const & LOG SFE = (2.93 $\pm$ 0.48) LOG (F$_{60}$/F$_{100}$)  $-$  (8.13 $\pm$ 0.08) & 0.37 & 0.76/0.74 & strong  \\
4  &  all  &  Var & LOG SFE = (3.38 $\pm$ 0.55) LOG (F$_{60}$/F$_{100}$)  $-$  (7.68 $\pm$ 0.11) & 0.49 & 0.73/0.72 & strong  \\
4  &  $>$1 $\frac{M_{\sun}}{yr}$  &  Var & LOG SFE = (3.60 $\pm$ 0.63) LOG (F$_{60}$/F$_{100}$)  $-$  (7.66 $\pm$ 0.11) & 0.49 & 0.71/0.71 & strong  \\
4  &  all  &  Const & LOG SFE = (0.23 $\pm$ 0.11) LOG SFR  $-$  (8.72 $\pm$ 0.16) & 0.53 & 0.40/0.33 & weak \\
4  &  $>$1 $\frac{M_{\sun}}{yr}$  &  Const & LOG SFE = (0.37 $\pm$ 0.14) LOG SFR  $-$  (8.95 $\pm$ 0.21) & 0.50 & 0.44/0.42 & weak \\
4  &  all  &  Var & LOG SFE = (0.61 $\pm$ 0.10) LOG SFR  $-$  (8.82 $\pm$ 0.15) & 0.50 & 0.75/0.71 & strong  \\
4  &  $>$1 $\frac{M_{\sun}}{yr}$  &  Var & LOG SFE = (0.88 $\pm$ 0.13) LOG SFR  $-$  (9.24 $\pm$ 0.19) & 0.43 & 0.77/0.78 & strong  \\
\\
\hline
\multicolumn{6}{c}{Comparisons with Volume} \\
\hline \\
9  &  all  &   & LOG VOLUME = (0.66 $\pm$ 0.10) LOG SFR  +  (67.01 $\pm$ 0.14) & 0.59 & 0.75/0.70 & strong  \\
9  &  $>$1 $\frac{M_{\sun}}{yr}$  &   & LOG VOLUME = (0.97 $\pm$ 0.15) LOG SFR  +  (66.54 $\pm$ 0.22) & 0.52 & 0.79/0.75 & strong  \\
10  &  all  &  Const & LOG VOLUME vs.\ LOG SFE & 0.85 & 0.16/0.06 & none \\
10  &  all  &  Var & LOG VOLUME = (0.57 $\pm$ 0.18) LOG SFE  +  (72.36 $\pm$ 1.48) & 0.75 & 0.53/0.47 & weak \\
10  &  all  &   & LOG VOLUME vs.\ LOG (F$_{60}$/F$_{100}$) & 0.79 & 0.34/0.29 & none \\
10  &  $>$1 $\frac{M_{\sun}}{yr}$  &   & LOG VOLUME vs.\ LOG (F$_{60}$/F$_{100}$) & 0.78 & 0.20/0.11 & none \\
11  &  all  &   & LOG VOLUME = (0.80 $\pm$ 0.12) LOG L$_{\rm X}$(gas)  +  (35.11 $\pm$ 4.85) & 0.57 & 0.79/0.72 & strong  \\
11  &  all  &   & LOG VOLUME = (-1.19 $\pm$ 0.39) LOG n$_{\rm e}$  +  (64.95 $\pm$ 0.9) & 0.75 & -0.39/-0.43 & weak anti \\
11  &  all  &   & LOG VOLUME = (0.85 $\pm$ 0.18) LOG L$_{\rm K}$  +  (58.23 $\pm$ 2.01) & 0.67 & 0.65/0.59 & strong  \\
11  &  $>$1 $\frac{M_{\sun}}{yr}$  &   & LOG VOLUME = (1.15 $\pm$ 0.23) LOG L$_{\rm K}$  +  (54.92 $\pm$ 2.58) & 0.60 & 0.65/0.66 & strong  \\
11  &  all  &   & LOG VOLUME = (0.40 $\pm$ 0.15) LOG (L$_{\rm FIR}$/L$_{\rm K}$)  +  (67.81 $\pm$ 0.13) & 0.77 & 0.49/0.37 & weak \\
none  &  all  &   & LOG VOLUME = (0.17 $\pm$ 0.07) ([3.6] $-$ [24])  +  (66.5 $\pm$ 0.48) & 0.77 & 0.44/0.36 & weak \\
none  &  all  &   & LOG (VOLUME/SFR) vs.\ LOG SFR & 0.59 & -0.34/-0.46 & none \\
none  &  $>$1 $\frac{M_{\sun}}{yr}$  &   & LOG (VOLUME/SFR) vs.\ LOG SFR & 0.52 & 0.02/-0.04 & none \\
none  &  all  &   & LOG (VOLUME/L$_{\rm K}$) = (0.32 $\pm$ 0.11) LOG SFR  +  (56.27 $\pm$ 0.14) & 0.61 & 0.40/0.42 & weak \\
none  &  $>$1 $\frac{M_{\sun}}{yr}$  &   & LOG (VOLUME/L$_{\rm K}$) = (0.52 $\pm$ 0.15) LOG SFR  +  (55.97 $\pm$ 0.21) & 0.51 & 0.51/0.53 & weak \\
none  &  all  &   & LOG (VOLUME/L$_{\rm K}$) vs.\ LOG L$_{\rm K}$ & 0.67 & 0.02/-0.13 & none \\
none  &  $>$1 $\frac{M_{\sun}}{yr}$  &   & LOG (VOLUME/L$_{\rm K}$) vs.\ LOG L$_{\rm K}$ & 0.60 & 0.16/0.12 & none \\
\\
\hline
\multicolumn{6}{c}{Comparisons with LOG (L$_{\rm X}$(gas)/SFR)} \\
\hline \\
none  &  all  &   & LOG (L$_{\rm X}$(gas)/SFR) vs.\ LOG SFR & 0.40 & -0.35/-0.53 & none \\
none  &  $>$1 $\frac{M_{\sun}}{yr}$  &   & LOG (L$_{\rm X}$(gas)/SFR) vs.\ LOG SFR & 0.30 & -0.35/-0.48 & none \\
none  &  all  &   & LOG (L$_{\rm X}$(gas)/SFR) = (-1.69 $\pm$ 0.37) LOG (F$_{60}$/F$_{100}$)  +  (39.6 $\pm$ 0.07) & 0.34 & -0.53/-0.58 & weak anti \\
none  &  $>$1 $\frac{M_{\sun}}{yr}$  &   & LOG (L$_{\rm X}$(gas)/SFR) = (-1.28 $\pm$ 0.38) LOG (F$_{60}$/F$_{100}$)  +  (39.64 $\pm$ 0.07) & 0.30 & -0.45/-0.50 & weak anti \\
none  &  all  &  Const & LOG (L$_{\rm X}$(gas)/SFR) = (-0.40 $\pm$ 0.10) LOG SFE  +  (36.39 $\pm$ 0.81) & 0.32 & -0.55/-0.58 & weak anti \\
none  &  $>$1 $\frac{M_{\sun}}{yr}$  &  Const & LOG (L$_{\rm X}$(gas)/SFR) = (-0.36 $\pm$ 0.09) LOG SFE  +  (36.76 $\pm$ 0.8) & 0.29 & -0.49/-0.56 & weak anti \\
none  &  all  &  Var & LOG (L$_{\rm X}$(gas)/SFR) = (-0.32 $\pm$ 0.08) LOG SFE  +  (37.22 $\pm$ 0.63) & 0.32 & -0.55/-0.57 & weak anti \\
none  &  $>$1 $\frac{M_{\sun}}{yr}$  &  Var & LOG (L$_{\rm X}$(gas)/SFR) = (-0.27 $\pm$ 0.07) LOG SFE  +  (37.58 $\pm$ 0.61) & 0.29 & -0.52/-0.55 & weak anti \\
\\
\hline
\multicolumn{6}{c}{Comparisons with LOG M$_{\rm X}$(gas) and LOG M$_{\rm X}$(gas)/SFR} \\
\hline \\
13  &  all  &   & LOG M$_{\rm X}$(gas) = (0.70 $\pm$ 0.07) LOG SFR  +  (7.6 $\pm$ 0.1) & 0.42 & 0.86/0.83 & strong  \\
13  &  $>$1 $\frac{M_{\sun}}{yr}$  &   & LOG M$_{\rm X}$(gas) = (0.88 $\pm$ 0.10) LOG SFR  +  (7.35 $\pm$ 0.14) & 0.34 & 0.86/0.84 & strong  \\
13  &  all  &   & LOG M$_{\rm X}$(gas) = (0.94 $\pm$ 0.14) LOG L$_{\rm K}$  $-$  (2.05 $\pm$ 1.58) & 0.53 & 0.69/0.72 & strong  \\
13  &  $>$1 $\frac{M_{\sun}}{yr}$  &   & LOG M$_{\rm X}$(gas) = (1.00 $\pm$ 0.17) LOG L$_{\rm K}$  $-$  (2.68 $\pm$ 1.94) & 0.45 & 0.67/0.71 & strong  \\
13  &  all  &   & LOG (M$_{\rm X}$(gas)/SFR) = (-0.30 $\pm$ 0.07) LOG SFR  +  (7.6 $\pm$ 0.1) & 0.42 & -0.38/-0.53 & weak anti \\
13  &  $>$1 $\frac{M_{\sun}}{yr}$  &   & LOG (M$_{\rm X}$(gas)/SFR) vs.\ LOG SFR & 0.34 & -0.12/-0.22 & none \\
13  &  all  &   & LOG (M$_{\rm X}$(gas)/SFR) vs.\ LOG L$_{\rm K}$ & 0.49 & 0.28/0.17 & none \\
13  &  $>$1 $\frac{M_{\sun}}{yr}$  &   & LOG (M$_{\rm X}$(gas)/SFR) = (0.16 $\pm$ 0.13) LOG L$_{\rm K}$  +  (5.37 $\pm$ 1.48) & 0.34 & 0.37/0.21 & weak \\
14  &  all  &  Const & LOG (M$_{\rm X}$(gas)/SFR) = (-0.37 $\pm$ 0.10) LOG SFE  +  (4.06 $\pm$ 0.84) & 0.33 & -0.52/-0.54 & weak anti \\
14  &  $>$1 $\frac{M_{\sun}}{yr}$  &  Const & LOG (M$_{\rm X}$(gas)/SFR) = (-0.33 $\pm$ 0.10) LOG SFE  +  (4.38 $\pm$ 0.85) & 0.31 & -0.45/-0.51 & weak anti \\
14  &  all  &  Var & LOG (M$_{\rm X}$(gas)/SFR) = (-0.26 $\pm$ 0.08) LOG SFE  +  (5.11 $\pm$ 0.68) & 0.35 & -0.42/-0.46 & weak anti \\
14  &  $>$1 $\frac{M_{\sun}}{yr}$  &  Var & LOG (M$_{\rm X}$(gas)/SFR) vs.\ LOG SFE & 0.33 & -0.34/-0.39 & none \\
14  &  all  &   & LOG (M$_{\rm X}$(gas)/SFR) = (-0.22 $\pm$ 0.03) ([3.6] $-$ [24])  +  (8.78 $\pm$ 0.19) & 0.31 & -0.66/-0.77 & strong anti \\
14  &  $>$1 $\frac{M_{\sun}}{yr}$  &   & LOG (M$_{\rm X}$(gas)/SFR) = (-0.17 $\pm$ 0.05) ([3.6] $-$ [24])  +  (8.41 $\pm$ 0.35) & 0.30 & -0.50/-0.52 & weak anti \\
14  &  all  &   & LOG (M$_{\rm X}$(gas)/SFR) = (-0.45 $\pm$ 0.07) LOG (L$_{\rm FIR}$/L$_{\rm K}$)  +  (7.13 $\pm$ 0.06) & 0.35 & -0.51/-0.71 & weak anti \\
14  &  $>$1 $\frac{M_{\sun}}{yr}$  &   & LOG (M$_{\rm X}$(gas)/SFR) vs.\ LOG (L$_{\rm FIR}$/L$_{\rm K}$) & 0.33 & -0.28/-0.33 & none \\
14  &  all  &   & LOG (M$_{\rm X}$(gas)/SFR) = (-1.66 $\pm$ 0.34) LOG (F$_{60}$/F$_{100}$)  +  (7.04 $\pm$ 0.07) & 0.31 & -0.59/-0.61 & strong anti \\
14  &  $>$1 $\frac{M_{\sun}}{yr}$  &   & LOG (M$_{\rm X}$(gas)/SFR) = (-1.33 $\pm$ 0.38) LOG (F$_{60}$/F$_{100}$)  +  (7.04 $\pm$ 0.07) & 0.30 & -0.48/-0.51 & weak anti \\
14  &  all  &   & LOG (M$_{\rm X}$(gas)/SFR) vs.\ LOG n$_{\rm e}$ & 0.46 & -0.34/-0.38 & none \\
14  &  $>$1 $\frac{M_{\sun}}{yr}$  &   & LOG (M$_{\rm X}$(gas)/SFR) = (-0.55 $\pm$ 0.20) LOG n$_{\rm e}$  +  (5.97 $\pm$ 0.46) & 0.32 & -0.39/-0.42 & weak anti \\
none  &  all  &   & LOG (M$_{\rm X}$(gas)/L$_{\rm K}$) = (0.53 $\pm$ 0.07) LOG (SFR/L$_{\rm K}$)  +  (2.54 $\pm$ 0.68) & 0.33 & 0.77/0.77 & strong  \\
none  &  $>$1 $\frac{M_{\sun}}{yr}$  &   & LOG (M$_{\rm X}$(gas)/L$_{\rm K}$) = (0.67 $\pm$ 0.11) LOG (SFR/L$_{\rm K}$)  +  (3.91 $\pm$ 1.09) & 0.31 & 0.73/0.73 & strong  \\
none  &  all  &   & LOG (M$_{\rm X}$(gas)/L$_{\rm K}$) = (0.37 $\pm$ 0.07) LOG SFR  $-$  (3.13 $\pm$ 0.1) & 0.41 & 0.63/0.73 & strong  \\
none  &  $>$1 $\frac{M_{\sun}}{yr}$  &   & LOG (M$_{\rm X}$(gas)/L$_{\rm K}$) = (0.42 $\pm$ 0.10) LOG SFR  $-$  (3.22 $\pm$ 0.15) & 0.37 & 0.62/0.73 & strong  \\
none  &  all  &   & LOG (M$_{\rm X}$(gas)/L$_{\rm K}$) vs.\ LOG L$_{\rm K}$ & 0.52 & 0.01/-0.07 & none \\
none  &  $>$1 $\frac{M_{\sun}}{yr}$  &   & LOG (M$_{\rm X}$(gas)/L$_{\rm K}$) vs.\ LOG L$_{\rm K}$ & 0.45 & 0.04/0.00 & none \\
none  &  all  &   & LOG (M$_{\rm X}$(gas)/SFR) = (0.47 $\pm$ 0.07) LOG (L$_{\rm K}$/SFR)  +  (2.54 $\pm$ 0.68) & 0.33 & 0.58/0.74 & strong  \\
none  &  $>$1 $\frac{M_{\sun}}{yr}$  &   & LOG (M$_{\rm X}$(gas)/SFR) = (0.33 $\pm$ 0.11) LOG (L$_{\rm K}$/SFR)  +  (3.91 $\pm$ 1.09) & 0.31 & 0.41/0.46 & weak \\
none  &  all  &  Const & LOG (M$_{\rm X}$(gas)/SFR) vs.\ LOG (M$_{\rm HI}$/M$_{\rm H_2}$) & 0.48 & -0.30/-0.27 & none \\
none  &  all  &  Var & LOG (M$_{\rm X}$(gas)/SFR) vs.\ LOG (M$_{\rm HI}$/M$_{\rm H_2}$) & 0.48 & -0.31/-0.27 & none \\
none  &  all  &   & LOG (M$_{\rm X}$(gas)/SFR) vs.\ LOG VOLUME & 0.49 & 0.17/0.16 & none \\
\\
\hline
\multicolumn{6}{c}{Comparisons with LOG (M$_{\rm X}$(gas)/(M$_{\rm H_2}$+M$_{\rm HI}$))} \\
\hline \\
15  &  all  &  Const & LOG (M$_{\rm X}$(gas)/(M$_{\rm H_2}$+M$_{\rm HI}$)) = (0.25 $\pm$ 0.10) LOG SFR  $-$  (2.05 $\pm$ 0.13) & 0.41 & 0.50/0.42 & weak \\
15  &  $>$1 $\frac{M_{\sun}}{yr}$  &  Const & LOG (M$_{\rm X}$(gas)/(M$_{\rm H_2}$+M$_{\rm HI}$)) = (0.58 $\pm$ 0.14) LOG SFR  $-$  (2.49 $\pm$ 0.19) & 0.37 & 0.71/0.65 & strong  \\
15  &  all  &  Var & LOG (M$_{\rm X}$(gas)/(M$_{\rm H_2}$+M$_{\rm HI}$)) = (0.40 $\pm$ 0.12) LOG SFR  $-$  (2.07 $\pm$ 0.15) & 0.48 & 0.65/0.54 & strong  \\
15  &  $>$1 $\frac{M_{\sun}}{yr}$  &  Var & LOG (M$_{\rm X}$(gas)/(M$_{\rm H_2}$+M$_{\rm HI}$)) = (0.82 $\pm$ 0.16) LOG SFR  $-$  (2.64 $\pm$ 0.21) & 0.41 & 0.80/0.73 & strong  \\
15  &  all  &  Const & LOG (M$_{\rm X}$(gas)/(M$_{\rm H_2}$+M$_{\rm HI}$)) vs.\ LOG SFE & 0.43 & 0.35/0.32 & none \\
15  &  $>$1 $\frac{M_{\sun}}{yr}$  &  Const & LOG (M$_{\rm X}$(gas)/(M$_{\rm H_2}$+M$_{\rm HI}$)) = (0.35 $\pm$ 0.17) LOG SFE  +  (1.21 $\pm$ 1.45) & 0.45 & 0.48/0.40 & weak \\
15  &  all  &  Var & LOG (M$_{\rm X}$(gas)/(M$_{\rm H_2}$+M$_{\rm HI}$)) = (0.52 $\pm$ 0.13) LOG SFE  +  (2.6 $\pm$ 1.07) & 0.45 & 0.61/0.61 & strong  \\
15  &  $>$1 $\frac{M_{\sun}}{yr}$  &  Var & LOG (M$_{\rm X}$(gas)/(M$_{\rm H_2}$+M$_{\rm HI}$)) = (0.60 $\pm$ 0.14) LOG SFE  +  (3.21 $\pm$ 1.16) & 0.45 & 0.70/0.66 & strong  \\
16  &  all  &  Const & LOG (M$_{\rm X}$(gas)/(M$_{\rm H_2}$+M$_{\rm HI}$)) vs.\ LOG L$_{\rm K}$ & 0.43 & 0.31/0.33 & none \\
16  &  $>$1 $\frac{M_{\sun}}{yr}$  &  Const & LOG (M$_{\rm X}$(gas)/(M$_{\rm H_2}$+M$_{\rm HI}$)) vs.\ LOG L$_{\rm K}$ & 0.46 & 0.31/0.33 & none \\
16  &  all  &  Var & LOG (M$_{\rm X}$(gas)/(M$_{\rm H_2}$+M$_{\rm HI}$)) = (0.49 $\pm$ 0.22) LOG L$_{\rm K}$  $-$  (7.12 $\pm$ 2.43) & 0.52 & 0.41/0.40 & weak \\
16  &  $>$1 $\frac{M_{\sun}}{yr}$  &  Var & LOG (M$_{\rm X}$(gas)/(M$_{\rm H_2}$+M$_{\rm HI}$)) = (0.60 $\pm$ 0.29) LOG L$_{\rm K}$  $-$  (8.29 $\pm$ 3.23) & 0.56 & 0.40/0.39 & weak \\
16  &  all  &  Const & LOG (M$_{\rm X}$(gas)/(M$_{\rm H_2}$+M$_{\rm HI}$)) vs.\ LOG (F$_{60}$/F$_{100}$) & 0.44 & 0.29/0.24 & none \\
16  &  $>$1 $\frac{M_{\sun}}{yr}$  &  Const & LOG (M$_{\rm X}$(gas)/(M$_{\rm H_2}$+M$_{\rm HI}$)) = (1.02 $\pm$ 0.66) LOG (F$_{60}$/F$_{100}$)  $-$  (1.68 $\pm$ 0.12) & 0.46 & 0.44/0.31 & weak \\
16  &  all  &  Var & LOG (M$_{\rm X}$(gas)/(M$_{\rm H_2}$+M$_{\rm HI}$)) vs.\ LOG (F$_{60}$/F$_{100}$) & 0.54 & 0.34/0.28 & none \\
16  &  $>$1 $\frac{M_{\sun}}{yr}$  &  Var & LOG (M$_{\rm X}$(gas)/(M$_{\rm H_2}$+M$_{\rm HI}$)) = (1.32 $\pm$ 0.82) LOG (F$_{60}$/F$_{100}$)  $-$  (1.5 $\pm$ 0.15) & 0.57 & 0.41/0.32 & weak \\
17  &  all  &  Const & LOG (M$_{\rm X}$(gas)/(M$_{\rm H_2}$+M$_{\rm HI}$)) vs.\ ([3.6] $-$ [24]) & 0.44 & 0.35/0.23 & none \\
17  &  $>$1 $\frac{M_{\sun}}{yr}$  &  Const & LOG (M$_{\rm X}$(gas)/(M$_{\rm H_2}$+M$_{\rm HI}$)) = (0.20 $\pm$ 0.08) ([3.6] $-$ [24])  $-$  (3.21 $\pm$ 0.61) & 0.44 & 0.54/0.44 & weak \\
17  &  all  &  Var & LOG (M$_{\rm X}$(gas)/(M$_{\rm H_2}$+M$_{\rm HI}$)) = (0.11 $\pm$ 0.07) ([3.6] $-$ [24])  $-$  (2.44 $\pm$ 0.46) & 0.54 & 0.44/0.31 & weak \\
17  &  $>$1 $\frac{M_{\sun}}{yr}$  &  Var & LOG (M$_{\rm X}$(gas)/(M$_{\rm H_2}$+M$_{\rm HI}$)) = (0.26 $\pm$ 0.10) ([3.6] $-$ [24])  $-$  (3.55 $\pm$ 0.74) & 0.53 & 0.55/0.47 & weak \\
17  &  all  &  Const & LOG (M$_{\rm X}$(gas)/(M$_{\rm H_2}$+M$_{\rm HI}$)) = (0.19 $\pm$ 0.13) LOG (L$_{\rm FIR}$/L$_{\rm K}$)  $-$  (1.73 $\pm$ 0.1) & 0.44 & 0.43/0.27 & weak \\
17  &  $>$1 $\frac{M_{\sun}}{yr}$  &  Const & LOG (M$_{\rm X}$(gas)/(M$_{\rm H_2}$+M$_{\rm HI}$)) = (0.53 $\pm$ 0.20) LOG (L$_{\rm FIR}$/L$_{\rm K}$)  $-$  (1.68 $\pm$ 0.1) & 0.42 & 0.61/0.49 & strong  \\
17  &  all  &  Var & LOG (M$_{\rm X}$(gas)/(M$_{\rm H_2}$+M$_{\rm HI}$)) = (0.33 $\pm$ 0.16) LOG (L$_{\rm FIR}$/L$_{\rm K}$)  $-$  (1.56 $\pm$ 0.12) & 0.53 & 0.54/0.36 & weak \\
17  &  $>$1 $\frac{M_{\sun}}{yr}$  &  Var & LOG (M$_{\rm X}$(gas)/(M$_{\rm H_2}$+M$_{\rm HI}$)) = (0.74 $\pm$ 0.24) LOG (L$_{\rm FIR}$/L$_{\rm K}$)  $-$  (1.49 $\pm$ 0.12) & 0.51 & 0.65/0.54 & strong  \\
\enddata
\end{deluxetable}

\end{document}